\documentclass[11pt]{article}
\usepackage{latexsym,amssymb,amstext,amsmath}
\usepackage{hyperref}
\allowdisplaybreaks 
\def\Slash#1{\rlap{\hbox{$\mskip 3 mu /$}}#1}      
\def\oneone{\rlap 1\mkern4mu{\rm l}} 

\begin{document}
%
\begin{titlepage}
\begin{flushright} \small
 Nikhef-2014-052 \\ ITP-UU-14/30 \\CPHT-RR097.1214
\end{flushright}
\bigskip

\begin{center}
 {\LARGE\bfseries  IIB Supergravity and the \\[2mm] 
   $\boldsymbol{\mathrm{E}_{6(6)}}$ covariant vector-tensor hierarchy} 
\\[10mm]
\textbf{Franz Ciceri$^a$, Bernard de Wit$^{a,b}$ and Oscar Varela$^{c,d}$}\\[5mm] 
\vskip 4mm
$^a${\em Nikhef Theory Group, Science Park 105, 1098 XG Amsterdam, The
  Netherlands}\\
$^b${\em Institute for Theoretical Physics, Utrecht
  University,} \\
  {\em Leuvenlaan 4, 3584 CE Utrecht, The Netherlands}\\
  $^c${\em Center for the Fundamental Laws of Nature,}\\
  {\em Harvard University, Cambridge, MA 02138, USA}  \\
  $^d${\em Centre de Physique Th\'eorique, Ecole Polytechnique, CNRS UMR 7644,}\\
  {\em 91128 Palaiseau Cedex, France}  \\[3mm]
 {\tt f.ciceri@nikhef.nl}\,,\;{\tt B.deWit@uu.nl}\,,\;{\tt ovarela@physics.harvard.edu} 
\end{center}

\vspace{3ex}

\begin{center}
{\bfseries Abstract}
\end{center}
\begin{quotation} \noindent IIB supergravity is reformulated with a
  manifest local $\mathrm{USp}(8)$ invariance that makes the embedding
  of five-dimensional maximal supergravities transparent. In this
  formulation the ten-dimensional theory exhibits all the 27 one-form
  fields and 22 of the 27 two-form fields that are required by the
  vector-tensor hierarchy of the five-dimensional theory. The missing
  5 two-form fields must transform in the same representation as a
  descendant of the ten-dimensional `dual graviton'.  The invariant
  $\mathrm{E}_{6(6)}$ symmetric tensor that appears in the
  vector-tensor hierarchy is reproduced. Generalized vielbeine are
  derived from the supersymmetry transformations of the vector fields,
  as well as consistent expressions for the $\mathrm{USp}(8)$
  covariant fermion fields. Implications are discussed for the
  consistency of the truncation of IIB supergravity compactified on
  the five-sphere to maximal gauged supergravity in five space-time
  dimensions with an $\mathrm{SO}(6)$ gauge group.
 \end{quotation}

\vfill

\end{titlepage}

\section{Introduction}
\label{sec:introduction}
\setcounter{equation}{0}
Maximal supergravity theories in various dimensions are known to
possess intruiging duality symmetries which can optionally be broken
by non-abelian gauge interactions. Many of these theories can be
described as truncations from eleven-dimensional supergravity or from
ten-dimensional IIB supergravity in the context of dimensional
compactification on an internal manifold of appropriate
dimensionality. Already at an early stage this raised the question
whether the higher-dimensional supergravities might somehow reflect
the exceptional duality symmetries that are present in their
lower-dimensional `descendants'. This question has a long history and
is also relevant for proving the existence of consistent truncations
to maximal supergravities, implying that any solution of the
lower-dimensional maximal supergravity can be uplifted to the
higher-dimensional one.

An early attempt to answer this question was based on a reformulation
of the full eleven-dimensional supergravity obtained by performing a
suitable Kaluza-Klein decomposition to four dimensions while retaining
the full dependence on the seven internal coordinates
\cite{deWit:1986mz}. The key element here was to ensure that the
resulting theory was invariant under the four-dimensional R-symmetry
group $\mathrm{SU}(8)$. This symmetry was {\it locally} realized with
respect to all the eleven coordinates, and it was introduced by a
gauge equivalent re-assembling of the original $\mathrm{Spin}(10,1)$
tangent space. The resulting supersymmetry transformation rules then
took a form that was almost identical to the four-dimensional ones,
which do indeed exhibit the typical characteristics of the
$\mathrm{E}_{7(7)}$ dualities, but now with fields that still depend
on all eleven space-time coordinates. Eventually this set-up made it
possible to establish the consistency of the $S^7$ truncation, meaning
that the whole field configuration of four-dimensional
$\mathrm{SO}(8)$ gauged supergravity can be uplifted as a submanifold
in the full eleven-dimensional theory by specifying the dependence of
the fields on the seven internal coordinates
\cite{deWit:1986iy,Nicolai:2011cy} .

Recently this approach was substantially extended by including the
supersymmetry transformations of dual fields, which opened a new
window to accessing the $\mathrm{E}_{7(7)}$ duality properties of the
full eleven-dimensional supergravity
\cite{deWit:2013ija,Godazgar:2013nma,Godazgar:2013dma,Godazgar:2013pfa}. Given
these recent insights, it is a natural question whether similar
structures can be derived for IIB supergravity in the context of a
$5+5$ split of the coordinates. In the present paper we confirm that
this is indeed the case and we present a detailed analysis to support
this. Qualitatively the results turn out to be rather similar to the
case of eleven-dimensional supergravity, but many new features
arise. In this case the tangent space is re-assembled such that the
theory is manifestly invariant under {\it local}
$\mathrm{USp}(8)$. This group contains the $\mathrm{USp}(4)$ subgroup
of the $10D$ tangent space group and the explicit $\mathrm{U}(1)$ of
IIB supergravity as subgroups. Obviously the
$\mathrm{SU}(1,1)\cong\mathrm{SL}(2)$ subgroup of $\mathrm{E}_{6(6)}$
is manifestly realized from the start. Another interesting aspect is
that the five-dimensional gauged supergravity theories, when described
in terms of the embedding tensor formalism \cite{deWit:2004nw},
involve 27 vector and 27 two-rank tensor fields which constitute the
beginning of an intricate vector-tensor hierarchy
\cite{deWit:2005hv,deWit:2008ta}. As we will discover in this paper,
these features are also present when one retains the dependence on the
extra internal coordinates for IIB supergravity, so that this
vector-tensor hierarchy does emerge in a ten-dimensional context. This
is undoubtedly related to the fact that in the recent work on an
$\mathrm{E}_{6(6)}$ {\it exceptional geometry} that incorporates both
11-dimensional and 10-dimensional IIB supergravity, the vector-tensor
hierarchy also plays a key role \cite{Hohm:2013vpa}. Irrespective of
these issues, the analysis presented in this paper has to address a
number of subtle technical complications that are absent in the
corresponding analysis of eleven-dimensional supergravity. Many of
those are caused by the fact that the field representation of IIB
supergravity is more reducible than that of the eleven-dimensional
one, while the supersymmetry is an extended one (i.e. $N=2$).

While it is clearly significant that the approach initiated in
\cite{deWit:1986mz} can be applied successfully to IIB supergravity,
we should also point out that a wider variety of alternative
approaches has been developed meanwhile. These approaches are also
aimed at understanding and/or exploiting the duality symmetries in the
context of M-theory and string theory, and sometimes involve
substantial extensions of the conventional supergravity
framework. Some of them make use of additional space-time coordinates
and extended geometrical structures or duality groups. One such
approach is based on {\it generalized geometry}
\cite{Hitchin:2004ut,Gualtieri:2003dx}, where one considers an
extended tangent space that captures all the bosonic degrees of
freedom, sometimes related to double field theory (see
e.g. \cite{Siegel:1993xq,Hull:2007zu,Hull:2009mi,Berman:2010is,Hohm:2010xe,
Coimbra:2011nw,  Berman:2011jh,Lee:2014mla}  and references quoted therein). There
exists also a variety of extended duality groups that have been
proposed in combination with a choice of an exceptional space-time,
such as in \cite{West:2001as,Damour:2002cu,Damour:2006xu,West:2011mm}.
The work in \cite{Hillmann:2009ci,Hohm:2013pua,Hohm:2013vpa,
  Hohm:2013uia} is based on extending the number of space-time
coordinates subject to an `exceptional geometry' so that the
higher-dimensional theory is manifestly duality invariant.

It is worth stressing that the work described in this paper is
exclusively based on the on-shell formulation of IIB supergravity, as
originally constructed in
\cite{Schwarz:1983wa,Schwarz:1983qr,Howe:1983sra}. As is well known
the compactification of ten-dimensional type-IIB supergravity on a
five-dimensional torus leads to five-dimensional maximal supergravity
\cite{Cremmer:1980gs} with a non-linearly realized $\mathrm{E}_{6(6)}$
symmetry, whose maximal compact subgroup $\mathrm{USp}(8)$ coincides
with the R-symmetry group. Compactification on a curved internal
manifold, such as the sphere $S^5$, will necessarily break some of the
symmetries mentioned above. In the case of $S^5$ one expects to obtain
the $\mathrm{SO}(6)$ gauging of maximal supergravity upon truncating
the massive modes, because the isometry group of $S^5$ equals
$\mathrm{SO}(6)$ \cite{Gunaydin:1985cu}. Various results on the
consistency of this truncation have already been reported in the
literature (see, e.g. \cite{Pilch:2000ue,Lee:2014mla}). From the
five-dimensional viewpoint, the breaking of the $\mathrm{E}_{6(6)}$
symmetry is understood as a result of the non-abelian gauge
interactions, because the $\mathrm{SO}(6)$ gauge group is embedded
into $\mathrm{E}_{6(6)}$.

As we discussed above, it is possible to reformulate the
higher-dimensional theory upon splitting the coordinates into 5
space-time and 5 internal coordinates, while retaining the full
dependence on the two sets of coordinates. To ensure that the theory
takes the form of the lower-dimensional theory with fields that depend
in addition on the five internal coordinates, one adopts a
gauge-equivalent version of the tangent space such that the tangent
space group will be restricted to the product group
$\mathrm{SO}(4,1)\times \mathrm{SO}(5)$. Subsequently one combines the
group $\mathrm{SO}(5)$ associated with the internal five-dimensional
tangent space with the manifest local $\mathrm{U}(1)$ group of IIB
supergravity. The crucial step is then to extend this product group to
$\mathrm{USp}(8)$, which is the R-symmetry group for five-dimensional
maximal supergravity. Hence we envisage
\begin{align}
  \label{eq:tangent-sequence-2b}
  \mathrm{Spin}(9,1)\times\mathrm{U}(1) \longrightarrow&\;
  \mathrm{Spin}(4,1) \times\big[\mathrm{USp}(4) \times \mathrm{U}(1)
  \big] \nonumber\\ 
  \longrightarrow&\; \mathrm{Spin}(4,1)\times \mathrm{USp}(8) \,,
\end{align}
where we now refer to the universal covering groups which are relevant
for the fermions. Initially only the $\mathrm{USp}(4) \times
\mathrm{U}(1)$ subgroup is realized as a {\it local} invariance that
involves all ten coordinates. In order to realize the full {\it local}
$\mathrm{USp}(8)$ invariance, it suffices to introduce a compensating
$\mathrm{USp}(8)/[\mathrm{USp}(4)\times\mathrm{U}(1)]$ phase factor.

The ensuing analysis will be more subtle for IIB supergravity
than for $11D$ supergravity. The latter contains a single fermion
field corresponding to the gravitino that decomposes directly into
$4D$ gravitini transforming in the $\boldsymbol{8}$ representation and
$4D$ spin-1/2 fermions transforming in the
$\boldsymbol{48}+\boldsymbol{8}$ representation of $\mathrm{Spin}(7)$.
As was first demonstrated in \cite{Cremmer:1979up}, these fields can
be reassembled upon extending the group $\mathrm{Spin}(7)$ to chiral
$\mathrm{SU}(8)$, so that the gravitini transform in the
$\boldsymbol{8}\oplus\overline{\boldsymbol{8}}$ representation and the
spin-1/2 fields in the
$\boldsymbol{56}\oplus\overline{\boldsymbol{56}}$ representation of
chiral $\mathrm{SU}(8)$.  The IIB fermion representation, on the
other hand, is already reducible in 10 dimensions and consists of a
complex gravitino and a complex dilatino field. The $\mathrm{USp}(4)$
tangent-space group can in principle be generalized for each of these
fields to $\mathrm{SU}(4)\cong\mathrm{SO}(6)$. Furthermore, the
fermions of the IIB theory transform under a {\it locally} realized
$\mathrm{U}(1)$. Therefore, the R-symmetry group of the $5D$ fermions
is extended from $\mathrm{SU}(4)$ to
$\mathrm{SU}(4)\times\mathrm{U}(1)$. For the gravitini this group can
be directly extended to the expected $\mathrm{USp}(8)$ R-symmetry
group, under which the gravitini will transform in the
$\boldsymbol{8}$ representation. However, for the spin-1/2 fermions
one must combine the gravitino associated with the internal space,
comprising 40 symplectic Majorana spinors, with the dilatino,
comprising 8 such spinors, into an irreducible $\boldsymbol{48}$
representation of the group $\mathrm{USp}(8)$. It is clear that
assembling the different IIB fermions into a single irreducible spinor
that transforms covariantly under $\mathrm{USp}(8)$, is a subtle task.

Therefore our strategy is to first identify the vector and tensor
gauge fields and their supersymmetry transformations, subject to the
vector-tensor hierarchy that is known from the embedding tensor
formulation of $5D$ maximal supergravity \cite{deWit:2004nw}.  Unlike
in the case of 11-dimensional supergravity one must also include the
tensor fields in the analysis, because in five dimensions the dynamical
degrees of freedom for generic gaugings are always carried by a
mixture of vector and tensor fields. Hence the vector-tensor hierarchy
plays a key role here at a much earlier stage of the analysis and it
is not sufficient to rely exclusively on a proper preparation of the
target space as indicated in \eqref{eq:tangent-sequence-2b}.  As it
turns out, five of the tensor fields are still unaccounted for, but
even without these missing tensors we have sufficient information to
determine the generalized vielbeine, the $\mathrm{USp}(8)$ covariant
spinor fields, and the supersymmetry transformations of the
generalized vielbeine.  Using the vector-tensor hierarchy as a guide,
one can incorporate the missing five tensor fields which turn out to
transform in a representation that coincides precisely with that of a
descendant of the $10D$ dual graviton
\cite{Curtright:1980yk,Hull:2001iu,Bekaert:2002uh}. Hence the dual
graviton emerges in the form of tensor fields, unlike in the
11-dimensional situation \cite{Godazgar:2013dma} where the dual
graviton resides in the vector sector. We present a basis for the
vector and tensor fields which is manifestly in agreement with the
$\mathrm{E}_{6(6)}$ assignments known from the $5D$ theory, which
involves the invariant three-rank symmetric tensor of that group.

In spite of many subtle differences, the gross features of the present
analysis are in agreement with those of 11-dimensional supergravity,
implying that the approach that has been adopted is sufficiently
robust to be applied to more complicated situations. The supersymmetry
transformations of the fields are covariant under local
$\mathrm{USp}(8)$ transformations. The results opens the way to study
many other detailed questions, such as the consistency of the
truncation to the $\mathrm{SO}(6)$ gauging of maximal five-dimensional
supergravity or other consistent truncations along the lines followed
in \cite{Godazgar:2013oba}. Also the precise relation with the
consistent Kaluza-Klein truncations using exceptional field theory
\cite{Hohm:2014qga} is worth pursuing, as well as many other issues
that have recently emerged.

This paper is organized as follows. In section
\ref{sec:results-2B-sugra} the relevant properties of IIB supergravity
are summarized and the conventions are defined. Subsequently, in
section \ref{sec:first-field-redefin}, the Kaluza-Klein decompositions
are carried out to ensure that the fields transform covariantly from
the viewpoint of the $5D$ space-time. Also the conversion to $5D$
spinors and gamma matrices is discussed as well as the proper
definitions of the $5D$ vector and tensor fields that emerge directly
from the $10D$ boson fields. As it turns out, further redefinitions on
the vector and tensor fields are required such that they transform
under supersymmetry in a way that is consistent with the vector-tensor
hierarchy.  In section \ref{sec:vector-tensor-hierarchy-dual} the dual
vector and tensor fields are introduced. Again their proper
identification is based on covariance in the $5D$ space-time and on
the vector-tensor hierarchy. As it turns out there are only 22 tensor
fields at this stage. It is then demonstrated how the missing fields
can emerge from a component of the $10D$ dual graviton. This enables
one to obtain the symmetric $\mathrm{E}_{6(6)}$ tensor that appears in
the transformation rules of the tensor fields. At this point the
supersymmetry transformations of the bosonic vector and tensor fields
clearly resemble the transformation rules encountered in the pure $5D$
theory as presented in \cite{deWit:2004nw}, including those related to
the vector-tensor hierarchy. By direct comparison between the
supersymmetry transformations of the vector fields arising from ten
dimensions and the five-dimensional ones, explicit expressions for the
generalized vielbeine are derived in section
\ref{sec:gen-vielbeine}. In addition the $\mathrm{USp}(8)$ covariant
definitions of the spinor fields are obtained, as well as the
supersymmetry transformations of the generalized vielbeine. A similar
strategy is then applied to the tensor fields, which leads to a
corresponding set of generalized vielbeine. Upon adopting suitable
normalizations of the vector and tensor fields one can show that this
new set of vielbeine constitutes the inverse of the generalized
vielbeine determined in the vector sector. In section
\ref{sec:fermion-transformations} the supersymmetry transformations of
the fermions are considered and it is shown that they take a
$\mathrm{USp}(8)$ covariant form. Finally, in
section~\ref{sec:cons-trunc-maxim} the question of the consistent
truncation to $\mathrm{SO}(6)$ gauged maximal $5D$ supergravity is
adressed. We include two appendices,
\ref{App:red-spinors-gamma-matrices} and
\ref{App:R-symm-assignm-fermions}, dealing with the definition and
decomposition of gamma matrices and the spinor and R-symmetry
representations associated with the various groups emerging upon
decomposing the tangent-space into two separate $5D$ subspaces.

\section{Summary of IIB supergravity}
\label{sec:results-2B-sugra}
\setcounter{equation}{0}
Here we summarize the relevant results for IIB supergravity in ten
space-time dimensions
\cite{Schwarz:1983wa,Schwarz:1983qr,Howe:1983sra}. The theory is
described in terms of a zehnbein $E_{M}{}^A$, a gravitino field
$\psi_M$, a spinor field $\lambda$, a complex three-rank tensor field
strength, $G_{MNP}$, a five-rank field strength $F_{MNPQR}$ subject to
a duality constraint, a complex
vector $P_M$ and a $\mathrm{U}(1)$ gauge field $Q_M$. The fermions are
complex and have opposite chirality,
\begin{equation}
  \label{eq:chiral-spin}
  \breve\Gamma_{11} \psi_M=\psi_M\,,\qquad \breve\Gamma_{11}\lambda
  =-\lambda\,, 
\end{equation}
where $\breve\Gamma_{11} =
\mathrm{i}\breve\Gamma_1\breve\Gamma_2\cdots\breve\Gamma_{10}$, with
$\breve\Gamma_A$ denoting the 10-dimensional gamma matrices. The
fermions transform under local phase transformations according to 
\begin{equation}
  \label{eq:phase-transf}
  \psi_M\to \mathrm{e}^{\mathrm{i}\Lambda/2} \,\psi_M\,,\qquad 
  \lambda\to \mathrm{e}^{3\mathrm{i}\Lambda/2} \,\lambda\,. 
\end{equation}
The zehnbein $E_M{}^A$ and the field strength $F_{MNPQR}$ are invariant
under $\mathrm{U}(1)$, unlike the other quantities, which transform as
follows,
\begin{equation}
  \label{eq:boson-phase-tr}
  G_{MNP}\to \mathrm{e}^{\mathrm{i}\Lambda} \,G_{MNP}\,,\qquad 
  P_{M}\to \mathrm{e}^{2\mathrm{i}\Lambda} \,P_{M}\,,\qquad  Q_M\to
  Q_M+\partial_M\Lambda\,. 
\end{equation}
The vectors $P_M$ and $Q_M$ satisfy the Maurer-Cartan equations
associated with the coset space $\mathrm{SU}(1,1)/\mathrm{U}(1)$,
which is parametrized by the scalar fields of the
theory,
\begin{equation}
  \label{eq:Maurer-Cartan}
  \partial_{[M}Q_{N]} = -\mathrm{i} P_{[M} \,\bar P_{N]}\,, \qquad
  \mathcal{D}_{[M} P_{N]} =0\,.
\end{equation}
In this section the derivative $\mathcal{D}_M$ is covariant with
respect to local Lorentz and local $\mathrm{U}(1)$ transformations. 

The coset representative can be expressed in terms of an
$\mathrm{SU}(1,1)$ doublet $\phi^\alpha$, ($\alpha=1,2$), transforming
under $\mathrm{U}(1)$ as
\begin{equation}
  \label{eq:phi-u1}
  \phi^\alpha\to \mathrm{e}^{\mathrm{i}\Lambda}\,\phi^\alpha\,,
\end{equation}
and subject to the $\mathrm{SU}(1,1)$ invariant constraint,
\begin{equation}
  \label{eq:alg-constr-phi}
  \vert\phi^1\vert^2 - \vert\phi^2\vert^2 =1\,. 
\end{equation}
In what follows we use the convenient notation $\phi_\alpha\equiv
\eta_{\alpha\beta} (\phi^\beta)^\ast$, with
$\eta_{\alpha\beta}=\mathrm{diag}(+1,-1)$, so that the above
constraint reads $\phi_\alpha \phi^\alpha=1$. In this convention the
vector fields take the following form,
\begin{align}
  \label{eq:P-Q-express}
  Q_M =&\, -\mathrm{i} \phi_\alpha \,\partial_M\phi^\alpha\,, \nonumber\\
   P_M=&\, \varepsilon_{\alpha\beta}
  \,\phi^\alpha\,\mathcal{D}_M\phi^\beta\,,  \nonumber \\
  \bar P_M=&\, -\varepsilon^{\alpha\beta}
  \,\phi_\alpha\,\mathcal{D}_M\phi_\beta\,,
\end{align}
where the Levi-Civita symbol is normalized by
$\varepsilon_{12}=\varepsilon^{12}=1$. Note that $\eta_{\alpha\beta}
\varepsilon^{\beta\gamma}\eta_{\gamma\delta} =
-\varepsilon_{\alpha\delta}$. We note the following useful identities,
\begin{equation}
  \label{eq:P-phi}
  \phi_\alpha\,\mathcal{D}_M \phi^\alpha =0\,,\qquad \phi_\alpha 
  P_M = \varepsilon_{\alpha\beta}   \,\mathcal{D}_M \phi^\beta\,,
  \qquad \phi^\alpha 
  \bar  P_M = -\varepsilon^{\alpha\beta}\,\mathcal{D}_M \phi_\beta \,.  
\end{equation}

Let us now turn to the tensor field strengths. The theory contains two
tensor fields $A^\alpha{\!}_{MN}$ transforming under
$\mathrm{SU}(1,1)\cong \mathrm{SL}(2)$. Here we use a pseudoreal basis
with $A^\alpha{\!}_{MN} = \varepsilon^{\alpha\beta}
(A_{MN})_\beta$, where the convention for lowering and raising of
indices is the same as for $\phi^\alpha$. Their field strengths are
defined as follows,
\begin{align}
  \label{eq:field-strenghts-tensors}
   3\,\partial_{[M} A^\alpha {\!}_{NP]} =&\,
   \phi^\alpha\,\bar G_{MNP} +
  \varepsilon^{\alpha\beta}\phi_\beta\,G_{MNP} \,, \nonumber\\
  G_{MNP}=&\, -3\,\varepsilon_{\alpha\beta}\, \phi^\alpha
  \,\partial_{[M} A^\beta{\!}_{NP]}  \,,\nonumber\\
  \bar G_{MNP}=&\, 3\,\phi_\alpha\,\partial_{[M}A^\alpha{\!}_{NP]}  \,.
\end{align}
The tensor fields are subject to rigid $\mathrm{SU}(1,1)$
transformations, just as the scalar fields $\phi^\alpha$, and to
tensor gauge transformations. The latter read
\begin{equation}
  \label{eq:gauge-2}
  \delta A^\alpha{\!}_{MN} = 2\,\partial_{[M} \Xi^\alpha{\!}_{N]} \,. 
\end{equation}
Furthermore we have a 4-rank antisymmetric gauge potential
$A_{MNPQ}$, which transforms under two types of gauge transformations 
\begin{equation}
  \label{eq:gauge-2-4}
  \delta A_{MNPQ} = 4\, \partial_{[M} \Lambda_{NPQ]} + \tfrac34 \mathrm{i}\,
  \varepsilon_{\alpha\beta} \,   \Xi^\alpha{\!}_{[M} \, \partial_{N}
  A^\beta {\!}_{PQ]} \,. 
\end{equation}
The corresponding 5-form field strength is defined by 
\begin{equation}
  \label{eq:5-form}
  F_{MNPQR} = 5\, \partial_{[M} A_{NPQR]} -\tfrac{15}{8}\mathrm{i}
  \varepsilon_{\alpha\beta} \, A^\alpha{\!}_{[MN}\,\partial_P
  A^\beta{\!}_{QR]} \,.
\end{equation}
The 3- and 5-rank field strengths satisfy the following Bianchi identities,
\begin{align}
  \label{eq:Bianchi}
  \mathcal{D}_{[M} G_{NPQ]} =&\,P_{[M}\,\bar G_{NPQ]} \,,  \nonumber\\
  \partial_{[M} F_{NPQRS]} =&\, -\tfrac5{12} \mathrm{i}
  \,G_{[MNP}\,\bar G_{QRS]} \,. 
\end{align}
In addition there is a constraint on the 5-index field strength which
involves the dual field strength, 
\begin{align}
  \label{eq:self-dual}
  \tfrac1{120} \mathrm{i} \,\varepsilon_{ABCDEFGHIJ}\,F^{FGHIJ}&\, =
  F_{ABCDE} -\tfrac1{8}\mathrm{i} \,\bar\psi_M \breve\Gamma^{[M}
  \breve\Gamma_{ABCDE}
  \breve\Gamma^{N]} \psi_N     \nonumber\\
  &\, +\tfrac1{16}\mathrm{i} \,\bar\lambda\, \breve\Gamma_{ABCDE}\,
  \lambda \,.
\end{align}
From the chirality of the fermion fields it follows that the fermionic
bilinears in \eqref{eq:self-dual} are anti-selfdual, which is
obviously required because otherwise \eqref{eq:self-dual} would
decompose into two independent constraints that would overconstrain
the system. Originally \eqref{eq:self-dual} was derived in superspace
\cite{Howe:1983sra}. Suppressing the fermionic terms would imply that
the bosonic field strength should be self-dual.  Note that the
constraint \eqref{eq:self-dual} is supersymmetric and it must
transform into the fermionic field equations. Upon combining it with
the Bianchi identity \eqref{eq:Bianchi}, one obtains the field
equations for $A_{MNPQ}$.

Let us now turn to the fermions $\psi_M$ and $\lambda$. 
The supersymmetry transformations for the spinor fields are as follows,
\begin{align}
  \label{eq:susy-fermions}
  \delta\psi_M =&\, \mathcal{D}_M \epsilon - \tfrac1{480}
  \mathrm{i}F_{NPQRS}\,\breve\Gamma^{NPQRS}\,\breve\Gamma_M\epsilon
  -\tfrac1{96} G_{NPQ} \big(\breve\Gamma_M \,\breve\Gamma^{NPQ}
  +2\,\breve\Gamma^{NPQ}\,\breve\Gamma_M\big) \epsilon^\mathrm{c}\,, \nonumber\\
  \delta\lambda=&\, -P_M \,\breve\Gamma^M\epsilon^\mathrm{c}
  -\tfrac1{24} G_{MNP} \breve\Gamma^{MNP} \epsilon\,,
\end{align}
where the quantities $\breve\Gamma^{MN\cdots}$ denote anti-symmetrized
products of $10D$ gamma matrices, and $\mathcal{D}_M\epsilon$ contains
the spin-connection field $\omega_M{}^{AB}$ and the $\mathrm{U}(1)$
connection $Q_M$,
\begin{equation}
  \label{eq:D-eps}
  \mathcal{D}_M\epsilon =\Big( \partial_M  -\tfrac14 \omega_M{}^{AB}
  \breve\Gamma_{AB} - \tfrac12 \mathrm{i} Q_M\Big) \epsilon\,. 
\end{equation}
Here $\epsilon$ is the space-time dependent spinor parameter of supersymmetry.  
In \eqref{eq:D-eps} we have introduced the Majorana conjugate of a $10D$ spinor
$\psi$, which is defined by
\begin{equation}
  \label{eq:doubling}
  \psi^\mathrm{c} = \breve C_\pm^{-1} \bar\psi^\mathrm{T}\,,\qquad
  \psi = \breve C_\pm^{-1}   \bar\psi^\mathrm{c}{}^\mathrm{T}\,. 
\end{equation}
Here $\breve C_\pm$ denotes the charge conjugation matrix in 10
space-time dimensions which can be either symmetric or
anti-symmetric. The gamma matrix conventions are discussed in detail
in appendix \ref{App:red-spinors-gamma-matrices}, but for the convenience
of the reader we note
\begin{equation}
  \label{eq:transprose-Gamma-sec2}
  \breve C_\pm\,\breve\Gamma_A \breve C_\pm^{-1} = \pm
  \breve\Gamma_A{}^\mathrm{T} \,,\quad 
  \breve C_\pm{}^\mathrm{T}= \pm \breve C_\pm\,, \quad \breve
  C_\pm{}^\dagger = \breve C_\pm^{-1}\,.
\end{equation}
We also note the following equation for spinor bilinears with strings
of gamma matrices,
\begin{equation}
  \label{eq:bilinear-mirror}
  \bar \chi\,\Gamma_{A_1} \cdots \Gamma_{A_n} \,\psi = -(\pm)^{n+1}
  \bar \psi^\mathrm{c} \,\Gamma_{A_n} \cdots \Gamma_{A_1}
  \,\chi^\mathrm{c} \,. 
\end{equation}

For type-IIB supergravity we have {\it chiral} spinors comprising 16 complex
components. One can show that $\psi$ and $\psi^\mathrm{c}$ have the
same chirality (see appendix \ref{App:red-spinors-gamma-matrices} for
details) and since the spinors are complex (so that
$\psi^\mathrm{c}\not=\psi$) one can adopt a pseudo-real representation
by combining $\psi$ and $\psi^\mathrm{c}$ into a 32-component chiral
spinor $\Psi= (\psi,\psi^\mathrm{c})$, subject to 
\begin{equation}
  \label{eq:pseudo-real-32}
  \Psi = \sigma_1 \,\breve C_\pm^{-1} \bar\Psi^\mathrm{T}\,,
\end{equation}
where $\sigma_1$ denotes the standard $2\times2$ Pauli spin matrix.
We need also the supersymmetry transformations for the bosons,
\begin{align}
  \label{eq:boson-susy}
  \delta E_M{}^A =&\, \tfrac12(\bar\epsilon \,\breve\Gamma^A\psi_M
  + \bar\epsilon^\mathrm{c}\,
  \breve\Gamma^A\psi_M^\mathrm{c}) \,, \nonumber \\
  \delta \phi^\alpha =&\, \tfrac12 \varepsilon^{\alpha\beta} \phi_\beta
  \,\bar\epsilon^\mathrm{c} \lambda  \,,\nonumber\\
  \delta A^\alpha{\!}_{MN}= &\, - \tfrac12 \phi^\alpha\big(\bar\lambda\, \breve\Gamma_{MN}
  \epsilon - 4\,\bar\epsilon\,
  \breve\Gamma_{[M}\psi_{N]}{\!}^\mathrm{c}\big)+ \tfrac12 
  \varepsilon^{\alpha\beta} \phi_\beta\,\big(\bar\epsilon\,
  \breve\Gamma_{MN}\lambda + 4\, \bar\psi_{[M}{\!}^\mathrm{c}\,
  \breve\Gamma_{N]}\epsilon\big) \,,\nonumber\\  
  \delta A_{MNPQ} =&\, \tfrac12\mathrm{i} \bar\epsilon
  \,\breve\Gamma_{[MNP}\psi_{Q]} + \tfrac12\mathrm{i} 
  \bar\psi_{[M} \breve\Gamma_{NPQ]} \epsilon + \tfrac38\mathrm{i} \,
  \varepsilon_{\alpha\beta} A^\alpha{\!}_{[MN} \,\delta
  A^\beta{\!}_{PQ]}\,. 
\end{align}

The above transformation rules \eqref{eq:susy-fermions} and
\eqref{eq:boson-susy} have been derived by imposing the supersymmetry algebra,
\begin{equation}
  \label{eq:susy-algebra}
  \big[ \delta(\epsilon_1)\,,\delta(\epsilon_2)\big] = \xi^MD_M +
  \delta_\Xi\big(\Xi^\alpha{\!}_{MN}\big) +
  \delta_\Lambda(\Lambda_{MNP})+  \cdots \,,  
\end{equation}
where 
\begin{align}
  \label{eq:susy-alg-para}
  \xi^M=&\, \tfrac12 \bar\epsilon_2\,\breve\Gamma^M\epsilon_1
  +\tfrac12 \bar\epsilon_2{\!}^\mathrm{c}\,
  \breve\Gamma^M\epsilon_1{\!}^\mathrm{c}\,,
  \nonumber\\
  \Xi^\alpha{\!}_M =&\, - \phi^\alpha \,\bar\epsilon_2 \breve\Gamma_M
  \epsilon_1{\!}^\mathrm{c} -\varepsilon^{\alpha\beta} \phi_\beta
  \,\bar\epsilon_2{\!}^\mathrm{c} \breve\Gamma_M\epsilon_1 \,, \nonumber\\
  \Lambda_{MNP}
  =&\,\tfrac18 \mathrm{i}\big(\bar{\epsilon}_{1}\breve\Gamma_{MNP}\epsilon_{2}-
  \bar{\epsilon}_{2}\breve\Gamma_{MNP}\epsilon_{1} \big)
   +\tfrac{3}{16}\mathrm{i}\big(\varepsilon_{\alpha\beta}\phi^{\alpha} A^{\beta}_{[MN}\,
   \bar{\epsilon}_{2}\breve\Gamma_{P]}\epsilon_{1}{\!}^\mathrm{c}
  +\phi_{\alpha}   A^{\alpha}_{[MN}\,
  \bar{\epsilon}_{2}{\!}^\mathrm{c}\breve\Gamma_{P]}\epsilon_{1}\big)\,,   
\end{align}
and where $\xi^MD_M$ denotes a fully covariantized space-time
diffeomorphism. 

For future use we also present the supersymmetry transformation rules
for the Majorana conjugate spinors,
\begin{align}
  \label{eq:susy-fermions-cc}
  \delta\psi_M{}^\mathrm{c} =&\, \mathcal{D}_M \epsilon^\mathrm{c} + \tfrac1{480}
  \mathrm{i}F_{NPQRS}\,\breve\Gamma^{NPQRS}\,\breve\Gamma_M\epsilon^\mathrm{c}
  -\tfrac1{96} 
  \bar G_{NPQ} \big(\breve\Gamma_M \,\breve\Gamma^{NPQ}
  +2\,\breve\Gamma^{NPQ}\,\breve\Gamma_M\big) \epsilon \,, \nonumber\\
  \delta\lambda^\mathrm{c}=&\, \pm  \bar P_M \breve\Gamma^M\epsilon
  \pm\tfrac1{24}\bar G_{MNP} \,\breve\Gamma^{MNP} \epsilon^\mathrm{c}\,. 
\end{align} 

To understand the various field equations it is convenient to first
consider the following $10D$ Lagrangian of IIB supergravity up to
terms of fourth-order in the fermion fields, ignoring for the moment
the constraint \eqref{eq:self-dual},
\begin{align}
  \label{eq:fermion-kin-10}
  \mathcal{L}=&\, -\tfrac12 E\,R -
  E\,\bar\psi_M\breve\Gamma^{MNP}\mathcal{D}_N\psi_P -\tfrac12
  E\,\bar\lambda \breve{\Slash{\mathcal{D}}} \lambda -E\,\vert
  P_M\vert^2 -\tfrac1{24} E\,\vert G_{MNP}\vert^2\nonumber\\
  &\, -\tfrac1{60} E\, (F_{MNPQR})^2 + \tfrac1{384} \,\varepsilon^{MNPQRSTUVW}
  \, \varepsilon_{\alpha\beta} \, \partial_M A_{NPQR}\, A_{ST}{\!}^\alpha
  \,\partial_U A_{VW}{\!}^\beta\nonumber\\
  &\, -\tfrac12 E\,\big[\bar\psi_M{\!}^\mathrm{c}\,\breve\Gamma^N
  \breve\Gamma^M\lambda\,\bar P_N
  +\bar\lambda  \,\breve\Gamma^M \breve\Gamma^N \psi_M{\!}^\mathrm{c} \,  P_N \big]   \nonumber\\
  &\,+ \tfrac1{240}\mathrm{i} E\,\bar\psi_M \breve\Gamma^{[M}
  \breve\Gamma^{ABCDE}
  \breve\Gamma^{N]} \psi_N \,F_{ABCDE} \nonumber\\
  &\, + \tfrac1{48} E\, \big[\bar\psi_M\, \breve\Gamma^{[M}
  \breve\Gamma^{ABC}\, \breve\Gamma^{N]} \psi_N{}^\mathrm{c} \,
  G_{ABC} + \bar\psi_M{\!}^\mathrm{c} \, \breve\Gamma^{[M}
  \breve\Gamma^{ABC}\,
  \breve\Gamma^{N]} \psi_N \, \bar G_{ABC} \big]\nonumber\\
  &\, +\tfrac1{48} E\, [\bar\psi_M \, \breve\Gamma^{ABC}\,
  \breve\Gamma^M\, \lambda\, \bar G_{ABC} - \bar\lambda\,
  \breve\Gamma^M\, \breve\Gamma^{ABC} \psi_M\,
  G^{ABC} \big]\nonumber\\
  &\, -\tfrac1{480} \mathrm{i} E\,\bar\lambda\, \breve\Gamma^{ABCDE}\,
  \lambda\, F_{ABCDE} + \cdots\,.
\end{align}
We have refrained from imposing the supersymmetric constraint
\eqref{eq:self-dual} so that it makes sense to include a term
proportional to $(F_{MNPQR})^2$, and furthermore we have included a
Chern-Simons term that is invariant under tensor gauge transformations
up to a total derivative. It is then straightforward to show that the
field equation for the 4-form field that follows from this Lagrangian
is consistent with the constraint \eqref{eq:self-dual} upon using the
second Bianchi identity \eqref{eq:Bianchi}. Here we should remind the
reader that there are extensive discussions in the literature about
manifestly covariant Lagrangians that imply self-duality constraints
for tensor fields (see, for instance, \cite{Dall'Agata:1997ju}, where
also the Chern-Simons terms is presented, and references cited
therein). However, these features are not relevant for our purpose. We
also recall that the field equations are already encoded in the
supersymmetry transformations, as supersymmetry is only realized
on-shell, so that one can determine most terms in
\eqref{eq:fermion-kin-10} by imposing super-covariance of the field
equations, just as was done in \cite{Schwarz:1983qr}. Our results are
also consistent with \cite {Howe:1983sra} where an on-shell superspace
treatment of IIB supergravity was presented.

For further convenience we list some of the field equations,
\begin{align}
  \label{eq:bosonic-field-eq}
  \mathcal{D}^MP_M +\tfrac1{24} G_{MNP}\,G^{MNP}&\, = 0\,,\nonumber \\[2mm]
  \mathcal{D}^MG_{MNP} + P^M\, \bar G_{MNP} - \tfrac23 \mathrm{i} F_{NPQRS}\,
  G^{QRS}&\, = 0 \,, \nonumber \\[2mm]
  R_{MN} +2\,P_{(M}\,\bar P_{N)} +\tfrac14 \big(\bar G_{PQ(M}
  \,G^{PQ}{}_{N)} -\tfrac1{12} g_{MN}\,\vert G_{PQR}\vert^2 \big)
  + \tfrac16 F_M{}^{PQRS}\,F_{NPQRS} &\,= 0 \,,\nonumber\\[2mm]
  \breve\Gamma^M\hat{D}_{M}\lambda+
  \tfrac1{240}\mathrm{i}\breve\Gamma^{NPQRS}\lambda\, F_{NPQRS}&\, =0\,,
  \nonumber\\[2mm] 
  \breve\Gamma^{MNP}\widehat{{D}_{N}\psi_{P}} \mp
  \tfrac12\breve\Gamma^Q\breve\Gamma^M\lambda^\mathrm{c}P_Q 
  -\tfrac{1}{48}\breve\Gamma^{QRS}\breve\Gamma^M\lambda\,\bar{G}_{QRS}&\,
  =0\,, 
\end{align}
where $\hat{D}_M\lambda$ denotes the supercovariant derivative of the
spinor $\lambda$ and $\widehat{D_{[M}\psi_{N]}}$ the supercovariant
curl of the gravitino. Here we suppressed higher-order fermion terms.

However, in section \ref{sec:vector-tensor-hierarchy-dual}, we will
need the field equations for the two-form fields including the terms
quadratic in the fermions. They follow directly from the Lagrangian
\eqref{eq:fermion-kin-10} and can be written as follows,
\begin{equation}
  \label{eq:2-form-f-eq}
  \partial_{[M}F_{NPQRSTU]\,\alpha} =  0\, ,
\end{equation}
where the seven-rank anti-symmetric tensors $F_{MNPQRST\,\alpha}$ are
equal to 
\begin{align}
  \label{eq:F-7-alpha}
  F_{\alpha\,MNPQRST}=&\, -\tfrac1{7} \mathrm{i}\,E\,
  \varepsilon_{MNPQRSTUVW} \,
  \big(\varepsilon_{\alpha\gamma} \,\phi^\gamma \phi_\beta + 
  \varepsilon_{\beta\gamma} \,\phi^\gamma \phi_\alpha
  \big)\,\partial^{U} A^{VW\,\beta} \nonumber\\[2mm]
  &\, -120\mathrm{i}\,\varepsilon_{\alpha\beta} \,A_{[MN}{\!}^\beta \big[  \partial_P
  A_{QRST]} - \tfrac{1}8\mathrm{i} 
  \varepsilon_{\gamma\delta}\, A_{PQ}{\!}^\gamma \,\partial_R
  A_{ST}{\!}^\delta  \big]\nonumber\\[2mm]
  & +\tfrac1{7} \varepsilon_{\alpha\beta} \phi^\beta\,
  \big[\bar\psi_U\, \breve\Gamma^{[U}
  \breve\Gamma_{MNPQRST}\, \breve\Gamma^{V]} \psi_V{}^\mathrm{c}  +\bar\lambda\,
  \breve\Gamma^U\, \breve\Gamma_{MNPQRST} \psi_U\big] \nonumber\\[2mm]
  & +\tfrac1{7}\phi_{\alpha}  \,
  \big[\bar\psi_U{}^\mathrm{c}\, \breve\Gamma^{[U}
  \breve\Gamma_{MNPQRST}\, \breve\Gamma^{V]} \psi_V  -\bar\psi_U\,
  \breve\Gamma_{MNPQRST} \breve\Gamma^U\, \lambda \big]   \,.
\end{align} 
Note that the normalization of this tensor is arbitrary but the phase
is dictated by the fact that its pseudo-reality condition is in line
with that of the other pseudo-real fields. 

\section{Kaluza-Klein decompositions and additional field
  redefinitions}  
\label{sec:first-field-redefin}
\setcounter{equation}{0}
The strategy in this paper is to describe IIB supergravity as a field
theory in a five-dimensional space-time, while still retaining the
dependence on the five additional coordinates that describe an
internal space. Hence the $10D$ coordinates are decomposed according
to $x^M\to (x^\mu, y^m)$, where $x^\mu$ are regarded as the space-time
coordinates and $y^m$ as the coordinates of the internal
manifold. Eventually, in a given background, the fields may be
decomposed in terms of a complete basis of functions of the internal
coordinates. For the $T^5$ background this is rather straightforward;
the spectrum of the tower of Kaluza-Klein supermultiplets for $S^5$
has been studied in \cite{Kim:1985ez,Gunaydin:1984fk}.
However, at this stage we will not be assuming any particular
space-time background and neither will we be truncating the theory in
any way. We are only reformulating the theory in a form that
emphasizes the five-dimensional space-time.

A crucial ingredient in this reformulation is provided by a change of
the tangent-space group, which we have already indicated in
\eqref{eq:tangent-sequence-2b}. First we impose a gauge choice,
reducing the $10D$ local Lorentz group to the product group
$\mathrm{SO}(4,1)\times \mathrm{SO}(5)$, whose universal covering
group equals $\mathrm{Spin}(4,1)\times \mathrm{USp}(4)$. The fermions
then transform according to the product representation of this group,
so that from a five-dimensional space-time perspective we are dealing
with four complex $\mathrm{Spin}(4,1)$ spinors, each carrying four
components. The fermions are subject to an extra local $\mathrm{U}(1)$
group, and the product group $\mathrm{USp}(4)\times \mathrm{U}(1)$
must be contained in the $5D$ R-symmetry group. Obviously we have to
convert the $10D$ gamma matrices to those appropriate for five
space-time dimensions, equiped with two sets of mutually commuting
gamma matrices, one associated with space-time and the other one with
the internal space. In due course we will also have to recombine the
spin-1/2 fermion fields into an irreducible representation of the
group $\mathrm{USp}(8)$, which is the R-symmetry group for eight 
symplectic Majorana supercharges in a $5D$ space-time. This last
redefinition will be considered in section \ref{sec:gen-vielbeine}.

The next step is to redefine the fields such that they transform
covariantly under the $5D$ space-time diffeomorphisms. These
Kaluza-Klein decompositions were systematically discussed in the
context of the $T^7$ reduction of $11D$ supergravity to $4D$
supergravity \cite{Cremmer:1979up}. Furthermore, we will find that the
vector and tensor fields require additional redefinitions  
beyond the Kaluza-Klein ones in order to generate transformations that
reflect the vector-tensor hierarchy \cite{deWit:2004nw}.

The standard Kaluza-Klein decompositions start with the vielbein
field and its inverse, which we write in triangular form by exploiting
the $10D$ local Lorentz transformations,
\begin{equation}
  \label{eq:kk-ansatz}
  E_M{}^A= \begin{pmatrix} \Delta^{-1/3} e_\mu{}^\alpha &  B_\mu{}^m\,
    e_m{}^a \\[6mm] 
    0 &  e_m{}^a 
    \end{pmatrix} \;,\qquad
    E_A{}^M= \begin{pmatrix} \Delta^{1/3}e_\alpha{}^\mu & -
      \Delta^{1/3}e_\alpha{}^\nu B_\nu{}^m \\[6mm] 
    0 & e_a{}^m 
    \end{pmatrix}\;.
\end{equation}
Here we used tangent-space indices $\alpha,\beta,\dots$ associated
with the $5D$ space-time and $a,b,\dots$ associated with the $5D$
internal space.\footnote{
  Note that we are also using indices $\alpha,\beta\ldots$ for the
  $\mathrm{SU}(1,1)$ indices on the scalar doublet and the tensor
  fields. This should not cause any confusion.} 
The scalar factor $\Delta$ is defined by,
\begin{equation}
  \label{eq:Delta}
  \Delta= \frac{\det [e_m{}^a(x,y)] }{\det
    [{\mathring{e}}_m{}^a(y)] } \,, 
\end{equation}
where ${\mathring{e}}_m{}^a$ is some reference
frame for the internal space parametrized by the coordinates
$y^m$. The rescaling of the f\"unfbein is such that the
gravitational coupling constants in $10D$ and $5D$ are related by 
$\kappa^{-2}\vert_{10D}= \kappa^{-2}\vert_{5D} \,\int \,\mathrm{d}^5y \,
\det [\mathring{e}_m{}^a]$, so that we are in the $5D$ Einstein
frame. 

An important feature of the gauge choice made in~\eqref{eq:kk-ansatz}
is that it must be preserved under supersymmetry. This requires to add
to the $10D$ supersymmetry transformations a uniform field-dependent
Lorentz transformation with a parameter equal to
\begin{equation}
  \label{eq:comp-Lorentz}
  \epsilon^{\alpha a}= -\epsilon^{a\alpha} = -\tfrac12 e_a{}^m\big(
  \bar\epsilon\,\breve\Gamma^\alpha \psi_m
  +\bar\epsilon^\mathrm{c}\,\breve\Gamma^\alpha \psi_m{}^\mathrm{c} \big) \,,
\end{equation}
where $\psi_a= e_a{}^m\,\psi_m$. The supersymmetry transformation of $e_a{}^m$
is not affected by the compensating Lorentz transformations, so that we have 
\begin{equation}
  \label{eq:delta-Delta}
  \delta\Delta= \tfrac12 \Delta \big(
  \bar\epsilon\,\breve\Gamma^a \psi_a
  +\bar\epsilon^\mathrm{c}\,\breve\Gamma^a \psi_a{}^\mathrm{c} \big)\,. 
\end{equation}
One can now determine the supersymmetry variation of the f\"unfbein
$e_\mu{}^\alpha$, taking into account the compensating Lorentz
transformation \eqref{eq:comp-Lorentz} and  the effect of the factor $\Delta$. Insisting on
the fact that $e_\mu{}^\alpha$ transforms into the $5D$ gravitino
field in the same way as before, one then derives a modified gravitino field,
\begin{equation}
  \label{eq:new-gravio}
  \psi_\mu{}^\mathrm{KK} \equiv \Delta^{1/6}\big[\psi_\mu -
  B_\mu{}^m\,\psi_m\big] +\tfrac13 \Delta^{-1/6} e_\mu{}^\alpha
  \,\breve\Gamma_\alpha \breve\Gamma^a\psi_a  \,, 
\end{equation}
and likewise for $\psi_\mu{}^\mathrm{c}$. This field transforms
covariantly under $5D$ space-time diffeomorphisms by virtue of the
presence of the field $B_\mu{\!}^m$. Accordingly we also perform
field-dependent scale transformations on the supersymmetry parameter,
the gravitino components $\psi_a$ and the dilatino,
\begin{equation}
  \label{eq:new-eps-psia}
  \epsilon^\mathrm{KK}= \Delta^{1/6} \epsilon\,,\qquad
  \psi_a{}^\mathrm{KK} = \Delta^{-1/6} \,e_a{}^m\,\psi_m\,, \qquad
  \lambda^\mathrm{KK} =  \Delta^{-1/6} \lambda\,. 
\end{equation}

Subsequently we must convert to different gamma matrices that
decompose into two commuting Clifford algebras corresponding to the
5-dimensional space-time and the 5-dimensional internal space, which
must both commute with $\breve\Gamma_{11}$ so that they will be
consistent with the $10D$ chirality restriction on the original
spinors. As mentioned previously every $10D$ spinor decomposes into
four complex $\mathrm{Spin}(4,1)$ spinors.  The gamma matrix
conversion is discussed in detail in appendix
\ref{App:red-spinors-gamma-matrices} and the results can be summarized
as follows. The $32\times32$ gamma matrices $\breve\Gamma_A$ can be
written as
\begin{equation}
  \label{eq:gamma-conversion}
  \breve\Gamma_\alpha = -\mathrm{i} \big(
  \hat\gamma_\alpha\,\tilde\Gamma\big)\,,\qquad
  \breve\Gamma_{a+5} = -\mathrm{i} \big(
  \hat\Gamma_a\,\tilde\gamma\big)\,,
\end{equation}
where $\breve\Gamma_{11}= \mathrm{i}\tilde\gamma\tilde\Gamma$ with
$\tilde\gamma$ and $\tilde\Gamma$ mutually anti-commuting hermitian
matrices that square to $\oneone_{32}$.  The tangent space indices in
the $5+5$ split were already defined below
\eqref{eq:kk-ansatz}.\footnote{ 
  We employed Pauli-K\"all\'en conventions where $x^\alpha$ equals
  $\mathrm{i}x^0$ for $\alpha=1$, so that all gamma matrices are
  hermitian. } 
Both $\hat\gamma^\alpha$ and $\hat \Gamma^a$ anti-commute with
$\tilde\gamma$ and $\tilde\Gamma$ (and therefore commute with
$\breve\Gamma_{11}$ as insisted on before). They generate two
commuting five-dimensional Clifford algebras. Furthermore, we will
insist on the Majorana condition $\hat C^{-1} \bar\psi^\mathrm{T} =
\psi^\mathrm{c}$ for all the $5D$ spinor fields, where $\hat C$ is
defined in terms of the $10D$ charge conjugation matrix in
\eqref{eq:charge-conj-10to5}. For the gravitino fields and the
supersymmetry parameters this leads to the following relations between
$10D$ and $5D$ fields,
\begin{align}
  \label{eq:firstresults}
  \psi\big\vert_{10D}^\mathrm{KK} = \psi\big\vert_{5D} \,,\qquad
  \psi{}^\mathrm{c}\big\vert_{10D}^\mathrm{KK}=
  \psi{}^\mathrm{c}\big\vert_{5D} \,,\qquad
  \bar\psi\big\vert_{10D}^\mathrm{KK}= -\mathrm{i}
  \bar\psi{}\big\vert_{5D} \tilde\Gamma \,,
\end{align}
where $\psi$ denotes either $\psi_M$ or $\epsilon$. 

For the dilatino field $\lambda$ the situation is somewhat different
in view of the fact that we wish to change its chirality by
absorbing the matrix $\tilde\Gamma$. This conversion is of course no
longer consistent with $10D$ Lorentz invariance, but it is convenient
to define all the spinor fields with the same (positive) chirality.
\begin{align}
  \label{eq:firstresults}
  \lambda\big\vert_{10D}^\mathrm{KK}=
  \tilde\Gamma\,\lambda\big\vert_{5D} \,,\qquad
  \lambda^\mathrm{c}\big\vert_{10D}^\mathrm{KK} =
  \mp\tilde\Gamma\,\lambda ^\mathrm{c}\big\vert_{5D} \,,\qquad
  \bar\lambda\big\vert_{10D}^\mathrm{KK} = \mathrm{i}\,
  \bar\lambda\big\vert_{5D} \,,
\end{align}
Once these modifications have been performed, one can simply restrict
oneself to the 16-dimensional subspace corresponding to the eigenspace
of $\breve\Gamma_{11}$ with eigenvalue $+1$. After this one drops the
carets on $\gamma_\alpha$ and $\Gamma_a$ and thus obtains a
description in term of 16-component complex spinors, with two mutually
commuting sets of gamma matrices $\gamma_\alpha$ and $\Gamma_a$. Note
that this is consistent with using the charge conjugation matrix
$\hat C$, which was introduced as a 32-dimensional matrix but which
commutes with the chirality operator (i.e. charge-conjugated fields carry
the same chirality). With these conversions the relation
\eqref{eq:new-gravio} for the $10D$ gravitino field $\psi_M$ with
$M=\mu$ in terms of the $5D$ fields reads
\begin{equation}
  \label{eq:new-gravitino}
  \Delta^{1/6} \psi_\mu= \psi_\mu{}^\mathrm{KK}  -\tfrac13\mathrm{i}
  \gamma_\mu \Gamma^m\psi_m{}^\mathrm{KK}+ \Delta^{1/3}
  B_\mu{}^m\,\psi_m{}^\mathrm{KK}   \,.
\end{equation}
Observe that here and henceforth $\gamma_\mu\equiv
e_\mu{}^\alpha\,\gamma_\alpha$ and $\Gamma_m\equiv e_m{}^a\,\Gamma_a$,
where the vielbein fields $e_\mu{}^\alpha $ and $e_m{}^a$ are defined
in \eqref{eq:kk-ansatz}.

In this way one finds the following transformation rules for the $5D$
fields emerging from $E_M{}^A$ as defined in \eqref{eq:new-gravio} and
\eqref{eq:new-eps-psia}, 
\begin{align}
  \label{eq:susy-var-e-B-e}
  \delta e_\mu{}^\alpha =&\, \tfrac12 \big[ \bar\epsilon\,
  \gamma^\alpha\psi_\mu + \bar\epsilon^\mathrm{c} 
  \gamma^\alpha\psi_\mu{\!}^\mathrm{c}\big]   \,,\nonumber \\
  \delta B_\mu{}^m = &\, \tfrac12\Delta^{-1/3} e_a{}^m
  \big[\mathrm{i} \big(\bar\epsilon\, \Gamma^a \psi_\mu
  +\bar\epsilon^\mathrm{c}\, \Gamma^a \psi_\mu{\!}^\mathrm{c}\big)
  \nonumber\\
  &\qquad\qquad\quad
  + \bar\epsilon\,\gamma_\mu (\delta^a{}_b +\tfrac13 \Gamma^a\Gamma_b)
  \psi^b +\bar\epsilon^\mathrm{c}\,\gamma_\mu (\delta^a{}_b +\tfrac13
  \Gamma^a\Gamma_b) \psi^b{}^\mathrm{c}  \big]  \,,\nonumber\\
  \delta e_m{}^a =&\, \tfrac12\mathrm{i}\big[ \bar\epsilon \,\Gamma^a
  \psi_m+   \bar\epsilon^\mathrm{c} \,\Gamma^a
  \psi_m{\!}^\mathrm{c}  \big] \,, 
\end{align}
up to an infinitesimal $5D$ local Lorentz transformation with a
parameter proportional to $\Gamma^m\psi_m$. Since we will be
suppressing terms of higher orders in the spinor fields, these
transformations will not play a role when evaluating the fermion
transformation rules later in this section. Here and in the following we are exclusively
considering the $5D$ fields, so that we have dropped the additional
labels.

We also evaluate the supersymmetry variations of the scalars and the
dilatini,
\begin{align}
  \label{eq:lambda-phi}
  \delta\phi^\alpha =&\, -\tfrac12\mathrm{i} \,
  \varepsilon^{\alpha\beta} \phi_\beta\, \bar\epsilon^\mathrm{c}
  \lambda\,, \nonumber\\[2mm]
  \delta\lambda =&\,  \Delta^{-1/3}\big[ -\mathrm{i}P_\alpha
  \gamma^\alpha   +P_a \Gamma^a\big]\epsilon^\mathrm{c} \nonumber\\
  &\, +\tfrac1{24} \Delta^{-1/3}\big[  G_{abc} \,\Gamma^{abc} -3\mathrm{i} \,
  G_{ab\alpha} \,\Gamma^{ab}\gamma^\alpha + 3\, G_{a\alpha\beta}\,
  \Gamma^a \gamma^{\alpha\beta} -\mathrm{i} \,
  G_{\alpha\beta\gamma} \, \gamma^{\alpha\beta\gamma}
  \big]\epsilon \,, \nonumber\\[2mm]
  \delta\lambda^\mathrm{c} =&\,  \Delta^{-1/3}\big[- \mathrm{i}\bar P_\alpha
  \gamma^\alpha   +\bar P_a \Gamma^a\big]\epsilon \nonumber\\
  &\, +\tfrac1{24} \Delta^{-1/3} \big[ \bar G_{abc} \,\Gamma^{abc} -3\mathrm{i}\, \bar
  G_{ab\alpha} \,\Gamma^{ab}\gamma^\alpha + 3\,\bar G_{a\alpha\beta}\,
  \Gamma^a \gamma^{\alpha\beta} -\mathrm{i} \,\bar
  G_{\alpha\beta\gamma} \, \gamma^{\alpha\beta\gamma} 
  \big]\epsilon^\mathrm{c} \,,
\end{align}
where the tensors $P$ and $G$ refer to the components of $P_A$ and
$G_{ABC}$, which are defined with $10D$ tangent-space indices.

Subsequently we derive the expressions for the supersymmetry variation
of the gravitino fields up to terms of higher order in the fermion
fields, which will now also involve the components of the field
strength $F_{ABCDE}$ and the spin-connection fields written with $10D$
tangent-space indices. We first list the gravitino fields that carry a
$5D$ space-time vector index,
\begin{align}
  \label{eq:spacetime-gravitini}
  \delta\psi_\mu =&\, \big[ \partial_ \mu -\tfrac16 \partial_\mu \ln\Delta
  -\Delta^{-1/3}e_\mu{}^\alpha\big( \tfrac14 
  \omega_\alpha{}^{\beta\gamma} \,\gamma_{\beta\gamma} +\tfrac12
  \mathrm{i}  \omega_\alpha{}^{\beta a}\, \Gamma_a \gamma_\beta
  +\tfrac14 \omega_\alpha{}^{ab}\,\Gamma_{ab} +\tfrac12\mathrm{i}
  Q_\alpha \big) \big]\epsilon \nonumber\\
  &\, -B_\mu{}^m  \big[ \partial_ m-\tfrac16 \partial_m \ln\Delta \big]\epsilon \nonumber \\
  &\,- \tfrac1{240} \mathrm{i} \Delta^{-1/3} \,\varepsilon^{abcde} \big[
  \mathrm{i} \, F_{abcde} -5 F_{\beta abcd} \gamma^\beta \Gamma_{e} -5
  \mathrm{i} \, F_{\beta\gamma abc} \gamma^{\beta\gamma}
  \Gamma_{de}   \big]  \gamma_\mu \epsilon \nonumber\\
  &\, -\tfrac1{96} \Delta^{-1/3}  \big[ - \mathrm{i}
  G_{bcd}\, \Gamma^{bcd} \gamma_\mu +3 G_{bc\alpha}\, \Gamma^{bc}
  \big(\gamma_\mu \gamma^{\alpha} +2\, \gamma^{\alpha}\gamma_\mu  
  \big) \nonumber\\
  &\,\qquad\qquad\quad + 3 \mathrm{i} G_{b\alpha\beta}\, \Gamma^{b}
  \big(\gamma_\mu \gamma^{\alpha\beta} -2\, \gamma^{\alpha\beta}
  \gamma_{\mu}\big) + G_{\alpha\beta\gamma}
  \big(  \gamma_\mu \gamma^{\alpha\beta\gamma} +2
  \gamma^{\alpha\beta\gamma}  \gamma_\mu  \big) \big]
  \,\epsilon^\mathrm{c} \nonumber \\ 
  & \, +\tfrac13 \mathrm{i} \Delta^{-1/6}  \gamma_\mu\,
  \Gamma^a \delta \psi_a \, , \nonumber\\[2mm]
  \delta\psi_\mu{}^\mathrm{c}  =&\,  \big[ \partial_ \mu
  -\tfrac16 \partial_\mu \ln\Delta 
  -\Delta^{-1/3}e_\mu{}^\alpha\big( \tfrac14  
  \omega_\alpha{}^{\beta\gamma} \,\gamma_{\beta\gamma} +\tfrac12
  \mathrm{i}  \omega_\alpha{}^{\beta a}\, \Gamma_a \gamma_\beta
  +\tfrac14 \omega_\alpha{}^{ab}\,\Gamma_{ab} -\tfrac12\mathrm{i}
  Q_\alpha \big) 
  \big]\epsilon^\mathrm{c}  \nonumber\\
  &\, -B_\mu{}^m  \big[ \partial_ m-\tfrac16 \partial_m \ln\Delta
  \big]\epsilon^\mathrm{c}  \nonumber \\ 
  &\,+ \tfrac1{240} \mathrm{i} \Delta^{-1/3} 
  \varepsilon^{abcde} \big[ 
  \mathrm{i} \, F_{abcde} -5 F_{\beta abcd} \gamma^\beta
  \Gamma_{e} -5 \mathrm{i} \, F_{\beta\gamma abc}
  \gamma^{\beta\gamma} \Gamma_{de} \big]
  \gamma_\mu \epsilon^\mathrm{c}  \nonumber\\
  &\, -\tfrac1{96} \Delta^{-1/3}  \big[ - \mathrm{i}
  \bar{G}_{bcd}\,    \Gamma^{bcd}  \gamma_\mu +3\,
  \bar{G}_{bc\alpha}\, \Gamma^{bc} \big(\gamma_\mu \gamma^\alpha
  +2\, \gamma^{\alpha}\gamma_\mu \big) \nonumber\\   
  &\,\qquad\qquad\quad + 3 \mathrm{i} \bar{G}_{b\alpha\beta}\,
  \Gamma^{b} \big(\gamma_\mu   \gamma^{\alpha\beta} -2\,
  \gamma^{\alpha\beta} \gamma_{\mu}\big) +
  \bar{G}_{\alpha\beta\gamma}  \big(  \gamma_\mu
  \gamma^{\alpha\beta\gamma} +2 \gamma^{\alpha\beta\gamma}
  \gamma_\mu  \big) \big] \,\epsilon  \nonumber \\ 
  & \, +\tfrac13 \mathrm{i} \Delta^{-1/6}  \gamma_\mu \,
  \Gamma^a \delta \psi_a{}^\mathrm{c} \, . 
\end{align}
where we made use of the self-duality condition on the field strength
\eqref{eq:self-dual} and the gamma matrices defined in appendix
\ref{App:red-spinors-gamma-matrices}, and in particular of
\eqref{eq:gamma-representations}, to simplify the terms involving the
various components of the field strength $F_{ABCDE}$. 

The transformation rules for the gravitini that carry a vector index of
the internal $5D$ space are given by 
\begin{align}
  \label{eq:internal-gravitini-var}
  \delta\psi_a =&\, \Delta^{-1/3} e_a{}^m \big[ \partial_m-\tfrac14
  \omega_m{}^{\alpha\beta} \,\gamma_{\alpha\beta} -\tfrac12 \mathrm{i}
  \omega_m{}^{\alpha a}\, \Gamma_a \gamma_\alpha
  -\tfrac14\omega_m{}^{ab}\,\Gamma_{ab} -\tfrac12\mathrm{i}Q_m
  -\tfrac16 \partial_m\ln\Delta\big]\epsilon \nonumber\\
  &\,+\tfrac1{240} \mathrm{i} \Delta^{-1/3} \varepsilon^{bcdef} \big[
  F_{bcdef} +5\mathrm{i}\, F_{\alpha bcde} \gamma^\alpha
  \Gamma_f - 5\, F_{\alpha\beta bcd} \gamma^{\alpha\beta} 
  \Gamma_{ef} 
  \big]
  \Gamma_a\epsilon \nonumber\\
  &\, -\tfrac1{96} \Delta^{-1/3} \big[ G_{bcd}\, \big(\Gamma_a
  \Gamma^{bcd} +2\, \Gamma^{bcd}\Gamma_a\big) - 3\mathrm{i} G_{bc\alpha}\,
  \gamma^\alpha \big(\Gamma_a \Gamma^{bc} -2\, \Gamma^{bc}\Gamma_a
  \big) \nonumber\\  
  &\,\qquad\qquad\quad + 3 G_{b\alpha\beta}\, \gamma^{\alpha\beta} \big(\Gamma_a
  \Gamma^{b} +2\, \Gamma^{b} \Gamma_a\big) +\mathrm{i}G_{\alpha\beta\gamma}
  \gamma^{\alpha\beta\gamma} \Gamma_a  \big] \,\epsilon^\mathrm{c}\, , \nonumber\\[2mm]
  \delta\psi_a{}^\mathrm{c} =&\, \Delta^{-1/3} e_a{}^m \big[ \partial_m-\tfrac14
  \omega_m{}^{\alpha\beta} \,\gamma_{\alpha\beta} -\tfrac12 \mathrm{i}
  \omega_m{}^{\alpha a}\, \Gamma_a \gamma_\alpha
  -\tfrac14\omega_m{}^{ab}\,\Gamma_{ab} +\tfrac12\mathrm{i}Q_m
  -\tfrac16 \partial_m\ln\Delta\big]\epsilon^\mathrm{c} \nonumber\\
  &\,- \tfrac1{240} \mathrm{i} \Delta^{-1/3} \varepsilon^{bcdef}\big[
  F_{bcdef} +5\mathrm{i}\, F_{\alpha bcde} \gamma^\alpha
  \Gamma_{f} -5\, F_{\alpha\beta bcd}\gamma^{\alpha\beta} 
  \Gamma^{ef} 
  \big]
  \Gamma_a\epsilon^\mathrm{c} \nonumber\\
  &\, -\tfrac1{96} \Delta^{-1/3} \big[ \bar{G}_{bcd}\, \big(\Gamma_a 
  \Gamma^{bcd} +2\, \Gamma^{bcd}\Gamma_a\big) - 3\mathrm{i}
  \bar{G}_{bc\alpha}\, 
  \gamma^\alpha \big(\Gamma_a \Gamma^{bc} -2\, \Gamma^{bc}\Gamma_a 
  \big) \nonumber\\  
  &\,\qquad\qquad\quad + 3 \bar{G}_{b\alpha\beta}\,
  \gamma^{\alpha\beta} \big(\Gamma_a 
  \Gamma^{b} +2\, \Gamma^{b} \Gamma_a\big) +\mathrm{i}
  \bar{G}_{\alpha\beta\gamma} 
  \gamma^{\alpha\beta\gamma} \Gamma_a  \big] \,\epsilon \, . 
\end{align}

The next topic concerns the rank-2 tensor fields $A^\alpha{\!}_{MN}$,
which decompose into twenty scalars $A^\alpha{\!}_{mn}$, ten $5D$
vectors $A^\alpha{\!}_{\mu m}$ and two $5D$ 2-rank tensors
$A^\alpha{\!}_{\mu\nu}$. Their consistent Kaluza-Klein definitions are
as follows,
\begin{align}
  \label{eq:KK-deco-2-tensor}
  A^\alpha{}_{mn} {\!}^\mathrm{KK} =&\, A^\alpha{}_{mn} \,, \nonumber\\
  A^\alpha{}_{\mu m}{\!}^\mathrm{KK} =&\,  A^\alpha{}_{\mu m}
  -B_\mu{}^p \,A^\alpha{}_{pm}  \,, \nonumber\\ 
 A^\alpha{}_{\mu\nu}{\!}^\mathrm{KK} =&\,  A^\alpha{}_{\mu\nu} +2\,
  B_{[\mu}{}^p \,A^\alpha{} _{\nu]p}  + B_\mu{}^p\,B_\nu{}^q
  A^\alpha{}_{pq}  \,. 
\end{align}
Their supersymmetry variations take the form, 
\begin{align}
  \label{eq:2-tensor-var}
  \delta A^\alpha{\!}_{mn} =&\, -\tfrac12\mathrm{i} \,\phi^\alpha
  \big[\bar\epsilon^\mathrm{c}\,\Gamma_{mn} \lambda^\mathrm{c} - 4\,
  \bar\epsilon\, \Gamma_{[m} \psi_{n]}{\!}^\mathrm{c} \big]
  -\tfrac12\mathrm{i} \,\varepsilon^{\alpha\beta} \phi_\beta
  \big[\bar\epsilon\,\Gamma_{mn} \lambda - 4\,
  \bar\epsilon^\mathrm{c} \, \Gamma_{[m} \psi_{n]} \big]\,,\nonumber\\[2mm]
   \delta A^\alpha{\!}_{\mu m} =&\, - \tfrac12\Delta^{-1/3}
   \phi^\alpha\big[ 2\mathrm{i}\,\bar\epsilon\, 
   \Gamma_m\psi_\mu{\!}^\mathrm{c} - 2\,\bar\epsilon\,\gamma_\mu
   (\delta_m{}^n -\tfrac13\Gamma_m\Gamma^n)\psi_n{\!}^\mathrm{c}
   +\bar\epsilon^\mathrm{c} \,\Gamma_m\gamma_\mu\lambda^\mathrm{c} \big] \nonumber\\
   &\,-\tfrac12 \Delta^{-1/3} \varepsilon^{\alpha\beta}\phi_\beta
   \big[2\mathrm{i}\,\bar\epsilon^\mathrm{c} \,\Gamma_m\psi_\mu -
   2\,\bar\epsilon^\mathrm{c} \gamma_\mu (\delta_m{}^n
   -\tfrac13\Gamma_m\Gamma^n)  \psi_n  +
   \bar\epsilon\,\Gamma_m\gamma_\mu\lambda\big]\nonumber\\ 
   &\, - \delta B_\mu{}^p\,A^\alpha{\!}_{pm} \,, \nonumber\\[2mm]
   \delta A^\alpha{\!}_{\mu\nu} =&\,
   -\tfrac12\Delta^{-2/3}\phi^\alpha\big[ - 4\,\bar\epsilon
   \,\gamma_{[\mu} \psi_{\nu]}{}^\mathrm{c} +\tfrac43\mathrm{i}
   \bar\epsilon\, \gamma_{\mu\nu} \Gamma^m\psi_m{\!}^\mathrm{c}
   +\mathrm{i}\,\bar\epsilon^\mathrm{c}
   \gamma_{\mu\nu}\lambda^\mathrm{c} \big] \nonumber\\
   &\,- \tfrac12\Delta^{-2/3}\varepsilon^{\alpha\beta}\phi_\beta
   \big[-4\,\bar\epsilon^\mathrm{c} \gamma_{[\mu} \psi_{\nu]}
   +\tfrac43\mathrm{i}\,\bar\epsilon^\mathrm{c}\gamma_{\mu\nu}\Gamma^m\psi_m
   +\mathrm{i}\,\bar\epsilon\,\gamma_{\mu\nu}\lambda\big]
   \nonumber\\ 
   &\, +2\, \delta B_{[\mu}{}^p \,A^\alpha{\!}_{\nu]p} \,,
\end{align}
where we have suppressed the $\mathrm{KK}$-label on both sides of the
equations. 

Subsequently we consider the 4-rank tensor $A_{MNPQ}$ which decomposes
into five $5D$ scalars $A_{mnpq}$, ten $5D$ vectors $A_{\mu mnp}$, ten
$5D$ 2-rank tensors $A_{\mu\nu mn}$, five $5D$ 3-rank tensors
$A_{\mu\nu\rho p}$ and one $5D$ 4-rank
tensor $A_{\mu\nu\rho\sigma}$. Their consistent definition is 
\begin{align}
  \label{eq:KK-deco-4-tensor}
  A_{mnpq}{\!}^\mathrm{KK} =&\, A_{mnpq} \,, \nonumber\\
  A_{\mu mnp}{\!}^\mathrm{KK} =&\,  A_{\mu mnp}
  -B_\mu{}^q \,A_{qmnp}  \,, \nonumber\\
  A_{\mu\nu mn}{\!}^\mathrm{KK}=&\,A_{\mu\nu mn}+2\,B_{[\mu}{}^q\,A_{\nu]qmn}
  +B_\mu{}^p\,B_\nu{}^q\,A_{pqmn}\,,\nonumber\\ 
  A_{\mu\nu\rho m}{\!}^\mathrm{KK}=&\,A_{\mu\nu\rho m}
  +3\,B_{[\mu}{}^p\, A_{\nu\rho]mp} +3\, B_{[\mu}{}^p\,B_{\nu}{}^q\,
  A_{\rho]mpq} - B_{\mu}{}^p\,B_{\nu}{}^q\,B_{\rho}{}^r\, A_{pqrm}\,,\nonumber\\
  A_{\mu\nu\rho\sigma}{\!}^\mathrm{KK}=&\,A_{\mu\nu\rho\sigma}
  +4\,B_{[\mu}{}^p\,A_{\nu\rho\sigma]p}+6\,B_{[\mu}{}^p\,B_{\nu}{}^q\,A_{\rho\sigma]pq}
  +4\, B_{[\mu}{}^p\,B_{\nu}{}^q\,B_{\rho}{}^r\,A_{\sigma]pqr}
  \nonumber\\ 
  &\, +B_{\mu}{}^p\,B_{\nu}{}^q\,B_{\rho}{}^r\,B_{\sigma}{}^s\,A_{pqrs}\,.  
 \end{align}
 The supersymmetry variations for these fields then take the following
 form,
 \begin{align}
  \label{eq:4-tensor-var}
  \delta A_{mnpq}=& -\tfrac12\bar\epsilon\,\Gamma_{[mnp}\psi_{q]}
  +\tfrac12\bar\epsilon^\mathrm{c}\Gamma_{[mnp}\psi_{q]}{\!}^\mathrm{c}+ \tfrac38
  \mathrm{i}\,\varepsilon_{\alpha\beta}A^\alpha{\!}_{[mn}\,\delta
  A^\beta{\!}_{pq]}\,,\nonumber\\[2mm]  
  \delta A_{\mu mnp}=&\,\tfrac18 \Delta^{-1/3}
  \big[\bar\epsilon\,\Gamma_{mnp}\psi_\mu +
  3\mathrm{i}\,\bar\epsilon\gamma_\mu\Gamma_{[mn}
  (\delta_{p]}{\!}^q-\tfrac19\Gamma_{p]}\Gamma^q)\psi_q\big] \nonumber\\     
  &\,+\tfrac18\Delta^{-1/3}\big[- \bar\epsilon^\mathrm{c}
  \Gamma_{mnp}\psi_\mu{\!}^\mathrm{c}   
  -3\mathrm{i}\,\bar\epsilon^\mathrm{c} \gamma_\mu \Gamma_{[mn}
  (\delta_{p]}{\!}^q-\tfrac19\Gamma_{p]}\Gamma^q )
  \psi_q{\!}^\mathrm{c}\big]\nonumber\\   
  &\, +\tfrac{3}{16}\mathrm{i}\varepsilon_{\alpha\beta}
  \big[A^\alpha{\!}_{\mu[m}\,\delta A^\beta{\!}_{np]} -\delta
  A^\alpha{\!}_{\mu[m} \,A^\beta{\!}_{np]}- \delta B_\mu{}^q
  A^\alpha{\!}_{q[m}\,A^\beta{\!}_{np]} \big]\nonumber\\ 
  & -\delta B_\mu{}^q\,A_{qmnp}\,, \nonumber\\[2mm]
 \delta A_{\mu\nu mn}=&\,
  \tfrac14\Delta^{-2/3}\big[\mathrm{i}\bar\epsilon\,\Gamma_{mn}
  \gamma_{[\mu}\psi_{\nu]} 
  -\bar\epsilon\,\gamma_{\mu\nu}\Gamma_{[m} 
  (\delta_{n]}{\!}^p-\tfrac13\Gamma_{n]}\Gamma^p)\psi_p\big]\nonumber\\
  &\,+\tfrac14\Delta^{-2/3}\big[-\mathrm{i}\bar\epsilon^\mathrm{c}\,
  \Gamma_{mn} \,\gamma_{[\mu}  \psi_{\nu]}{\!}^\mathrm{c}    
  +\bar\epsilon^\mathrm{c}\gamma_{\mu\nu} \Gamma_{[m}
  (\delta_{n]}{\!}^p-\tfrac13\Gamma_{n]}\Gamma^{p}) 
  \psi_p{\!}^\mathrm{c} \big]\nonumber\\   
  &\,+\tfrac{1}{16}\mathrm{i}\varepsilon_{\alpha\beta}
  \big[A^\alpha{\!}_{\mu\nu}\,\delta A^\beta{\!}_{mn} 
  +A^\alpha{\!}_{mn} \,\delta A^\beta{\!}_{\mu\nu} 
  -4\, A^\alpha{\!}_{[\mu[m}\,\delta A^\beta{\!}_{\nu]n]} \big]\nonumber\\
   &\,+\tfrac18\mathrm{i}\varepsilon_{\alpha\beta}\,\delta B_{[\mu}{}^p
   \big[A^\alpha{\!}_{\nu]p}\, A^\beta{\!}_{mn}-2\,A^\alpha{\!}_{\nu][m}\,
    A^\beta{\!}_{n]p}\big]\nonumber\\
    &\,+2\,\delta B_{[\mu}{}^p\, A_{\nu]pmn}\,,\nonumber \\[2mm]
    \delta A_{\mu\nu\rho m}=&\,\tfrac18\Delta^{-1}
    \big[3\,\bar\epsilon\,\Gamma_{m}\gamma_{[\mu\nu}\psi_{\rho]}
    +\mathrm{i}\bar\epsilon\,\gamma_{\mu\nu\rho}
    (\delta_m{\!}^p-\Gamma_m\Gamma^p)\psi_p\big]\nonumber\\
    &\,+\tfrac18\Delta^{-1}\big[-3\,\bar\epsilon^\mathrm{c}\, 
    \Gamma_{m}\gamma_{[\mu\nu}\psi_{\rho]}{\!}^\mathrm{c} 
     -\mathrm{i}\bar\epsilon^\mathrm{c}\gamma_{\mu\nu\rho}
    (\delta_m{\!}^p-\Gamma_m\Gamma^p) \psi_p{\!}^\mathrm{c} \big]
     \nonumber\\
    &\,+\tfrac{3}{16}\mathrm{i}\,\varepsilon_{\alpha\beta}\,
    \big[A^\alpha{\!}_{[\mu\nu} \,\delta A^\beta{\!}_{\rho]m}
    -\delta A^\alpha{\!}_{[\mu\nu}\, A^\beta{\!}_{\rho]m}\big] \nonumber\\
    &\,+\tfrac{3}{16}\mathrm{i}\,\varepsilon_{\alpha\beta} \, \delta
    B_{[\mu}{}^p \,\big[A^\alpha{\!}_{\nu\rho]}\, A^\beta_{pm}
    +2A^\alpha{\!}_{\nu m}\, A^\beta{\!}_{\rho]p}\big] \nonumber\\
    &\,+3\,\delta B_{[\mu}{}^p\,A_{\nu\rho]mp}\,,\nonumber\\
    \delta A_{\mu\nu\rho\sigma}=&\,\tfrac12\Delta^{-{4}/{3}}
    \big[\mathrm{i}\bar\epsilon\,\gamma_{[\mu\nu\rho}\psi_{\sigma]}
    +\tfrac13\bar\epsilon\,\gamma_{\mu\nu\rho\sigma}
    \Gamma^p\psi_p\big]\nonumber\\
    &\,+\tfrac12\Delta^{-{4}/{3}}\big[-\mathrm{i}\bar\epsilon^\mathrm{c}
    \gamma_{[\mu\nu\rho}\psi_{\sigma]}{\!}^\mathrm{c} 
    -\tfrac13\bar\epsilon^\mathrm{c}\gamma_{\mu\nu\rho\sigma}
    \Gamma^p\psi_p{\!}^\mathrm{c}  \big]\nonumber\\
    &\,+\tfrac38\mathrm{i}\varepsilon_{\alpha\beta} \,
    A^\alpha{\!}_{[\mu\nu}\,\delta A^\beta{\!}_{\rho\sigma]}
    -\tfrac34\mathrm{i}\varepsilon_{\alpha\beta} \,\delta
    B_{[\mu}{}^p\, A^\alpha{\!}_{\nu\rho}\, A^\beta{\!}_{\sigma]p}\nonumber\\
    &\,+4\,\delta B_{[\mu}{}^p\, A_{\nu\rho\sigma]p}\,,
\end{align}
where again we suppressed the $\mathrm{KK}$-label on both sides of
these equations. 

Let us review the various fields that we have obtained and compare
them with the fields that are generically contained in maximal $5D$
supergravity. First of all we have the f\"unfbein field
$e_\mu{}^\alpha$ and the eight independent gravitini fields consisting
of the fields $(\psi_\mu,\psi_\mu{\!}^\mathrm{c})$. Furthermore there
are 48 spin-1/2 fields consisting of $(\psi_a,\psi_a{\!}^\mathrm{c})$,
and $(\lambda,\lambda^\mathrm{c})$.

Then there are 42 scalar fields, consisting of $e_m{}^a$,
$\phi^\alpha$, $A^\alpha{\!}_{mn}$ and $A_{mnpq}$. The field $e_m{}^a$
corresponds to 15 scalars and the fields $\phi^\alpha$ to 2 scalars
upon subtracting the degrees of freedom associated with tangent space
transformations of the internal space and local $\mathrm{U}(1)$
transformations. The fields $A^\alpha{\!}_{mn}$ and $A_{mnpq}$
describe 20 and 5 scalars, respectively. The total number of scalars
is thus equal to the dimension of the
$\mathrm{E}_{6(6)}/\mathrm{USp}(8)$ coset space that parametrizes the
scalars in $5D$ maximal supergravity.

To appreciate the systematics of the vector and tensor fields we introduce
the following (re)definitions. The 25 vector fields that we have obtained
at this stage will be denoted by 
\begin{align}
  \label{eq:C-vectors}
  C_\mu{}^m =&\, B_\mu{}^m\,,\nonumber\\
  C_\mu{}^\alpha{\!}_m =&\, A^\alpha{\!}_{\mu m}{\!}^\mathrm{KK} \,,\nonumber\\
  C_{\mu\,mnp} =&\, A_{\mu mnp}{\!}^\mathrm{KK}  -\tfrac3{16} \mathrm{i}
  \varepsilon_{\alpha\beta} A^\alpha{\!}_{\mu[m}{\!}^\mathrm{KK}
  \,A^\beta{\!}_{np]}   \,,  
\end{align}
where the extra term in the definition of $C_{\mu\,mnp}$ has been
included such that its supersymmetry variation will not contain the
vector field. Observe also that in the above result we have suppressed
the $\mathrm{KK}$-label for the scalar field $A^\beta{\!}_{ np}$;
henceforth we will do this consistently for both $A^\beta{\!}_{np}$
and $A_{mnpq}$. The fields $C_\mu{}^m$ and $C_{\mu\,mnp}$ can be
combined into the 15-dimensional anti-symmetric representation of
$\mathrm{SL}(6)$. The remaining vector fields $C_\mu{}^\alpha{\!}_m$
transform as five doublets under
$\mathrm{SU}(1,1)\cong\mathrm{SL}(2)$. As compared to the vector
fields of $5D$ maximal supergravity, we should expect six such
doublets. As we will show in the next section, the extra doublet will
emerge from a dual tensor field, $A^\alpha{\!}_{MNPQRS}$, which leads
to the fields $A^\alpha{\!}_{\mu mnpqr}$.  In view of the self-duality
constraint \eqref{eq:self-dual}, we do not expect any tensor fields
dual to $A_{MNPQ}$. 

The 25 vector fields \eqref{eq:C-vectors} transform as follows, 
\begin{align}
  \label{eq:delta-C-vectors}
    \delta C_\mu{}^m =&\, \tfrac12\Delta^{-1/3} e_a{}^m
  \big[\mathrm{i} \big(\bar\epsilon\, \Gamma^a \psi_\mu
  +\bar\epsilon^\mathrm{c}\, \Gamma^a \psi_\mu{\!}^\mathrm{c}\big)
  \nonumber\\ 
&\qquad\qquad\quad
  + \bar\epsilon\,\gamma_\mu (\delta^a{}_b +\tfrac13 \Gamma^a\Gamma_b)
  \psi^b +\bar\epsilon^\mathrm{c}\,\gamma_\mu (\delta^a{}_b +\tfrac13
  \Gamma^a\Gamma_b) \psi^b{}^\mathrm{c}  \big]  \,,\nonumber\\[2mm]
   \delta C_\mu{}^\alpha{\!}_m =&\,  - \tfrac12\Delta^{-1/3}
   \phi^\alpha\big[ 2\mathrm{i}\,\bar\epsilon\, 
   \Gamma_m\psi_\mu{\!}^\mathrm{c} - 2\,\bar\epsilon\,\gamma_\mu
   (\delta_m{}^n -\tfrac13\Gamma_m\Gamma^n)\psi_n{\!}^\mathrm{c}
   +\bar\epsilon^\mathrm{c} \,\Gamma_m\gamma_\mu\lambda^\mathrm{c} \big] \nonumber\\
   &\,-\tfrac12 \Delta^{-1/3} \varepsilon^{\alpha\beta}\phi_\beta
   \big[2\mathrm{i}\,\bar\epsilon^\mathrm{c} \,\Gamma_m\psi_\mu -
   2\,\bar\epsilon^\mathrm{c} \gamma_\mu (\delta_m{}^n
   -\tfrac13\Gamma_m\Gamma^n)  \psi_n  +
   \bar\epsilon\,\Gamma_m\gamma_\mu\lambda\big]\nonumber\\ 
   &\,+ \tfrac12 \mathrm{i}\,\Delta^{-1/3} A^\alpha{\!}_{mp} 
   \big[ \bar\epsilon\, \Gamma^p \psi_\mu
   +\bar\epsilon^\mathrm{c}\, \Gamma^p \psi_\mu{\!}^\mathrm{c}\big]
   \nonumber\\ 
   \,& + \tfrac12\Delta^{-1/3} A^\alpha{\!}_{mp}  \big[
   \bar\epsilon\,\gamma_\mu (e_a{}^p  +\tfrac13 \Gamma^p\Gamma_a)
  \psi^a +\bar\epsilon^\mathrm{c}\,\gamma_\mu (e_a{}^p +\tfrac13
  \Gamma^p\Gamma_a) \psi^a{}^\mathrm{c}  \big]  \,,\nonumber\\[2mm] 
  \delta C_{\mu\,mnp} =&\, \tfrac18 \Delta^{-1/3}
  \big[\bar\epsilon\,\Gamma_{mnp}\psi_\mu +
  3\mathrm{i}\,\bar\epsilon\gamma_\mu\Gamma_{[mn}
  (\delta_{p]}{}^q-\tfrac19\Gamma_{p]}\Gamma^q)\psi_q\big] \nonumber\\
  &\,+\tfrac18\Delta^{-1/3}\big[-\bar\epsilon^\mathrm{c}\,\Gamma_{mnp}
  \psi_\mu{\!}^\mathrm{c} -3\mathrm{i}\,\bar\epsilon^\mathrm{c}
  \gamma_\mu \Gamma_{[mn} (\delta_{p]}{}^q-\tfrac19\Gamma_{p]}\Gamma^{q})
   \psi_q{\!}^\mathrm{c} \big]\nonumber\\
  &\, - \tfrac3{16}\mathrm{i}\,\Delta^{-1/3} \varepsilon_{\alpha\beta}
  \,A^\alpha{\!}_{[mn} \phi^\beta\big[ 2\mathrm{i}\,\bar\epsilon\,
  \Gamma_{p]}\psi_\mu{\!}^\mathrm{c} - 2\,\bar\epsilon\,\gamma_\mu
  (\delta_{p]}{}^n -\tfrac13\Gamma_{p]}\Gamma^n)\psi_n{\!}^\mathrm{c}
  +\bar\epsilon^\mathrm{c} \,\Gamma_{p]}\gamma_\mu\lambda^\mathrm{c}
  \big] \nonumber\\
  &\,+\tfrac3{16}\mathrm{i}\, \Delta^{-1/3} \,A^\alpha{\!}_{[mn}
  \,\phi_\alpha \big[2\mathrm{i}\,\bar\epsilon^\mathrm{c}
  \,\Gamma_{p]}\psi_\mu - 2\,\bar\epsilon^\mathrm{c} \gamma_\mu
  (\delta_{p]}{}^n -\tfrac13\Gamma_{p]}\Gamma^n) \psi_n +
  \bar\epsilon\,\Gamma_{p]}\gamma_\mu\lambda\big]\nonumber\\
  & +\tfrac12\mathrm{i} \,\Delta^{-1/3} \big[A_{mnpq}
  +\tfrac{3}{16}\mathrm{i}\varepsilon_{\alpha\beta}
  A^\alpha{\!}_{[mn}\,A^\beta{\!}_{p]q} \big] \big[\bar\epsilon\,
  \Gamma^q \psi_\mu +\bar\epsilon^\mathrm{c}\, \Gamma^q
  \psi_\mu{\!}^\mathrm{c}\big]
  \nonumber\\
  & +\tfrac12\Delta^{-1/3} \big[A_{mnpq}
  +\tfrac{3}{16}\mathrm{i}\varepsilon_{\alpha\beta}
  A^\alpha{\!}_{[mn}\,A^\beta{\!}_{p]q}  \big] \nonumber \\
  &\qquad\qquad\quad \times\big[\bar\epsilon\,\gamma_\mu (e_b{}^q
  +\tfrac13 \Gamma^q\Gamma_b) \psi^b
  +\bar\epsilon^\mathrm{c}\,\gamma_\mu (e_b{}^q +\tfrac13
  \Gamma^q\Gamma_b) \psi^b{}^\mathrm{c} \big] \,.
\end{align}

Furthermore we have identified 12 two-rank tensor fields, which we
define by 
\begin{align}
  \label{eq:C-2-tensors}
  C_{\mu\nu}{}^\alpha =&\, A^\alpha{\!}_{\mu\nu}{\!}^\mathrm{KK}
  -C_{[\mu}{\!}^p \, C_{\nu]}{}^\alpha{}_p \,,   \nonumber\\
  C_{\mu\nu\,mn}=&\, A_{\mu\nu mn}{\!}^\mathrm{KK}-\tfrac1{16} \mathrm{i}
  \varepsilon_{\alpha\beta} A^\alpha{\!}_{\mu\nu}{\!}^\mathrm{KK}\,A^\beta{\!}_{mn} -
  C_{[\mu}{\!}^p \, C_{\nu]pmn} \,. 
\end{align}
The supersymmetry transformations of these tensors are expressed by 
\begin{align}
  \label{eq:delta-C-2-tensors}
  &\delta C_{\mu\nu}{}^\alpha +C_{[\mu}{\!}^p \, \delta
  C_{\nu]}{}^\alpha{\!}_p+ C _{[\mu}{}^\alpha{\!}_p \,\delta
  C_{\nu]}{}^p  \nonumber\\
  &=    -\tfrac12\Delta^{-2/3}\phi^\alpha\big[ - 4\,\bar\epsilon
   \,\gamma_{[\mu} \psi_{\nu]}{}^\mathrm{c} +\tfrac43\mathrm{i}
   \bar\epsilon\, \gamma_{\mu\nu} \Gamma^m\psi_m{\!}^\mathrm{c}
   +\mathrm{i}\,\bar\epsilon^\mathrm{c}
   \gamma_{\mu\nu}\lambda^\mathrm{c} \big] \nonumber\\
   &\,\quad - \tfrac12\Delta^{-2/3}\varepsilon^{\alpha\beta}\phi_\beta
   \big[-4\,\bar\epsilon^\mathrm{c} \gamma_{[\mu} \psi_{\nu]}
   +\tfrac43\mathrm{i}\,\bar\epsilon^\mathrm{c}\gamma_{\mu\nu}\Gamma^m\psi_m
   +\mathrm{i}\,\bar\epsilon\,\gamma_{\mu\nu}\lambda\big]\,,
   \nonumber\\[2mm]
  &\delta C_{\mu\nu\,mn} + C_{[\mu}{\!}^p \, \delta C_{\nu]pmn}+  
  C_{[\mu \,pmn} \,\delta C_{\nu]}{}^p +\tfrac{1}{4}\mathrm{i}\varepsilon_{\alpha\beta}
  \, C_{[\mu}{\!}^\alpha{\!}_{[m}\,  \delta
  C_{\nu]}{\!}^\beta{\!}_{n]} \nonumber \\
 & = 
  \tfrac14\Delta^{-2/3}\big[\mathrm{i}\bar\epsilon\,\Gamma_{mn}
  \gamma_{[\mu}\psi_{\nu]} 
  -\bar\epsilon\,\gamma_{\mu\nu}\Gamma_{[m} 
  (\delta_{n]}^p-\tfrac13\Gamma_{n]}\Gamma^p)\psi_p\big]\nonumber\\
  &\quad+ \tfrac14\Delta^{-2/3}\big[-\mathrm{i}\bar\epsilon^\mathrm{c}\,
  \Gamma_{mn} \,\gamma_{[\mu}  \psi_{\nu]}{\!}^\mathrm{c}    
  + \bar\epsilon^\mathrm{c}\gamma_{\mu\nu} \Gamma_{[m}
  (\delta_{n]}{\!}^p-\tfrac13\Gamma_{n]}\Gamma^{p}) 
  \psi_p{\!}^\mathrm{c} \big]\nonumber\\   
   & \quad -\tfrac1{16}\mathrm{i}\,\Delta^{-2/3}\varepsilon_{\alpha\beta}
   \,A^\alpha{\!}_{mn} \, \phi^\beta\big[ - 4\,\bar\epsilon
   \,\gamma_{[\mu} \psi_{\nu]}{}^\mathrm{c} +\tfrac43\mathrm{i}
   \bar\epsilon\, \gamma_{\mu\nu} \Gamma^m\psi_m{\!}^\mathrm{c}
   +\mathrm{i}\,\bar\epsilon^\mathrm{c}
   \gamma_{\mu\nu}\lambda^\mathrm{c} \big] \nonumber\\
   &\quad +\tfrac1{16}\mathrm{i}\,\Delta^{-2/3} \,A^\alpha{\!}_{mn}\,\phi_\alpha  
   \big[-4\,\bar\epsilon^\mathrm{c} \gamma_{[\mu} \psi_{\nu]}
   +\tfrac43\mathrm{i}\,\bar\epsilon^\mathrm{c}\gamma_{\mu\nu}\Gamma^m\psi_m
   +\mathrm{i}\,\bar\epsilon\,\gamma_{\mu\nu}\lambda\big]  \,.
\end{align}
These transformation rules are in line with what is known from the
vector-tensor hierarchy that appears in the context of the embedding
tensor formalism \cite{deWit:2004nw,deWit:2008ta}. We have actually
verified that also the variation of the three-rank tensor fields,
$A_{\mu\nu\rho m}{\!}^\mathrm{KK}$ listed in
\eqref{eq:KK-deco-4-tensor} will exhibit the same structure upon
introducing a suitable modification. Since we will not be considering
tensors of rank higer than two, we refrain from giving further
details.

At this point the number of tensor fields is less than the 27 fields
that one expects on the basis of $5D$ maximal supergravity in the
context of the embedding tensor formalism. Ten extra 2-rank tensors
$A^\alpha{\!}_{\mu\nu mnpq}$ will be provided by the dual field,
$A^\alpha{\!}_{MNPQRS}$, which will bring the total of 2-rank tensors
to 22. The dual vectors and tensors are evaluated in the next section.

\section{Dual fields and the vector-tensor hierarchy}
\label{sec:vector-tensor-hierarchy-dual}
\setcounter{equation}{0}
In \eqref{eq:2-form-f-eq} we presented the field equation for the
tensor fields $A^\alpha{\!}_{MN}$ written as a Bianchi identity of
the seven-rank field strength $F_{\alpha\,MNPQRST}$ defined
in \eqref{eq:F-7-alpha}. The field equation thus implies that
this field strength can be written in terms of a dual six-form field
$A_{\alpha\,MNPQRS}$ according to
\begin{equation}
  \label{eq:6-form}
  F_{\alpha\,MNPQRST}= 6\,\partial_{[M} A_{\alpha\,NPQRST]}\,. 
\end{equation}
It is not possible to derive an expression for $A_{\alpha\,MNPQRS}$ in
closed form, but it is possible to determine how this field transforms
under supersymmetry. Obviously, the Bianchi identity
\eqref{eq:2-form-f-eq} should transform under supersymmetry into
fermionic equations which are of at most first order in
derivatives. Therefore one expects that $F_{\alpha\,MNPQRST}$
transforms into fermionic field equations and into terms that carry
explicit space-time derivatives such that they can be identified as 
the result of the supersymmetry variation of the dual
six-form. Because the field equations are supercovariant all the
contributions of the variation of the six-form can be identified from
the terms that are proportional to the derivative of the supersymmetry
parameters. The consistency of this approach can easily be verified
and it leads to the following result,
\begin{align}
   \label{eq:deltaA6-alpha}
   \delta A_{\alpha\,MNPQRS}= &\,
   \varepsilon_{\alpha\beta} \phi^\beta
   \big(\tfrac16\bar\lambda\,\breve\Gamma_{MNPQRS}\epsilon+
   2\,\bar\epsilon\,\breve\Gamma_{[MNPQR}\psi_{S]}{}^\mathrm{c}\big)
   \nonumber\\
   &\, -\phi_\alpha\big(\tfrac16 \bar\epsilon\,
   \breve\Gamma_{MNPQRS}\lambda -2\,\bar \psi_{[M}{\!}^\mathrm{c}\,
   \breve\Gamma_{NPQRS]}\epsilon\big) \nonumber\\
   &\, -20\mathrm{i}\,\varepsilon_{\alpha\beta}A^\beta{\!}_{[MN} \,\big( \delta
   A_{PQRS]} -\tfrac18 \mathrm{i} \, \varepsilon_{\gamma\delta}
   A ^\gamma{\!}_{PQ} \,\delta A ^\delta{\!}_{RS]}\big) \,.
\end{align}
In particular we note the dual fields $A^\alpha{\!}_{\mu mnpqr}$ and
$A^\alpha{\!}_{\mu\nu mnpq}$, which constitute two $5D$ vector fields
and twelve $5D$ tensor fields transforming under
$\mathrm{SU}(1,1)$. We first consider the transformation rule of the
vector field $A^\alpha{\!}_{\mu mnpqr}$, which takes the following
form,
\begin{align}
   \label{eq:deltaA6-alpha-muKK}
   \delta &A_{\alpha\,\mu mnpqr}=\nonumber\\
   &\, -\tfrac13\mathrm{i}\, \Delta^{-1/3}
   \varepsilon_{\alpha\beta} \phi^\beta 
   \big[\bar\epsilon\,\Gamma_{mnpqr} \psi_\mu{\!}^\mathrm{c} 
   +5\mathrm{i}\,   \bar\epsilon\, \gamma_\mu(\Gamma_{[mnpq}\delta_{r]}{}^s -
   \tfrac1{15}\Gamma_{r]} \Gamma^s) \psi_s{\!}^\mathrm{c}
   +\tfrac12\mathrm{i} \bar\epsilon^\mathrm{c}\,\gamma_\mu
   \Gamma_{mnpqr}\lambda^\mathrm{c} \big]
   \nonumber\\
   &\, -\tfrac13\mathrm{i}\, \Delta^{-1/3} \phi_\alpha
   \big[\bar\epsilon^\mathrm{c} \,\Gamma_{mnpqr}\psi_\mu
   +5\mathrm{i}\,\bar\epsilon^\mathrm{c} \,
   \gamma_\mu\Gamma_{[mnpq}(\delta_{r]}{\!}^s -\tfrac1{15} \Gamma_{r]}
   \Gamma^s )\psi_{s} +\tfrac12\mathrm{i}
   \bar\epsilon\,\gamma_\mu\Gamma_{mnpqr} \lambda \big]
   \nonumber\\
   &\, -\tfrac{20}3\mathrm{i}\,\varepsilon_{\alpha\beta}\big(
   A^\beta{\!}_{\mu[m} \,\delta A_{npqr]} -2\, A^\beta{\!}_{[mn}
   \,\delta A_{pqr]\mu}\big)
   \nonumber\\
   &\,- \tfrac56 
   \varepsilon_{\alpha\beta}\,\varepsilon_{\gamma\delta} \big( 2\,
   A^{(\beta}{\!}_{\mu[m} A^{\gamma)} {\!}_{np} \,\delta A
   ^\delta{\!}_{qr]} - A^\beta{\!}_{[mn} A^\gamma{\!}_{pq} \,\delta A
   ^\delta{\!}_{r]\mu} \big) \nonumber \\
   & \, + \tfrac{40}{3}\mathrm{i}\,  \varepsilon_{\alpha \beta} \ \delta
   B_\mu{\!}^s A^\beta{\!}_{[mn}   
   \big( A_{pqr]s} - \tfrac{1}{16}\mathrm{i} \varepsilon_{\gamma
     \delta} A^\gamma_{pq} \,A^\delta_{r]s} \big) \,,
\end{align}
where on the right-hand side all the fields have been subject to
Kaluza-Klein redefinitions. The field $A_{\alpha\,\mu mnpqr}$ already
transforms consistently as a vector in the $5D$ space-time because
tensors anti-symmetric in more than five internal-space indices must
vanish. The consistency of the above result is confirmed by the fact
that no terms are generated proportional to the Kaluza-Klein
vector field $B_\mu{}^m$, simply because the corresponding terms are
fully anti-symmetric in six internal-space indices and therefore vanish.  

However, from the perspective of the vector-tensor hierarchy further
redefinitions are required, as the supersymmetry variations should not
contain any vector fields, but at most variations of vector fields. A
preliminary analysis suggests to add modifications that are quadratic
and cubic terms in the four- and two-form fields but here we
have to make sure that also the modification itself transforms
consistently as a vector in the $5D$ space-time. This leads us to the
following redefinition,
\begin{equation}
  \label{eq:dualC-vector}
  C_{\mu\,\alpha mnpqr} = A_{\alpha\,\mu mnpqr} + \tfrac{20}3\mathrm{i}
  \,\varepsilon_{\alpha\beta} \,C_\mu{}^\beta{}_{[m} \,A_{npqr]}
  -\tfrac56  \varepsilon_{\alpha\beta}\,\varepsilon_{\gamma\delta}
  \,A^\beta{\!}_{[mn} \, C{\!}_\mu{}^\gamma{}_p \,A^\delta{\!}_{qr]} \,,
\end{equation}
where $C_\mu{}^\alpha{}_m$ is a proper vector field defined in
\eqref{eq:C-vectors}. Under supersymmetry the field $C_{\mu\,\alpha
  mnpqr}$ transforms in the required way,
\begin{align}
  \label{eq:var-dual-vector-C}
  \delta C_{\mu\,\alpha mnpqr}=&\,
  -\tfrac13 \mathrm{i}\,\Delta^{-1/3} \varepsilon_{\alpha\beta}
   \phi^\beta\big[ \bar\epsilon\,\Gamma_{mnpqr}
   \psi_\mu{\!}^\mathrm{c} +5\mathrm{i}\, \bar\epsilon\,
   \gamma_\mu(\Gamma_{[mnpq}\delta_{r]}{}^s - \tfrac1{15}\Gamma_{r]}
   \Gamma^s) \psi_s{\!}^\mathrm{c} +\tfrac12\mathrm{i}
   \bar\epsilon^\mathrm{c}\,\gamma_\mu
   \Gamma_{mnpqr}\lambda^\mathrm{c} \big]
   \nonumber\\
   &\, -\tfrac13 \mathrm{i}\,\Delta^{-1/3} \phi_\alpha \big[\bar
   \epsilon^\mathrm{c} \,\Gamma_{mnpqr}\psi_\mu
   +5\mathrm{i}\,\bar\epsilon^\mathrm{c} \,
   \gamma_\mu\Gamma_{[mnpq}(\delta_{r]}{\!}^s -\tfrac1{15} \Gamma_{r]}
   \Gamma^s )\psi_{s}
   +\tfrac12\mathrm{i} \bar\epsilon\,\gamma_\mu\Gamma_{mnpqr} \lambda
   \big] 
   \nonumber\\
  &\,+\tfrac{20}{3}\mathrm{i} \,\varepsilon_{\alpha\beta}\big[ \delta
  C_\mu{}^\beta{}_{[m} \,  A_{npqr]}
  - 2\, \delta  C_{\mu[mnp}\, A^\beta{\!}_{qr]}
  -2\,\, \delta C_\mu{}^s
  A_{s[mnp}\, A ^\beta{\!}_{qr]} \big] \nonumber\\
  &\, +\tfrac52 \varepsilon_{\alpha\beta}\,
  \varepsilon_{\gamma\delta} \big[ \delta  C_\mu{}^\gamma{}_{[m} \, 
  A^\delta{\!}_{np} \, A^\beta{\!}_{qr]}  +\tfrac13 \delta
  C_\mu{}^s \,A ^\gamma{\!}_{s[m} \, A ^\beta{\!}_{np}\,  A
  ^\delta{\!}_{qr]} \big] \,,
\end{align}
where, for conciseness, we refrained from substituting the explicit
expressions for $C_\mu{\!}^m$, $\delta C_\mu{\!}^\alpha{\!}_{m}$ and $\delta
C_{\mu \,mnp}$ in the right-hand of the last equation.

Subsequently we consider the tensor field $A_{\alpha\,\mu\nu
  mnpq}$. To ensure that this field transforms as a proper $5D$ tensor
one performs the standard Kaluza-Klein redefinition,
\begin{equation}
  \label{KK-dual-tensor} 
A_{\alpha\, \mu\nu mnpq}{\!}^\mathrm{KK}=A_{\alpha\,\mu\nu mnpq}+2
B_{[\mu}{}^r\, A_{\alpha \nu] mnpqr}\,. 
\end{equation}
This modified tensor field transforms as
\begin{align}
   \label{eq:deltaA6-alpha-mu-nu}
   \delta &A_{\alpha\,\mu\nu mnpq}  = \nonumber\\
   &\, -\tfrac23 \mathrm{i}\,\Delta^{-2/3}\varepsilon_{\alpha\beta}
   \phi^\beta\big[ \mathrm{i} \bar\epsilon \,\Gamma_{mnpq}
   \gamma_{[\mu} \psi_{\nu]}{}^\mathrm{c} - 2\,
   \bar\epsilon\,\gamma_{\mu\nu}
   \Gamma_{[mnp}(\delta_{q]}{\!}^r-\tfrac16 \Gamma_{q]}\Gamma^r)
   \psi_{r}{}^\mathrm{c} -\tfrac14\bar\epsilon^\mathrm{c}
   \,\gamma_{\mu\nu} \Gamma_{mnpq} \lambda^\mathrm{c}
   \big] \nonumber\\
 &\, -\tfrac23\mathrm{i}\, \Delta^{-2/3}
   \phi_\alpha \big[ \mathrm{i} \bar\epsilon^\mathrm{c} \,\Gamma_{mnpq}
   \gamma_{[\mu} \psi_{\nu]}- 2\,
   \bar\epsilon^\mathrm{c}\,\gamma_{\mu\nu}
   \Gamma_{[mnp}(\delta_{q]}{\!}^r-\tfrac16 \Gamma_{q]}\Gamma^r)
   \psi_{r} -\tfrac14\bar\epsilon
   \,\gamma_{\mu\nu} \Gamma_{mnpq} \lambda 
   \big] \nonumber\\
   &\, -\tfrac43\mathrm{i}\,\varepsilon_{\alpha\beta}\big[ A^\beta{\!}_{\mu\nu}
   \,\delta A_{mnpq} - 8\,A^\beta{\!}_{[\mu[m} \,\delta A_{\nu]npq]}
   + 6\, A^\beta{\!}_{[mn} \,\delta A_{ pq] \mu\nu}    \big] \nonumber\\
   &\, -\tfrac16 \varepsilon_{\alpha\beta} \,
   \varepsilon_{\gamma\delta} \big( 2\, A^{(\beta}{\!}_{\mu\nu} \, A
   ^{\gamma)} {\!}_{[mn} \,\delta A ^\delta{\!}_{pq]} +
   A^\beta{\!}_{[mn} \,
   A ^\gamma{\!}_{pq]} \,\delta A ^\delta{\!}_{\mu\nu}  \big) \nonumber\\
   &\, +\tfrac23 \varepsilon_{\alpha\beta} \,
   \varepsilon_{\gamma\delta} \big( A^\beta{\!}_{[\mu[m} \, A
   ^\gamma{\!}_{\nu]n} \,\delta A ^\delta{\!}_{pq]} + 2\,
   A^{(\beta}{\!}_{[\mu[m} \, A ^{\gamma)}{\!}_{np} \, \delta
   A^\delta{\!}_{\nu]q]} \big)\nonumber\\
  &\, +\tfrac{16}3\mathrm{i}  \,\varepsilon_{\alpha\beta} \,  \delta B_{[\mu}{}^{r} \,  \big(
   2 A^\beta{\!}_{\nu] [m} \, A_{npq]r} + 3\, A_{\nu] r [mn} \, A^\beta{\!}_{pq]}
    \big) 
   \nonumber\\
   &\,+ \tfrac{1}{3} \,
   \varepsilon_{\alpha\beta}\,\varepsilon_{\gamma\delta}  \,  \delta
   B_{[\mu}{}^{r} \, \big( 4 \, 
   A^{(\beta}{\!}_{ \nu ] [m} A^{\gamma)} {\!}_{np} \,
   A^\delta{\!}_{q] r} 
   + A^\delta{\!}_{\nu] r} A^\beta{\!}_{[mn} \, A
   ^\gamma{\!}_{pq]} \big)  \nonumber\\
   &\,+ 2\,\delta B_{[\mu}{}^r\, A_{\alpha\nu]mnpqr} \,,
\end{align}
where we again dropped  KK-label on both sides of the equation. 

Again this result is not consistent with regard to the vector-tensor
hierarchy so that further redefinitions of the tensor field are
required. As it turns out, they take the following form,
\begin{align}
  \label{eq:dualC-tensor}
  C_{\mu\nu\,\alpha mnpq}=&\, A_{\mu\nu\,\alpha mnpq}
  + \tfrac43\mathrm{i}\,
  \varepsilon_{\alpha\beta}\,A^\beta{\!}_{\mu\nu}\,A_{mnpq}
  \nonumber\\
  &\,
  -\tfrac1{6} 
  \varepsilon_{\alpha\beta}\,\varepsilon_{\gamma\delta}
  \,\big[ A^\gamma{\!}_{\mu\nu} \,  A^\beta{\!}_{[mn}\,
  A^\delta{\!}_{pq]}  -8\, C _{[\mu}{\!}^\beta{\!}_{[m} \,
  C_{\nu]}{\!}^\gamma{\!}_{n}\, A^\delta{\!}_{pq]}\big]  \nonumber\\
  &\,
  -\tfrac{16}{3}\mathrm{i}\,\varepsilon_{\alpha\beta}\,C_{[\mu}{}^\beta{\!}_{[m}
  \, C_{\nu]npq]}-C_{[\mu}{}^r C_{\nu]\,\alpha mnpqr}\,,
\end{align}
where on the the right-hand side the KK-labels have again been
suppressed.  The transformation rule of $C_{\mu\nu\,\alpha mnpq}$
takes the form
\begin{align}
  \label{eq:var-dual-tensor-C}
  \delta& C_{\mu\nu\,\alpha mnpq} -\tfrac{16}{3}\mathrm{i}\,
  \varepsilon_{\alpha\beta}\,\big[ C_{[\mu}{}^\beta{\!}_{[m} \,
    \delta C_{\nu]npq]} + C_{[\mu[npq}\, \delta
    C_{\nu]}{}^\beta{\!}_{m]} \big] + 
    C_{[\mu}{}^r\,\delta C_{\nu]\,\alpha mnpqr}+
    C_{[\mu\,\alpha mnpqr}\, \delta C_{\nu]}{}^r \nonumber\\[1mm] 
    =
    &\, -\tfrac23\mathrm{i}\, \Delta^{-2/3}\varepsilon_{\alpha\beta}
    \phi^\beta\Big[\big[ \mathrm{i} \bar\epsilon \,\Gamma_{mnpq}
    \gamma_{[\mu} \psi_{\nu]}{}^\mathrm{c} - 2\,
    \bar\epsilon\,\gamma_{\mu\nu}
    \Gamma_{[mnp}(\delta_{q]}{\!}^r-\tfrac16 \Gamma_{q]}\Gamma^r)
    \psi_{r}{}^\mathrm{c} -\tfrac14\bar\epsilon^\mathrm{c}
    \,\gamma_{\mu\nu} \Gamma_{mnpq} \lambda^\mathrm{c}
    \big] \nonumber\\
    &\,\qquad\qquad\qquad\qquad  +A_{mnpq} \big[ - 4\,\bar\epsilon
   \,\gamma_{[\mu} \psi_{\nu]}{}^\mathrm{c} +\tfrac43\mathrm{i}
   \bar\epsilon\, \gamma_{\mu\nu} \Gamma^m\psi_m{\!}^\mathrm{c}
   +\mathrm{i}\,\bar\epsilon^\mathrm{c}
   \gamma_{\mu\nu}\lambda^\mathrm{c} \big] \Big]\nonumber\\
   &\, -\tfrac23\mathrm{i}\, \Delta^{-2/3} \phi_\alpha \Big[ \big[ \mathrm{i}
   \bar\epsilon^\mathrm{c} \,\Gamma_{mnpq} \gamma_{[\mu} \psi_{\nu]}-
   2\, \bar\epsilon^\mathrm{c}\,\gamma_{\mu\nu}
   \Gamma_{[mnp}(\delta_{q]}{\!}^r-\tfrac16 \Gamma_{q]}\Gamma^r)
   \psi_{r} -\tfrac14\bar\epsilon \,\gamma_{\mu\nu} \Gamma_{mnpq}
   \lambda     \big] \nonumber\\
   &\,\qquad\qquad\quad \quad -A_{mnpq} \big[-4\,\bar\epsilon^\mathrm{c}
   \gamma_{[\mu} \psi_{\nu]}
   +\tfrac43\mathrm{i}\,\bar\epsilon^\mathrm{c}
   \gamma_{\mu\nu}\Gamma^m\psi_m 
   +\mathrm{i}\,\bar\epsilon\,\gamma_{\mu\nu}\lambda\big]\Big]
   \nonumber\\
     &\, - 2 \mathrm{i} \, \Delta^{-2/3} \,\varepsilon_{\alpha\beta}\,
    A^\beta{\!}_{[mn}  \Big[ \big[\mathrm{i}\bar\epsilon\,\Gamma_{pq]}
   \gamma_{[\mu}\psi_{\nu]} -\bar\epsilon\,\gamma_{\mu\nu}\Gamma_{p}
   (\delta_{q]}{}^r-\tfrac13\Gamma_{q]}\Gamma^r)\psi_r\big]\nonumber\\  
   &\,\qquad\qquad\qquad\qquad\quad -\big[\mathrm{i}\bar\epsilon^\mathrm{c}\,
  \Gamma_{pq]} \,\gamma_{[\mu}  \psi_{\nu]}{\!}^\mathrm{c}    
  - \bar\epsilon^\mathrm{c}\gamma_{\mu\nu} \Gamma_{p}
  (\delta_{q]}{\!}^r-\tfrac13\Gamma_{q]}\Gamma^{r}) 
  \psi_r{\!}^\mathrm{c} \big]\Big] \nonumber\\ 
      &\, - \mathrm{i}\, \Delta^{-2/3} \varepsilon_{\alpha\beta}\,
     A^\beta{\!}_{[mn}  \Big[\varepsilon_{\gamma\delta}
   \,A^\gamma{\!}_{pq]} \, \phi^\delta\big[  \mathrm{i}\,\bar\epsilon
   \,\gamma_{[\mu} \psi_{\nu]}{}^\mathrm{c} +\tfrac13 
   \bar\epsilon\, \gamma_{\mu\nu} \Gamma^r\psi_r{\!}^\mathrm{c}
   +\tfrac14 \bar\epsilon^\mathrm{c}
   \gamma_{\mu\nu}\lambda^\mathrm{c} \big] \nonumber\\
   &\,\qquad\qquad\qquad\qquad\quad - A^\gamma{\!}_{pq]}\,\phi_\gamma
   \big[\mathrm{i}\,\bar\epsilon^\mathrm{c}   \gamma_{[\mu} \psi_{\nu]}
   +\tfrac13\,\bar\epsilon^\mathrm{c}\gamma_{\mu\nu}\Gamma^r\psi_r
   +\tfrac14\bar\epsilon\,\gamma_{\mu\nu}\lambda\big]\Big] \,.
\end{align}

To conclude this section let us summarize the situation regarding the
vector and tensor fields. We have identified precisely 27 vector fields,
namely,
\begin{equation}
  \label{eq:vectors-27}
  C_\mu{}^M = \big\{ C_\mu{\!}^m\,, \,  C_{\mu\, mnp} \,, 
  \,C_\mu{\!}^\alpha{\!}_m \, ,\,  C_{\mu\,\alpha mnpqr} \big\}   \,. 
\end{equation}
For the tensor fields the situation is somewhat different. First of
all, we expect 27 tensor fields whereas previously we found only 22
fields. Secondly, we note that the tensor fields, which we will denote
by $C_{\mu\nu\, Q}$, carry different indices. The vector-tensor
hierarchy implies that there must be 5 additional tensor fields and
furthermore requires the existence of a constant tensor $d_{Q,MN}$,
symmetric in $(M,N)$, in order to obtain the characteristic term
$d_{Q,MN} \, C_{[\mu}{}^{M}\, \delta C_{\nu]}{}^{N}$ in $\delta
C_{\mu\nu\,Q}$. Assuming that the overall covariance of this
expression must be preserved and that precisely five additional fields
are needed, one deduces that these five fields can be precisely
represented by new fields $C_{\mu\nu\,m;npqrs}$, where the array
$[npqrs]$ is fully antisymmetric. Hence the decomposition of the 27
tensors  takes the following form, in direct analogy with
\eqref{eq:vectors-27}, 
\begin{equation}
  \label{eq:tensors-27}
  C_{\mu\nu\,Q} = \big\{ C_{\mu\nu\,m;npqrs} \,, \,C_{\mu\nu\,mn} \,
  ,\,  C_{\mu\nu\,\alpha mnpq}  \,, \,C_{\mu\nu}{\!}^\alpha  \big\}   \,. 
\end{equation}
The new field $C_{\mu\nu\,m;npqrs}$ indeed has the representation that
is expected from the dualization of $10D$ gravity
\cite{Curtright:1980yk,Hull:2001iu} (although this dualization can not
be fully understood at the non-linear level in $10D$
\cite{Bekaert:2002uh}). 

The systematics of the vector and tensor fields can be improved upon
converting to dual representations by extracting the anti-symmetric
tensors $\mathring{e}\,\varepsilon_{mnpqr}$ and/or
$\varepsilon_{\alpha\beta}$. Note that the first tensor depends only
on the reference background of the internal space, because of the
definition $\mathring{e}(y) \equiv\det[\,\mathring{e}_m{}^a(y)]$, and
not on the space-time coordinates $x^\mu$. Hence these conversions
have no bearing on the supersymmetry transformations nor the vector
and tensor gauge transformations. Now consider the following
redefinitions for the vector fields,
\begin{equation}
  \label{eq:index-changes-vector}
    \begin{array}{rcl} 
    C_\mu{\!}^m &\!\!=\!\!&  C_\mu{\!}^m\,,\\[2mm]
    C_\mu{\!}^\alpha{\!}_m &\!\!=\!\!& \mathrm{i}\,\varepsilon^{\alpha\beta} \,
    C_{\mu\,\beta m} \,,
  \end{array}
  \quad
  \begin{array}{rcl} 
  C_{\mu\,mnp} &\!\!=\!\!& \tfrac1{128} \sqrt{5}\,\mathring{e}\,
  \varepsilon_{mnpqr} 
  \,C_\mu{\!}^{qr}\,, \\[2mm]
  C_{\mu\,\alpha mnpqr} &\!\!=\!\!& -\tfrac16\sqrt{5} 
  \mathring{e} \,\varepsilon_{mnpqr} 
  \,C_{\mu \,\alpha} \,. 
  \end{array}
\end{equation}
For the tensor fields the corresponding redefinitions are 
\begin{equation}
  \label{eq:index-changes-tensor}
   \begin{array}{rcl} 
     C_{\mu\nu\,m;npqrs} &\!\!\propto \!\!&  \mathring{e} \,\varepsilon_{npqrs}\, 
     C_{\mu\nu\,m} \,, \\[2mm]
     C_{\mu\nu\,\alpha mnpq} &\!\!=\!\!& \frac16\sqrt{5}\mathrm{i}
     \mathring{e}\, 
     \varepsilon_{mnpqr} \,\varepsilon_{\alpha\beta} \,
     C_{\mu\nu}{\!}^{\beta r}  \,,
   \end{array}
   \quad
   \begin{array}{rcl} 
     C_{\mu\nu\,mn} &\!\!=\!\!& C_{\mu\nu\,mn} \,, \\[2mm]
     C_{\mu\nu}{\!}^\alpha &\!\!=\!\!& C_{\mu\nu}{\!}^\alpha  \,.
    \end{array}
\end{equation}
Now the vector and tensor fields can be written as $C_\mu{}^M$ and
$C_{\mu\nu\,M}$, respectively, where the indices $M$ decompose
according to ${}^M=\big\{ {}^m\,,\,{}^{mn}\,,\, {}_{\alpha\,m}\,, \,
{}_{\alpha}\big\}$ and ${}_M=\big\{ {}_m\,,\,{}_{mn}\,,\,
{}^{\alpha\,m}\,, \, {}^{\alpha}\big\}$, respectively. Here we observe
that the normalization of the vector and tensor fields is at this
point completely arbitrary.  Nevertheless, identifying the (upper) index $M$ on
$C_\mu{}^M$ with the $\overline{\mathbf{27}}$ representation of
$\mathrm{E}_{6(6)}$ and the (lower) index $M$ on $C_{\mu\nu M}$ as the
$\mathbf{27}$ representation, then the decompositions
\eqref{eq:index-changes-vector} and \eqref{eq:index-changes-tensor}
correspond to the branchings
\begin{align}
    \label{eq:27-branching}
    \overline{\mathbf{27}}  \;&\stackrel{\mathrm{SL}(2) \times
      \mathrm{SL}(6)}{\longrightarrow}\;  
    (\mathbf{1} ,\overline{\mathbf{15}} )+(\mathbf{2}
    ,{\mathbf{6}} ) 
    \;\stackrel{\mathrm{SL}(2) \times
      \mathrm{SO}(5)}{\longrightarrow}\;  (\mathbf{1}
    ,\mathbf{5} ) + (\mathbf{1} ,\mathbf{10} ) + (\mathbf{2}
    ,\mathbf{5}) + (\mathbf{2} ,\mathbf{1})\,, \nonumber\\
    {\mathbf{27}}  \;&\stackrel{\mathrm{SL}(2) \times
      \mathrm{SL}(6)}{\longrightarrow}\;   
    (\mathbf{1} ,\mathbf{15} )+(\mathbf{2}, \overline{\mathbf{6}} )
    \;\stackrel{\mathrm{SL}(2) \times
      \mathrm{SO}(5)}{\longrightarrow}\;  (\mathbf{1}
    ,\mathbf{5} ) + (\mathbf{1} ,\mathbf{10} ) + (\mathbf{2}
    ,\mathbf{5}) + (\mathbf{2} ,\mathbf{1})\,. 
\end{align}

At this point it makes sense to compare our results for variations of
the tensor fields to the corresponding expressions known from maximal
$5D$ supergravity \cite{deWit:2004nw}. In the latter case these
variations are encoded in the symmetric three-rank $\mathrm{E}_{6(6)}$
invariant tensor $d_{MNP}$,
\begin{align}
  \label{eq:delta-C-M}
  \delta C_{\mu\nu \,M} - 2\, d_{MNP}\, C_{[\mu}{}^N\,\delta
  C_{\nu]}{}^P \,. 
\end{align}
Expressions such as these are characteristic for the vector-tensor
hierarchy. Obviously the tensor $d_{MNP}$ decomposes into three
$\mathrm{SL}(2) \times \mathrm{SO}(5)$ invariant components, 
\begin{align}
  \label{eq:decomp-d}
  d_{MNP} \propto \left\{ 
  \begin{array}{rcl} 
  d( {}_{mn} \vert {}^{\alpha p} \vert {}^{\beta q} )&\!\!=\!\!&
  \delta_{mn}{}^{pq}\, \varepsilon^{\alpha\beta}\,,\\[2mm]
  d ({}_{mn}\vert {}_{pq} \vert {}_r )&\!\!=\!\!& \mathring{e}\,
  \varepsilon_{mnpqr}  \,, \\[2mm]
  d({}_{m} \vert {}^{\alpha n} \vert {}^{\beta} )&\!\!=\!\!&
  \delta_{m}{}^{n}\, \varepsilon^{\alpha\beta}\,,  
\end{array}  
\right. 
\end{align}
where normalization factors are not specified because they can be
changed by rescaling the normalization of the vector and tensor
fields. Nevertheless the fact that a single symmetric tensor $d_{MNP}$
must encode the variations above for all the fields does pose certain
restrictions on the relative normalizations of vectors and tensor
fields, especially because the product of the normalization of a
tensor and its corresponding dual vector is constrained, just as in
the maximal $5D$ theory \cite{deWit:2004nw}. We return to this issue
in the next section, but note that this normalization condition has
been incorporated when adopting the rescalings of the vector and
tensor fields in \eqref{eq:index-changes-vector} and
\eqref{eq:index-changes-tensor}, repectively. It then turns out that
the following expressions for the independent components of the
combined variations \eqref{eq:delta-C-M} must be equivalent to the
following,
\begin{align}
  \label{eq:dC+CdC}
   &\delta C_{\mu\nu}{}^{\alpha m} - \tfrac18\mathrm{i}\, \varepsilon^{\alpha\beta}
   \big[ C_{[\mu\, \beta n} \,\delta C_{\nu]}{}^{mn} +
   C_{[\mu}{}^{mn}\, \delta C_{\nu]\, \beta n}\big]  -\mathrm{i}\,
   \varepsilon^{\alpha\beta} \big[C_{[\mu}{}^m \, \delta
   C_{\nu] \beta} +C_{[\mu \beta} \,\delta
   C_{\nu] }{}^m   \big]\,, \nonumber\\[2mm]
   &\delta C_{\mu\nu}{}^\alpha +\mathrm{i}\,\varepsilon^{\alpha\beta}\big[
   C_{[\mu}{}^m\,\delta C_{\nu]\,\beta m} +C_{[\mu\,\beta m} \,\delta
   C_{\nu]}{}^m \big] \,,\nonumber\\[2mm]
   &\delta C_{\mu\nu\,mn} +
   \tfrac1{128}\sqrt{5}\,\mathring{e}\,\varepsilon_{mnpqr} \big[
   C_{[\mu}{}^p \,\delta C_{\nu]}{}^{qr} + C_{[\mu}{}^{qr} \,\delta
   C_{\nu]}{}^p\big] -\tfrac14\mathrm{i} \, \varepsilon^{\alpha\beta}
   \,C_{[\mu \alpha [m} \,\delta C_{\nu] \beta n]}\,,\nonumber\\[2mm]
   &\delta C_{\mu\nu\,m} - \mathrm{i} \,\varepsilon^{\alpha\beta}
   \big[ C_{[\mu\,\alpha m} \,\delta C_{\nu]\,\beta} - C_{[\mu\,\alpha}
   \,\delta C_{\nu]\,\beta m} \big] + \tfrac1{256}\sqrt{5}
   \mathring{e}\,\varepsilon_{mnpqr} \,C_{[\mu}{}^{np} \,\delta
   C_{\nu]}{}^{qr}  \,,
\end{align}
where the last line is not derived directly from the $10D$
supergravity as the tensor field $C_{\mu\nu\,m}$ is associated with
the elusive dual graviton. Nethertheless it is remarkable that one can
also derive the coefficients in the variation of $C_{\mu\nu\,m}$ by
comparing to the $5D$ vector-tensor hierarchy.

\section{Generalized vielbeine and 
  $\boldsymbol{\mathrm{USp}(8)}$ covariant spinors}
\label{sec:gen-vielbeine}
\setcounter{equation}{0}
The spinor fields $\psi_\mu$, $\psi_\mu{\!}^\mathrm{c}$, $\psi_a$,
$\psi_a{\!}^\mathrm{c}$, $\lambda$ and $\lambda^\mathrm{c}$, which
were defined in section \ref{sec:first-field-redefin}, obviously
transform under the $\mathrm{Spin}(4,1)\times\mathrm{USp}(4)$ subgroup
of the $10D$ tangent space group $\mathrm{Spin}(9,1)$. Hence every
$10D$ spinor consists of four complex $\mathrm{Spin}(4,1)$ spinors
which rotate among each other under $\mathrm{USp}(4)$
transformations. In the following we will not consider the
$\mathrm{Spin}(4,1)$ aspects but concentrate on the extension of the
$\mathrm{USp}(4)$ transformations to the full automorphism group of
the $5D$ space-time Clifford algebra. This so-called R-symmetry group contains
also the $\mathrm{U}(1)$ group of IIB supergravity (which can be
regarded as the $10D$ R-symmetry group) and it can be further extended
by realizing that the spinors can actually transform under
$\mathrm{SU}(4)\cong \mathrm{SO}(6)$ (for instance, by regarding them
as chiral spinors of $\mathrm{SO}(6)$). It is then convenient to
introduce corresponding $\mathrm{SO}(6)$ gamma matrices as well, which
requires to combine the spinors with their charge conjugates,
i.e. $\psi_\mu$ with $\psi_\mu{\!}^\mathrm{c}$, and likewise, $\psi_a$
with $\psi_a{\!}^\mathrm{c}$, and $\lambda$ with
$\lambda^\mathrm{c}$. This is described in detail in
appendix~\ref{App:R-symm-assignm-fermions}. The $\mathrm{SO}(6)$ gamma
matrices will be denoted by $\boldsymbol{\Gamma}_{\hat{a}}$, with
$\hat{a}=1,\ldots,6$, and act on the eight-component pseudo-real
spinors. We may then introduce the chirality operator
$\boldsymbol{\Gamma}_7 \equiv \mathrm{i}
\boldsymbol{\Gamma}_1\boldsymbol{\Gamma}_2\cdots
\boldsymbol{\Gamma}_6$, which decomposes as
$\boldsymbol{\Gamma}_7=\oneone_4 \otimes \sigma_3$, so that the
$\mathrm{SO}(6)$ chirality of the charge conjugate fermions is
opposite to the original ones. Here we are using a basis where the
positive-chirality (negative-chirality) components carry positive
(negative) $\mathrm{U}(1)$ charge.  In this section and henceforth we
will be using these 8-component spinor arrays whenever possible
(labeled by indices $A=1,\ldots,8$) and they will simply be denoted by
$\psi_\mu{\!}^A$, $\psi_a{\!}^A$ and $\lambda^A$. Each of these
spinors are then $5D$ symplectic Majorana spinors, i.e.,
\begin{equation}
  \label{eq:sympl-majorana}
  C^{-1} \bar\psi_A{}^\mathrm{T}  = \Omega_{AB}\,\psi^B\,,
\end{equation}
where $C$ is the charge conjugation matrix in five space-time
dimensions and $\Omega$ is the anti-symmetric $\mathrm{USp}(8)$
invariant tensor. 

The appearance of $\Omega$ indicates that the full R-symmetry group is
equal to $\mathrm{USp}(8)$, as is to be expected for $5D$
spinors. Indeed, the gravitini $\psi_\mu{}^A$ transform consistently
in the $\boldsymbol{8}$ representation of this extended R-symmetry
group. However, the fields $\psi_a$ and $\lambda$ cannot possibly
transform in the $\boldsymbol{8}$ representation, in view of the fact
that the $\mathrm{U}(1)$ charges of the fields $\psi_a{\!}^A$ and
$\lambda^A$ are equal to $\pm1/2$ and $\pm3/2$,
respectively. Therefore those fields must transform in a different
representation of the $\mathrm{USp}(8)$ group. In view of the values
for the $\mathrm{U}(1)$ charges and the fact that $\psi_a{\!}^A$ and
$\lambda^A$ define precisely 48 $5D$ symplectic Majorana spinors,
these fields must combine into the $\boldsymbol{48}$ representation of
the group $\mathrm{USp}(8)$. At this point we should recall that only
the $\mathrm{USp}(4)\times\mathrm{U}(1)$ subgroup is realized as a
local gauge invariance, as they originate from the symmetries of $10D$
IIB supergravity that were already realized as local ones. As we have
stressed in the introduction, the full $\mathrm{USp}(8)$ R-symmetry
group can be realized locally upon introducing a compensating phase
factor belonging to $\mathrm{USp}(8)/[\mathrm{USp}(4)\times
\mathrm{U}(1)]$. We will postpone the introduction of this phase
factor till later, so that the present calculations will describe the
results subject to a gauge condition that sets the compensating phase
factor equal to unity. However, it is important to realize that the
local transformations depend on both sets of coordinates, $x^\mu$ and
$y^m$. This is the reason why we adopted the indices $A,B,\ldots$ for
the spinors in this case, while in the maximal $5D$ supergravity, the
spinors will carry indices $i,j,\ldots$ with local R-symmetry
transformations that depend only on the space-time coordinates
$x^\mu$. This issue will be important in
section~\ref{sec:cons-trunc-maxim}, when considering the {\it
  truncation} of $10D$ supergravity to $5D$,

In the previous section we have identified 27 vector fields
$C_\mu{}^M$ as listed in \eqref{eq:index-changes-vector}, which
transform under supersymmetry into the symplectic Majorana spinors
$\psi_\mu{\!}^A$, $\psi_a{\!}^A$ and $\lambda^A$. As it turns out the
supersymmetry variations of these fields can be written in the same
way as the variations of the vector fields in $5D$ maximal
supergravity \cite{deWit:2004nw},
\begin{equation}  
  \label{eq:VecVar}
  \delta C_{\mu}{\!}^{M}=  2 \,\big[\mathrm{i}\bar\Omega^{AC}\,
  \bar\epsilon_{C}\,\psi_\mu{\!}^B 
  +\bar\epsilon_C\,\gamma_\mu\chi^{ABC}\big] \mathcal{V}_{AB}{\!}^M\,, 
\end{equation}
except that, as explained above, we changed the $\mathrm{USp}(8)$ indices
from $i,j,\ldots$ to $A,B,\ldots$.  Here $\Omega^{AB}$ is the
symplectic $\mathrm{USp}(8)$ invariant tensor introduced aboved and the
$\mathcal{V}_{AB}{}^M$ depend on the 42 scalar fields. All these
fields depend on coordinates $x^\mu$ and $y^m$. In the pure $5D$
theory the corresponding quantities $\mathcal{V}_{ij}{}^M$ are defined
in terms of the $\mathrm{E}_{6(6)}/\mathrm{USp}(8)$ coset representative. The
transformations \eqref{eq:VecVar} are consistent with the
$\mathrm{USp}(8)$ R-symmetry group and the anti-symmetric traceless
spinors $\chi^{ABC}$ are symplectic Majorana spinors, satisfying
\begin{equation}
  \label{eq:sympl-majorana-tri}
  C^{-1} \bar\chi_{ABC}{}^\mathrm{T}   =
  \Omega_{AD}\,\Omega_{BE}\,\Omega_{CF}\,\chi^{DEF}\,,
\end{equation}
in direct correspondence with the $5D$ theory \cite{deWit:2004nw}.
Because of the anti-symmetry in $[ABC]$ and the condition
$\Omega_{AB}\,\chi^{ABC}=0$, this representation is irreducible. Hence
the spinor $\chi^{ABC}$ should be linearly related to the spinors
$\psi_a{\!}^A$ and $\lambda^A$. Indeed, as we demonstrate in appendix
\ref{App:R-symm-assignm-fermions} (c.f. \eqref{eq:8+48-branching}) the
branching of the $\boldsymbol{8}$ and $\boldsymbol{48}$
$\mathrm{USp}(8)$ representations of the fermions with respect to the
$\mathrm{SU}(4)\times\mathrm{U}(1)$ subgroup accounts precisely for
the fermion fields $\psi_\mu{\!}^A$, $\psi_a{\!}^A$ and $\lambda^A$
including their $\mathrm{U}(1)$ charge assignments.

The supersymmetry transformation rules for the vector fields
$C_\mu{}^M$ in terms of the spinors $\psi_\mu{\!}^A$, $\psi_a{\!}^A$,
$\lambda^A$ based on IIB supergravity follow from
\eqref{eq:delta-C-vectors} and \eqref{eq:var-dual-vector-C} upon
taking into account the redefinitions
\eqref{eq:index-changes-vector}. By comparing these expressions to
\eqref{eq:VecVar} we obtain explicit representations of the so-called
{\it generalized vielbeine} $\mathcal{V}_{AB}{}^M$, which depend on
all $10D$ coordinates. Furthermore we can deduce the explicit relation
between the $\mathrm{USp}(8)$ covariant spinor field $\chi^{ABC}$ and
the fields $\psi_a{\!}^A$ and $\lambda^A$.  In the same fashion one
can evaluate the supersymmetry transformations of the tensor fields, a
topic that will be dealt with at the end of this section.

Matrices in spinor space can be decomposed into direct products of the
$5D$ gamma matrices $\gamma^\mu$ and the $\mathrm{SO}(6)$ gamma
matrices.  The latter products can be conveniently decomposed into 28
anti-symmetric matrices $\Omega$, $\Omega\, \boldsymbol{\Gamma}_{\hat
  a}$, $\Omega\, \boldsymbol{\Gamma}_{\hat a}\boldsymbol{\Gamma}_7$
and $\Omega \,\boldsymbol{\Gamma}_{\hat a\hat
  b}\boldsymbol{\Gamma}_7$, and 36 symmetric matrices $\Omega\,
\boldsymbol{\Gamma}_7$, $\Omega\, \boldsymbol{\Gamma}_{\hat a\hat b}$
and $\Omega\,\boldsymbol{\Gamma}_{\hat a\hat b\hat c}$. The latter are
proportional to the anti-hermitian generators of $\mathrm{USp}(8)$
(note that the matrices $\boldsymbol{\Gamma}_{\hat a\hat b}$ are the
generators of the group $\mathrm{SU}(4)\cong \mathrm{SO}(6)$).  Before
obtaining a representation of the generalized vielbeine
$\mathcal{V}_{AB}{}^M$ we note that the $\mathrm{USp}(8)$
transformations of the spinors $\psi_\mu{\!}^A$ and $\epsilon^A$ have
been defined in appendix \ref{App:R-symm-assignm-fermions}, and they
imply that the bilinears $\Omega^{AC}\, \bar\epsilon_{C}\psi_\mu{}^B$
transform in the $\mathbf{27}$ representation of $\mathrm{USp}(8)$. Since
the vector fields are not subject to the R-symmetry, it follows that
the generalized vielbeine $\mathcal{V}_{AB}{}^M$ transform in the same
representation, so that they can be expanded in the corresponding
gamma matrix combinations,
\begin{align} 
  \label{eq:Vielbeinexpansion}
  \mathcal{V}_{AB}{}^M = &\, \mathcal{V}_a{}^M
  \big(\Omega\,\boldsymbol{\Gamma}^{a}\big){}_{AB}    
  +  \mathcal{V}_6{}^M \big (\Omega\,
  \boldsymbol{\Gamma}^{6}\big){}_{AB}   
  + \tilde{\mathcal{V}}_a{}^M \big(\Omega\, \boldsymbol{\Gamma}^{a} 
  \boldsymbol{\Gamma}_7 \big){}_{AB}  
  + \tilde{\mathcal{V}}_6{}^M \big (\Omega \,\boldsymbol{\Gamma}^{6} 
  \boldsymbol{\Gamma}_7 \big){}_{AB}  
  \nonumber \\ 
  &\, 
  + \mathcal{V}_{ab}{}^M \big (\Omega \,\boldsymbol{\Gamma}^{ab} 
  \boldsymbol{\Gamma}_7 \big){}_{AB}   
  + 2\, \mathcal{V}_{a6}{}^M \big(\Omega\,\boldsymbol{\Gamma}^{a6}
  \boldsymbol{\Gamma}_7 \big){}_{AB}  \,, 
\end{align} 
which defines the branching of the $\boldsymbol{27}$ representation of
$\mathrm{USp}(8)$ with respect to $\mathrm{SO}(5)$ (which directly follows via the
branching with respect to $\mathrm{SO}(6)$),
\begin{equation}
  \label{eq:27-branching}
  \mathbf{27}\;\stackrel{\mathrm{SO}(6)}{\longrightarrow} \; \mathbf{6}
  + \bar{ \mathbf{6}}   + \mathbf{15}  
  \; \stackrel{\mathrm{SO}(5)}{\longrightarrow}\;  \mathbf{1}
  + \mathbf{5} + \mathbf{1} + \mathbf{5} + \mathbf{10}   + \mathbf{5}
  \,. 
\end{equation}
The generalized vielbeine can now be directly determined from the
supersymmetry transformations of the vector fields, which leads to
\begin{align}
  \label{eq:gen-vielbeine}
  \mathcal{V}_{AB}{}^m=&\, -\tfrac14\mathrm{i}\,\Delta^{-1/3}\,e_a{}^m\,
  \big(\Phi^\mathrm{T}
  \Omega\,\boldsymbol{\Gamma}^{a6}\boldsymbol{\Gamma}_7\,\Phi \big){}_{AB}
  \,,  \nonumber\\[2mm]
  \mathcal{V}_{AB}{}^{mn}=&\,
   - \tfrac45 \sqrt{5} \mathrm{i}\,\Delta^{2/3} \, 
  \big(\Phi^\mathrm{T}\Omega\,\boldsymbol{\Gamma}^{mn}
  \boldsymbol{\Gamma}_7\,\Phi\big){}_{AB}\nonumber\\  
  &\,
  +\tfrac{4}{5}\sqrt{5} \,\mathring{e}{}^{-1} \varepsilon^{mnpqr} 
   A^\alpha{\!}_{pq} \,\mathcal{V}{}_{AB\,\alpha r} 
  \nonumber \\ 
  &\,+ \tfrac{32}{15}\sqrt{5} \, \mathring{e}{}^{-1}
  \varepsilon^{mnpqr}  \big[A_{pqrs}
 - \tfrac{3  }{16}  \mathrm{i}
  \varepsilon_{\alpha\beta}A^\alpha{\!}_{pq}A^\beta{\!}_{rs}\big]\, \mathcal{V}{}_{AB}{}^s
   \,, \nonumber\\[2mm]  
  \mathcal{V}_{AB\,\alpha m} =&\, 
   \tfrac14\mathrm{i}\,\Delta^{-1/3}\,\big[
   (\phi_\alpha-\varepsilon_{\alpha\beta}\phi^\beta )\, 
  \big(\Phi^\mathrm{T}\Omega\,\boldsymbol{\Gamma}_{m} \,\Phi \big){}_{AB}   
  +(\phi_\alpha  +\varepsilon_{\alpha\beta}\phi^\beta)\,\big(\Phi^\mathrm{T}\Omega\,
  \boldsymbol{\Gamma}_{m}\boldsymbol{\Gamma}_{7} \,\Phi\big)_{AB}\big]  
  \nonumber \\ 
  &\, +\mathrm{i}\, \varepsilon_{\alpha\beta} A^\beta{\!}_{mn}
  \,\mathcal{V}_{AB}{}^n  \,,  \nonumber\\[2mm] 
  {\cal V}_{AB\,\alpha}  =  &\, \tfrac1{10}\sqrt{5}\mathrm{i}   \, \Delta^{2/3} 
  \big[ (\phi_\alpha - \varepsilon_{\alpha \beta}\phi^\beta )
  \big(\Phi^\mathrm{T}\Omega\,\boldsymbol{\Gamma}_{6} \,\Phi
  \big){}_{AB} + (\phi_\alpha +\varepsilon_{\alpha \beta}\phi^\beta )
  \big(\Phi^\mathrm{T}\Omega\,
  \boldsymbol{\Gamma}_{6}\boldsymbol{\Gamma}_{7}
  \,\Phi \big){}_{AB}    \big] \nonumber \\
    &\,+\tfrac1{16}\mathrm{i}\, \varepsilon_{\alpha\beta} A^\beta{\!}_{mn}\,
    \mathcal{V}_{AB}{}^{mn} \nonumber\\
    &\,-\tfrac{1}{15} \sqrt{5} \, \mathring{e}{}^{-1}
    \varepsilon^{mnpqr} \big[A_{mnpq} \, \mathcal{V}_{AB\alpha r}
    +2\mathrm{i}\,\varepsilon_{\alpha\beta} \, A ^\beta{\!}_{mn}\,
    A_{pqrs} \,\mathcal{V}_{AB}{\!}^s \big] \nonumber\\
    &\, -\tfrac1{40}\sqrt{5}\mathrm{i}  \,\varepsilon_{\alpha\beta}\,
    \mathring{e}{}^{-1} \varepsilon^{mnpqr} \big[ \,
    A^\beta{\!}_{mn} \, A^\gamma{\!}_{pq} \,\mathcal{V}_{AB\, \gamma r}
    -\tfrac13 \mathrm{i}\, \varepsilon_{\gamma\delta} \,A
    ^\gamma{\!}_{sm} \, A^\delta{\!}_{np}\,A^\beta{\!}_{qr}\,
    \mathcal{V}_{AB}{\!}^s \big]
    \,.
\end{align} 
In the above equations we have now included the compensating phase
factors $\Phi^A{\!}_B$ that were discussed earlier, which enable the
$\mathrm{USp}(8)$ R-symmetry group to be realized locally. The phase
factors are simply generated by a redefinition of the fermion fields,
as $\Phi\in \mathrm{USp}(8)$ is assumed to transform under the action
of $\mathrm{USp}(8)$ from the right and under
$\mathrm{USp}(4)\times\mathrm{U}(1)$ from the left, so that fermion
fields $\Phi^\dagger \Psi$, where $\Psi$ denotes the original fields
in a proper basis, transform indeed under this local group. Previously
we have assumed the gauge condition $\Phi=\oneone$ which suffices to
carry out most of the various calculations. In fact, we will continue
to use this gauge condition in most of what follows. The phase factors
can always be introduced later to elevate the R-symmetry group to a
local invariance group, just as what was done long ago for
$11D$ supergravity~\cite{deWit:1986mz}.

The next task is to establish the relation between the
$\mathrm{USp}(8)$ covariant spinors $\chi^{ABC}$ and the spinors
originating from $10D$, $\psi_a{\!}^A$ and $\lambda^A$. Comparing the terms
proportional to these fields in the supersymmetry variations of the
vector fields, one finds the following set of equations,
\begin{align}
  \label{eq:psi-lambda=chi}
  \psi_a{}^A =&\, -\mathrm{i} \big[\chi^{ABC} \,\delta_a{\!}^b
  -\tfrac18 (\boldsymbol{\Gamma}_a \boldsymbol{\Gamma}^b)^A{\!}_D \,
  \chi^{DBC} \big] \,\big[\Omega \,\boldsymbol{\Gamma}_{b6}
  \boldsymbol{\Gamma}_7\big]_{BC}
  \,, \nonumber\\
  \big[(\oneone \pm \boldsymbol{\Gamma}_7)
  \boldsymbol{\Gamma}_{a6}\big]{}^A{\!}_D \,
  \lambda^D =&\, \pm\mathrm{i}\, \big[\Omega\,\boldsymbol{\Gamma}_a
  (\oneone \pm 
  \boldsymbol{\Gamma}_7)\big]_{BC} \, \big(\oneone \pm
  \boldsymbol{\Gamma}_7\big){}^A{\!}_D\,\chi^{DBC} \,,\nonumber\\
  \big(\oneone \pm \boldsymbol{\Gamma}_7\big){}^A{\!}_D \, \lambda^D 
  =&\, \pm\mathrm{i}\, \big(\Omega\,\boldsymbol{\Gamma}_6 (\oneone \pm
  \boldsymbol{\Gamma}_7)\big)_{BC} \, \big(\oneone \pm
  \boldsymbol{\Gamma}_7\big){}^A{\!}_D\,\chi^{DBC} \,,\nonumber\\
  \big[(\boldsymbol{\Gamma}_{[ab} (\delta_{c]}{\!}^d \oneone -\tfrac19
  \boldsymbol{\Gamma}_{c]} \boldsymbol{\Gamma}^d )
  \boldsymbol{\Gamma}_7\big]{}^A{\!}_D\, 
  \psi_d{}^D =&\, -\tfrac16 \varepsilon_{abcde}
  \big(\Omega\,\boldsymbol{\Gamma}^{de}
  \boldsymbol{\Gamma}_7\big)_{BC}\,   \chi^{ABC} \,,\nonumber\\ 
  \big[(\Omega\,\boldsymbol{\Gamma}_a \big(\oneone \pm
  \boldsymbol{\Gamma}_7)\big]{}_{BC} \, \big(\oneone \mp
  \boldsymbol{\Gamma}_7\big)^A{\!}_D\,\chi^{DBC} =&\,\pm
  2\mathrm{i}\,\big[\big(\oneone \mp \boldsymbol{\Gamma}_7\big)
  \boldsymbol{\Gamma}_6 \, \big(\delta_a{\!}^b\oneone -\tfrac13
  \boldsymbol{\Gamma}_a\boldsymbol{\Gamma}^b\big)\big]{}^A{\!}_D\,
  \psi_b{}^D \,,
  \nonumber\\
  \big[(\oneone \pm  \boldsymbol{\Gamma}_7)
  \boldsymbol{\Gamma}^a\big]{}^A{\!}_D \,
  \psi_a{}^D  =&\, \mp\tfrac34\mathrm{i}\,
  \big[\Omega\,\boldsymbol{\Gamma}_6 (\oneone \mp
  \boldsymbol{\Gamma}_7)\big]{}_{BC} \,\big(\oneone \pm
  \boldsymbol{\Gamma}_7\big){}^A{\!}_D \,\chi^{DBC} \,.
\end{align}
These are the relations that determine the (linear) relation between
the spinors $\psi_a{\!}^A$ and $\lambda^A$ and the $\mathrm{USp}(8)$
covariant spinors $\chi^{ABC}$.  Just as in $11D$ supergravity, where
the expression for the $4D$ spinors $\chi^{ABC}$ as first given in
\cite{Cremmer:1979up} is only unique up to Fierz reordering, there are
various different ways to express the solution for $\chi^{ABC}$. One
solution follows by substituting the $\mathrm{SO}(6)$ covariant
parametrization derived in appendix \ref{App:R-symm-assignm-fermions}
into \eqref{eq:psi-lambda=chi}, which then leads to
\eqref{eq:solution-for-chi}. However, given that the ansatz for
$\chi^{ABC}$ is not unique, one might wonder whether there exists an
alternative version of this solution that may be even more
concise. Indeed we have found such a solution taking the form
\begin{align}
  \label{eq:solution-for-chi-alt}
  \chi^{ABC}=&\, -\tfrac3{8} \mathrm{i} \Big[
  \big(\boldsymbol{\Gamma}_6\,\bar\Omega\big){}^{[AB}
  \,\big(\boldsymbol{\Gamma}_7 \lambda\big){}^{C]} +
  \big(\boldsymbol{\Gamma}_7
  \boldsymbol{\Gamma}_6\,\bar\Omega\big){}^{[AB} \,\lambda^{C]} \Big]
  \nonumber\\
  &\, -\tfrac34\mathrm{i}
    \,\big(\boldsymbol{\Gamma}^a
  \boldsymbol{\Gamma}_6\boldsymbol{\Gamma}_7 \,\bar\Omega\big){}^{[AB}
  \,\psi_a{}^{C]}  -\tfrac14\mathrm{i} \,
     \bar\Omega^{[AB} \,\big(\boldsymbol{\Gamma}_6
     \boldsymbol{\Gamma}_7 \boldsymbol{\Gamma}^a\psi_a \big){}^{C]} \,.
\end{align}
which also satisfies \eqref{eq:psi-lambda=chi}. Its equivalence to
\eqref{eq:solution-for-chi} has been confirmed by demonstrating that
both solutions are related by Fierz reordering to a single expression
that involves eight different structures.  This result satisfies the
reality condition \eqref{eq:sympl-majorana-tri} and vanishes upon
contraction with $\Omega_{AB}$. Note also that the above expression
should in principle have been contracted with three different phase
factors $\Phi^\dagger$ as was discussed above. For clarity of the
presentation we have set $\Phi=\oneone$.

Subsequently we derive a formula for the
supersymmetry transformations of the generalized vielbeine
$\mathcal{V}_{AB}{}^M$. For maximal $5D$
supergravity \cite{deWit:2004nw} there exists the following
expression (with indices $i,j,\ldots$
replaced again by $A,B,\ldots$),
\begin{align}
  \label{eq:susy-gen-vielbein}
  \delta\mathcal{V}_{AB}{}^M =&\, -\mathrm{i} \big[ 4\,\Omega_{G[A}
  \,\bar\chi_{BCD]}\,\epsilon^G
  +3\,\,\Omega_{[AB}\,\bar\chi_{CD]G} \,\epsilon^G \big]\, \bar\Omega^{CE}
  \bar\Omega^{DF} \, \mathcal{V}_{EF}{}^M \nonumber\\
  = &\, \mathrm{i} \,\Omega_{AC} \Omega_{BD}
  \,\big[ 4\, \bar\Omega^{G[C} \,\bar \epsilon_G \,\chi^{DEF]} 
  +3\,\bar\Omega^{[CD}\,\bar\epsilon_G \,\chi^{EF]G} \big]
  \,\mathcal{V}_{EF}{}^M \,. 
\end{align}
This result is expected to be identical to the result that one obtains
by calculating the variations of the generalized vielbeine
\eqref{eq:gen-vielbeine} induced by the supersymmetry transformations
of the scalar fields,
\begin{align}
  \label{eq:variations-scalar}
  \delta e_m{\,}^a =&\, \tfrac12 e_m{\!}^b
  \,\bar\epsilon\,\boldsymbol{\Gamma}^{a6} \boldsymbol{\Gamma}_7
  \psi_b\,,  \nonumber\\ 
  \delta \phi^\alpha =&\,-\tfrac14 \varepsilon^{\alpha\beta}  \phi_\beta
  \,\bar\epsilon\,\boldsymbol{\Gamma}_6 (\oneone+
  \boldsymbol{\Gamma}_7) \lambda \,, 
   \nonumber\\ 
  \delta \phi_\alpha =&\,-\tfrac14 \varepsilon_ {\alpha\beta}  \phi^\beta
  \,\bar\epsilon\,\boldsymbol{\Gamma}_6 (\oneone- 
  \boldsymbol{\Gamma}_7) \lambda \,, \nonumber\\ 
  \delta A^\alpha{\!}_{mn} =&\, -\tfrac14 \mathrm{i}\, e_m{\!}^a\,e_n{\!}^b\,
  (\phi^\alpha+\varepsilon^{\alpha\beta}\phi_\beta) \,\bar\epsilon\,
  (\boldsymbol{\Gamma}_{ab} \lambda -4\,
  \boldsymbol{\Gamma}_{[a}\psi_{b]} ) \nonumber\\
  &\, +\tfrac14 \mathrm{i}\, e_m{\!}^a\,e_n{\!}^b\,
  (\phi^\alpha-\varepsilon^{\alpha\beta}\phi_\beta) \,\bar\epsilon\,
  (\boldsymbol{\Gamma}_{ab} \boldsymbol{\Gamma}_7 \lambda -4\,
  \boldsymbol{\Gamma}_{[a}\boldsymbol{\Gamma}_7\psi_{b]} ) \nonumber\\  
  \delta A_{mnpq}  =&\, -\tfrac12 \mathrm{i} \, e_m{\!}^a\,e_n{\!}^b \,
  e_p{\!}^c \, e_q{\!}^d  
  \,\bar\epsilon\, \boldsymbol{\Gamma}_6\boldsymbol{\Gamma}_{[abc}
  \psi_{d]}  +\tfrac38\mathrm{i} \varepsilon_{\alpha\beta}
  A^\alpha{\!}_{[mn} \,\delta A^\beta{\!}_{pq]} \,. 
\end{align}

Based on the similar construction for $11D$ supergravity
\cite{deWit:1986mz}, we expect the supersymmetry transformations of
the vielbeine induced by the variations \eqref{eq:variations-scalar}
to coincide with \eqref{eq:susy-gen-vielbein} up to a uniform
$\mathrm{USp}(8)$ transformation. By very laborious calculations we
have been able to demonstrate that this expectation is correct so that
\eqref{eq:susy-gen-vielbein} can be regarded as the supersymmetry
transformation rule for the vielbeine. More precisely, the results induced
by \eqref{eq:variations-scalar} take the form 
\begin{equation}
  \label{eq:susy-vielbeine}
  \delta\mathcal{V}_{AB}{\!}^M =
  \delta\mathcal{V}_{AB}{\!}^M\big\vert_{\eqref{eq:susy-gen-vielbein}} 
    - \Lambda^C{\!}_{[A}\, \mathcal{V}_{B]C}{\!}^M 
\end{equation}
where $\Lambda^A{\!}_B$ is the field-dependent infinitesimal
$\mathrm{USp}(8)$ transformation given by
\begin{align}
  \label{eq:field-dep-usp8}
  \Lambda^A{\!}_B= &\,
  -\tfrac{1}{16}\bar{\epsilon}\,\boldsymbol{\Gamma}_7
  [\boldsymbol{\Gamma}_{ab}  
  \lambda +4\,\boldsymbol{\Gamma}_{[a}  \psi_{b]}]
\,\big(\boldsymbol{\Gamma}^{ab6}\big){}^A{\!}_B \nonumber\\
  &\,+\tfrac{1}{48}\bar{\epsilon}\,\boldsymbol{\Gamma}_7
  [\boldsymbol{\Gamma}_{abc6} \lambda
  +2\,\boldsymbol{\Gamma}_{abcd6} \psi^d] \,
  \big(\boldsymbol{\Gamma}^{abc}){}^A{\!}_B\nonumber\\   
  &  +\tfrac14 \bar{\epsilon}\,\boldsymbol{\Gamma}_7
  \boldsymbol{\Gamma}_{ac}  \psi^c\,
  \big(\boldsymbol{\Gamma}^{a6}\big){}^A{\!}_B 
  +\tfrac14\bar{\epsilon}\,\boldsymbol{\Gamma}_7 \boldsymbol{\Gamma}_{6[a}
  \psi_{b]}\,\big(\boldsymbol{\Gamma}^{ab}){}^A{\!}_B  \,.  
\end{align} 

We now proceed with the supersymmetry transformations of the tensor
fields $C_{\mu\nu\,mn}$, $C_{\mu\nu}{\!}^{\alpha{m}}$ and
$C_{\mu\nu}{}^{\alpha}$ that were defined in \eqref{eq:tensors-27},
following the same approach as for the vector fields. Their
supersymmetry transformations follow upon substituting the results
specified in \eqref{eq:delta-C-2-tensors} and
\eqref{eq:var-dual-tensor-C}. Subsequently we compare them to the
five-dimensional transformation rules for the tensor fields
\cite{deWit:2004nw} with the indices adjusted as before,
\begin{align}
  \label{eq:5D-delta-tensor}
  \delta C_{\mu\nu\,M} - &2\,d_{MNP} \,C_{[\mu}{\!}^N \,\delta
  C_{\nu]}{\!}^P\nonumber\\
  = &\,\tfrac45\sqrt{5}\, \mathcal{V}_M{\!}^{AB} \big[
  2\,\bar\psi_{[\mu A}\,\gamma_{\nu]} \epsilon^C\,\Omega_{BC}
  -\mathrm{i} \bar\chi_{ABC} \,\gamma_{\mu\nu}
  \epsilon^C\big]\nonumber\\
  = &\,-\tfrac45\sqrt{5}\, \mathcal{V}_M{\!}^{AB} \big[ 2\,
  \Omega_{AC} \,\bar\epsilon_B \,\gamma_{[\mu} \psi_{\nu]}{\!}^C
  +\mathrm{i} \Omega_{AD}\Omega_{BE}\,\bar\epsilon_C \,\gamma_{\mu\nu}
  \chi^{DEC} \big] \,.
\end{align}
In $5D$ maximal gauged supergravity the tensor fields constitute a
$\boldsymbol{27}$ representation of $\mathrm{E}_{6(6)}$.  From IIB
supergravity we have initially identified only 22 different tensor
fields. The missing five tensors $C_{\mu\nu\,m}$ have been identified
as originating from a component of the $10D$ dual graviton. The second
term on the left-hand side of \eqref{eq:5D-delta-tensor} has already
been specified in \eqref{eq:dC+CdC}.

From the terms in \eqref{eq:5D-delta-tensor} proportional to
$\psi_{\mu}{\!}^C$ one can directly obtain the following expressions
for the 22 components of $\mathcal{V}_M{}^{ij}$, by making use of the
supersymmetry transformations of the corresponding tensors derived in
the previous sections,
\begin{align}
  \label{eq:inverse-vielbeine} 
  \mathcal{V}_{mn}{}^{AB}=&\,-\tfrac1{32} \sqrt{5}
  \mathrm{i}\,\Delta^{-2/3}e_m{}^a e_n{}^b 
  \big(\Phi^\dagger\,\boldsymbol{\Gamma}_{ab}
  \boldsymbol{\Gamma}_7\bar{\Omega} 
  \, \bar\Phi\big){}^{AB} + \tfrac18\mathrm{i}\,
  \varepsilon_{\alpha\beta} \, A^\alpha{\!}_{mn}    
  \,\mathcal{V}^{\beta\,AB}\,, 
  \nonumber\\[2mm]
  \mathcal{V}^{\alpha m\,AB} =&\, -\tfrac14 \mathrm{i} \Delta^{1/3} \,e_a{}^m\,
  \big[\big(\phi^\alpha-\varepsilon^{\alpha\beta}\phi^\beta\big)\,
  \big(\Phi^\dagger\,\boldsymbol{\Gamma}^{e}  \bar\Omega \,
  \bar\Phi\big){}^{AB} \nonumber\\
  &\,\qquad \qquad\qquad\qquad
  -\big(\phi^\alpha+\varepsilon^{\alpha\beta}\phi_\beta\big)\, 
    \big(\Phi^\dagger\boldsymbol{\Gamma}^{e} 
    \boldsymbol{\Gamma}_7\bar\Omega \, \bar\Phi \big){}^{AB}\big]
    \nonumber\\
    &\, +\tfrac1{15}\sqrt{5} \,\mathring{e}{\!}^{-1}
    \varepsilon^{mnpqr} \big[A_{npqr}  \,\mathcal{V}^{\alpha\,AB}
    +\tfrac38 \mathrm{i}\, A^\alpha{\!}_{np} 
    A^\beta{\!}_{qr} \,\varepsilon_{\beta\gamma}
    \mathcal{V}^{\gamma\,AB} -6\,  A ^\alpha {\!}_{np}
    \mathcal{V}_{qr}{}^{AB}  \big] \,,  \nonumber\\[2mm]
  \mathcal{V}^\alpha{}^{AB}= &\,
  - \tfrac18\sqrt{5}\mathrm{i}\,\Delta^{-2/3}
  \big[\big(\phi^\alpha-\varepsilon^{\alpha\beta}\phi_{\beta}\big)\,
  \big(\Phi^\dagger\,\boldsymbol{\Gamma}_{6}\bar\Omega \,
  \bar\Phi\big){}^{AB} \nonumber\\
  &\,\qquad\qquad\qquad\qquad
  -\big(\phi^\alpha+\varepsilon^{\alpha\beta}\phi_{\beta}\big)\, 
  \big(\Phi^\dagger\,\boldsymbol{\Gamma}_{6}
  \boldsymbol{\Gamma}_{7}\bar\Omega \,
  \bar\Phi\big){}^{AB}\big]\,,
\end{align} 
where we have again included the phase factors $\Phi$. Before
discussing how to obtain the missing components of
$\mathcal{V}_M{\!}^{AB}$ that are associated with the dual graviton,
we first consider the contractions of the form
$\mathcal{V}_M{\!}^{AB}\,\mathcal{V}_{AB}{\!}^N$ making use of the
expressions \eqref{eq:gen-vielbeine} and
\eqref{eq:inverse-vielbeine}. As it turns out the only non-zero
contractions are given by
\begin{align}
  \label{eq:V-V-contracted}
  \mathcal{V}_{mn}{\!}^{AB} \, \mathcal{V}_{AB}{\,}^{pq} =&\, 2\,
  \delta_{mn}{}^{pq} \,,        \nonumber\\ 
  \mathcal{V}^{\alpha\,AB} \, \mathcal{V}_{AB\, \beta}  =&\,
  \delta^\alpha{\!}_\beta \,,   \nonumber\\ 
  \mathcal{V}^{\alpha{m}\,AB} \, \mathcal{V}_{AB\,\beta{n}}\
  =&\,  \delta^\alpha{\!}_\beta\,\delta^m{\!}_n \,,
\end{align}
suggesting that 
\begin{equation}
  \label{eq:inverse-vielbeine-conjecture}
  \mathcal{V}_M{\!}^{AB}\,\mathcal{V}_{AB}{\!}^N= \delta_M{\!}^N\,. 
\end{equation}
This condition is actually identical to the one that holds in $5D$
maximal gauged supergravity. In the same spirit as before, we may
assume that \eqref{eq:inverse-vielbeine-conjecture} holds in this case
as well, and this then enables us to also determine the five missing
components $\mathcal{V}_m{\!}^{AB}$, 
\begin{align}
  \label{eq:dual-grav-vielbein}
  \mathcal{V}_m{\!}^{AB}=&\,-\tfrac12\mathrm{i}\,\Delta^{1/3}
  e_m{}^a  \big(\Phi^\dagger \boldsymbol{\Gamma}_{a6}
  \boldsymbol{\Gamma}_7\bar\Omega\bar\Phi\big)^{AB}  
  \nonumber\\ 
  &\,+\tfrac{16}{15}\sqrt{5} \,\mathring{e}{\!}^{-1}\varepsilon^{npqrs}
  \big[A_{mqrs}-\tfrac{3}{16}\mathrm{i}\,\varepsilon_{\alpha\beta}
  A^\alpha{\!}_{qr}A^\beta{\!}_{sm}\big] 
  \mathcal{V}_{np}{}^{ij} -\mathrm{i}A ^\alpha {\!}_{mn}\,\varepsilon_{\alpha\beta}\,
  \mathcal{V}^{\beta{n}\,AB}\nonumber\\ 
  &\,+\tfrac{1}{15} \sqrt{5} \mathrm{i}\,\mathring{e}{}^{-1}
  \varepsilon^{npqrs}\,
  \varepsilon_{\alpha\beta}\big[A_{npqr}\,A^\beta{\!}_{sm}
  -\tfrac1{8} \mathrm{i} \,\varepsilon_{\gamma\delta} 
    A ^\beta{\!}_{np}  A^\gamma{\!}_{qr} A^\delta{\!}_{sm} \big] \mathcal{V}^{\alpha\,AB} \,.
\end{align}
Note that the conditions \eqref{eq:inverse-vielbeine-conjecture}
implies that also the supersymmetry transformations of the
$\mathcal{V}_M{\!}^{AB}$ are determined and take the same form as the
corresponding supersymmetry transformations in $5D$ maximal
supergravity. Needless to say, the results obtained from the vector
fields on the covariant spinors $\chi^{ABC}$ can be verified also from
the perpective of the transformations of the tensor fields. The
results turn out to be mutually consistent.

This completes the evaluation of the bosons and their supersymmetry
transformations. We have succeeded in identifying these fields from
IIB supergravity such that the results resemble as closely as possible
the structure of the $5D$ maximal gauged supergravities
\cite{deWit:2004nw} while retaining the full dependence on all ten
coordinates. For the fields associated with the dual graviton, we
obtained their supersymmetry transformations by requiring them to be
consistent with the global structure exhibited for the other
fields. In this way the results exhibit covariance with respect to the
duality group $\mathrm{E}_{6(6)}$, although the IIB theory is not in
any way invariant under this group. This is further confirmed by the
fact that the following representation of the invariant tensor
$d_{MNP}$ which was noted for maximal $5D$ supergravity
\cite{deWit:2004nw},
\begin{equation}
  \label{eq:d-tensor}
  d_{MNP} = \tfrac25\sqrt{5} \,\mathcal{V}_M{\!}^{AB}
  \,\mathcal{V}_M{\!}^{CD}\,\mathcal{V}_M{\!}^{EF} \,\Omega_{BC}
  \,\Omega_{DE} \,\Omega_{FA}  \,, 
\end{equation}
is also satisfied here, as this expression precisely reproduces the
tensor $d_{MNP}$ as specified in \eqref{eq:dC+CdC}.

As a final comment we note that the generalized vielbeine are
pseudo-real. This property is inherited form the (pseudo-)reality of
the tensors and the fermionic bilinears. We remind the reader that
taking complex conjugates of vielbeine that carry the
$\mathrm{SU}(1,1)$ requires the contraction with a two-dimensional
metric $\eta_{\alpha\beta} = \mathrm{diag}(+1,-1)$ in order to obtain
a covariant quantity (see section \ref{sec:results-2B-sugra}).

\section{The fermion transformation rules}
\label{sec:fermion-transformations}
\setcounter{equation}{0}
In the previous sections we concentrated mostly on the supersymmetry
transformations of the vector and tensor fields. Their supersymmetry
transformations take the form of $\mathrm{USp}(8)$ invariant
contractions between covariant spinor bilinears with the generalized
vielbeine. This is consistent with the fact that the vector and tensor
fields are invariant under the R-symmetry. Also the space-time
f\"unfbein is invariant under $\mathrm{USp}(8)$, and so is its
supersymmetry transformations. The scalar fields do not transform
covariantly (c.f. \eqref{eq:variations-scalar}), but indirectly they
do respect the $\mathrm{USp}(8)$ symmetry as their supersymmetry
transformations induce covariant variations on the generalized
vielbeine. In view of the above it is therefore of interest to
consider the supersymmetry transformations of the fermion fields,
$\psi_\mu{}^A$ and $\chi^{ABC}$, to verify whether they will also take
a $\mathrm{USp}(8)$ covariant form. These results will not only
complement the previous results, but they will enable one to properly
identify various bosonic $\mathrm{USp}(8)$ tensors. Here we follow the
same strategy as was applied to $11D$ supergravity \cite{deWit:1986mz}. As it will
turn out, the global structure of the results of the ensuing analysis
is rather similar.

The analysis starts with the fermionic transformation rules given in
\eqref{eq:lambda-phi}-\eqref{eq:internal-gravitini-var}, but now
written with eight-component symplectic Majorana spinors and
$\mathrm{SO}(6)$ gamma matrices. We start by presenting  the
spin-$1/2$ fields, $\psi_a{\!}^A$ and $\lambda^A$, which transform as
follows under supersymmetry (up to terms of higher order in the
fermions),
\begin{align}
  \label{eq:grav-a-var}
  \delta \psi_{a} =&\,
  \Delta^{-1/3}e_a{}^m\big[\partial_m
  -\tfrac14\omega_m{}^{\alpha\beta}\, \gamma_{\alpha\beta}
  -\tfrac16\partial_m\ln \Delta\big]\epsilon\nonumber\\
&\,
  -\tfrac12 \Delta^{-1/3} \big[
  \omega_a{}^{\alpha b}\,\gamma_\alpha \boldsymbol{\Gamma}_{b6}
  \boldsymbol{\Gamma}_7
  +\tfrac12\omega_a{}^{bc} \boldsymbol{\Gamma}_{bc}
  + \mathrm{i}   Q_a\boldsymbol{\Gamma}_7
  \big]\epsilon\nonumber\\
  &\,+\tfrac{1}{240}\Delta^{-1/3}\varepsilon^{bcdef}\big[F_{bcdef}
  \boldsymbol{\Gamma}_{a6} 
   - 5 \gamma^\alpha  F_{\alpha bcde}
   \boldsymbol{\Gamma}_{f}\boldsymbol{\Gamma}_{a} \boldsymbol{\Gamma}_7
   -5 \gamma^{\alpha\beta} F_{\alpha\beta bcd}
   \boldsymbol{\Gamma}_{ef}\boldsymbol{\Gamma}_{a6} 
   \big]    \epsilon\nonumber\\
   &\,-\tfrac{1}{96}\mathrm{i}\Delta^{-1/3}
   \big[(G_{bcd}\mathbb{P}_{+}
   -\bar  G_{bcd}\mathbb{P}_{-})
   (\boldsymbol{\Gamma}_a\boldsymbol{\Gamma}^{bcd} 
   +2\,\boldsymbol{\Gamma}^{bcd}\boldsymbol{\Gamma}_a)
   \boldsymbol{\Gamma}_6 \nonumber\\
   &\, \qquad \qquad\qquad -3 \,\gamma^{\alpha} 
      (G_{bc\alpha}\mathbb{P}_{+}
      +\bar G_{bc\alpha}\mathbb{P}_{-})
      (\boldsymbol{\Gamma}_{a}\boldsymbol{\Gamma}^{bc} 
   -2\,\boldsymbol{\Gamma}^{bc}\boldsymbol{\Gamma}_{a})
   \big]\epsilon\nonumber\\ 
   &\,-\tfrac{1}{96}\mathrm{i}\Delta^{-1/3} \gamma^{\alpha\beta} \big[3
   (G_{\alpha\beta{b}}\mathbb{P}_{+} -\bar
   G_{\alpha\beta{b}}\mathbb{P}_{-})
   (\boldsymbol{\Gamma}_a\boldsymbol{\Gamma}^b+2\,
   \boldsymbol{\Gamma}^b\boldsymbol{\Gamma}_a) \boldsymbol{\Gamma}_6
   \nonumber\\ 
   &\,\qquad\qquad\qquad - \tfrac12\varepsilon_{\alpha\beta\gamma\delta\tau}
   (G^{\gamma\delta\tau}\mathbb{P}_{+} +\bar
   G^{\gamma\delta\tau}\mathbb{P}_{-}) \boldsymbol{\Gamma}_{a} \big]
   \epsilon\,, \nonumber \\[2mm]
  \delta\lambda =&\,  \Delta^{-1/3}\big[ \gamma^\alpha
  ( P_\alpha \,\mathbb{P}_+ -\bar P_\alpha\, \mathbb{P}_-)\boldsymbol{\Gamma}_6    
    +( P_a \, \mathbb{P}_+ + \bar P_a \, \mathbb{P}_-)
    \boldsymbol{\Gamma}^{a}  \big]\epsilon \nonumber\\
  &\, +\tfrac1{8} \Delta^{-1/3}\big[ -\tfrac13\mathrm{i} (G_{abc}
  \,\mathbb{P}_+ - \bar 
  G_{abc} \,\mathbb{P}_-) \boldsymbol{\Gamma}^{abc6}  
  -\mathrm{i} \,
  \gamma^\alpha (G_{\alpha ab} \, \mathbb{P}_+ +\bar G_{\alpha ab}
    \, \mathbb{P}_-) \boldsymbol{\Gamma}^{ab} \nonumber\\
    &\,\qquad\qquad - \mathrm{i}\, \gamma^{\alpha\beta} ( G_{\alpha\beta
      a}\, \mathbb{P}_+ - \bar G_{\alpha\beta a}\, \mathbb{P}_-)
    \boldsymbol{\Gamma}^{a6} +\tfrac16\mathrm{i} \,\gamma^{\alpha\beta}
    \varepsilon_{\alpha\beta\gamma\delta\tau}  (G^{\gamma\delta\tau}
    \,\mathbb{P}_+ +\bar G^{\gamma\delta\tau} 
    \,\mathbb{P}_-) \big]\epsilon\,,
\end{align}
where we employed the $\mathrm{SO}(6)$ chiral projection operators
$\mathbb{P}_{\pm}=\tfrac12\big(\oneone\pm\boldsymbol{\Gamma}_7\big)$.
To verify that these results are consistent with $\mathrm{USp}(8)$
R-symmetry is subtle and requires us to first combine the two
equations \eqref{eq:grav-a-var} into the covariant tri-spinor
variation $\delta\chi^{ABC}$. For this one makes use of
\eqref{eq:solution-for-chi-alt}. Since this is rather involved, let us 
first proceed to the gravitino variation and return to the spin-$1/2$
variations at the end of the section.

The supersymmetry transformations of the gravitino fields $\psi_\mu{\!}^A$ take the
following form, where we have ordered the various terms
in a particular way in view of what will follow,
\begin{align}
  \label{eq:gravitino-var} 
  \delta \psi_{\mu}=&\, \big[\partial_\mu-B_\mu{}^m\partial_m
  -\tfrac16(\partial_\mu-B_\mu{}^m\partial_m)\ln\Delta -
  \tfrac14\Delta^{-1/3}e_\mu{}^\alpha\big(  \omega_\alpha{}^{\beta\gamma}
  \gamma_{\beta\gamma} + \tfrac23 \gamma_\alpha \gamma_\beta
  \, \omega_{a}{}^{a\beta} \big) \big]   \epsilon\nonumber\\[1mm]
  &- \tfrac12
  \Delta^{-1/3}e_\mu{}^\alpha\big[\mathrm{i}
  Q_\alpha\boldsymbol{\Gamma}_7 +\tfrac12\omega_\alpha{}^{ab}
  \boldsymbol{\Gamma}_{ab} -\tfrac{1}{12}\varepsilon^{abcde} F_{\alpha
    bcde}\boldsymbol{\Gamma}_{a6} 
  \nonumber\\
  &\,\qquad\qquad\qquad\quad +\tfrac14\mathrm{i}(G_{\alpha ab}
  \mathbb{P}_{+} -\bar G_{\alpha ab}\mathbb{P}_{-})
  \boldsymbol{\Gamma}^{ab6}\big]\epsilon
  \nonumber\\[1mm]
  &\,+\tfrac{1}{24}
  \Delta^{-1/3}e_\mu{}^\alpha(\gamma_{\alpha}{}^{\beta\gamma}
  -4\delta_\alpha{}^\beta\gamma^\gamma)\,\big[\mathrm{i}
  (G_{\beta\gamma a}\mathbb{P}_{+}
  +\bar G_{\beta\gamma a}\mathbb{P}_{-}) \boldsymbol{\Gamma}^a  \nonumber\\
  &\,\qquad\qquad\qquad\qquad\qquad
  +\tfrac{1}{6}\mathrm{i}\varepsilon_{\beta\gamma\delta\tau\lambda}
  (G^{\delta\tau\lambda}\mathbb{P}_{+}-\bar
  G^{\delta\tau\lambda}\mathbb{P}_{-}) \boldsymbol{\Gamma}^6
  \nonumber\\
  &\,\qquad\qquad\qquad\qquad\qquad -2\, \omega_{a\,\beta\gamma}
  \,\boldsymbol{\Gamma}^{a6}\boldsymbol{\Gamma}_7
  -\tfrac{1}{3}\varepsilon^{abcde}F_{\beta\gamma abc}
  \boldsymbol{\Gamma}_{de}\boldsymbol{\Gamma}_7 \big]
  \epsilon\nonumber\\[1mm]
  &\,+\tfrac13\Delta^{-1/3} e_\mu{}^\alpha\gamma_\alpha \,
  \boldsymbol{\Gamma}^{m6}\boldsymbol{\Gamma}_7  
   \big[\partial_m
  -\tfrac16\partial_m \ln\Delta -\tfrac12 \mathrm{i}
  Q_m\boldsymbol{\Gamma}_7  -\tfrac14\omega_m{}^{bc}
  \boldsymbol{\Gamma}_{bc}   \nonumber\\
  &\,\qquad\qquad\qquad\quad +\tfrac{1}{600}\varepsilon^{bcdef}
  F_{bcdef}\boldsymbol{\Gamma}_{m6} -\tfrac1{24}\mathrm{i}
  (G_{mbc}\mathbb{P}_{+}-\bar G_{mbc}\mathbb{P}_{-}\big)
  \boldsymbol{\Gamma}^{bc6} \big]\epsilon  \nonumber\\[1mm]
  &\, 
  +\tfrac12 \Delta^{-1/3}e_\mu{}^\alpha \big[(
  \omega_{\alpha\,a\beta} - \omega_{a\,\alpha\beta} )
  \gamma^\beta \, \boldsymbol{\Gamma}^{a6}\boldsymbol{\Gamma}_7
    +\tfrac13 \gamma_\alpha \gamma^\beta
  \, \omega_{a\,\beta b}\, \boldsymbol{\Gamma}^{ab} \big] \epsilon \,, 
\end{align}
where we haved used the same notation as in \eqref{eq:grav-a-var} and
have suppressed terms of higher-order in the fermion fields.  It is
worth noting at this point that some terms already combine into
representations of $\mathrm{USp}(8)$. In particular, the terms in the
second bracket span the $\boldsymbol{36}$ and thus take values in
$\mathfrak{usp}(8)$ and those in the third bracket span the
$\overline{\boldsymbol{27}}$ representation of $\mathrm{USp}(8)$.  The
structure of the last two brackets is more subtle and will be
discussed momentarily.

Following \cite{deWit:1986mz}, the next step is to expand the
components of the $10D$ spin connection about the reference background
of the internal $5D$ space characterized by the f\"unfbein
$\mathring{e}_m{}^a(y)$. To this purpose, we write the spin
connection in terms of the anholonomity coefficients, which depend on
the zehnbein and its derivatives,
\begin{align}
  \label{eq:omega-10}
    \omega_{MAB}=&\,\tfrac12 E_M{}^C(\Omega_{ABC}-
    \Omega_{BCA}-\Omega_{CAB})\,, \nonumber\\
  \Omega_{AB}{}^C=&\,2\,E_{[A}{}^ME_{B]}{}^N\,\partial_M E_{N}{}^C\,.
\end{align}
Writing the internal f\"unfbein as 
\begin{align}
  \label{eq:def-S}
  e_m{}^a(x,y)=\mathring{e}_m{}^b(y)\, S_b{}^a(x,y)\,, \qquad
  e^m{}_a(x,y)=S^{-1}{\!}_a{}^b(x,y)\, \mathring{e}^m{}_b(y)\,, 
\end{align} 
such that $\Delta=\det[S_a{}^b]$, one can evaluate the components of
$\Omega_{AB}{}^C$ making use of \eqref{eq:kk-ansatz},
\begin{align} 
  \label{eq:Omega-decomp}
   \Omega_{\alpha \beta}{}^\gamma =&\,   2\,\Delta^{1/3} \big[
    e_{[\alpha}{}^\mu e_{\beta]}{}^\nu \, {\cal D}_\mu e_\nu{}^\gamma 
    - \tfrac{1}{3}  e_{[\alpha}{}^\mu\,
    \delta_{\beta]}{\!}^\gamma\,  {\cal D}_\mu \ln \Delta\big]  \, , \nonumber \\
    \Omega_{\alpha \beta}{}^c =&\, 2\,\Delta^{2/3} e_{[\alpha}{}^\mu
    e_{\beta]}{}^\nu \, \mathring e_m{}^b S_b{}^c
    \,  {\cal D}_\mu B_\nu{}^m \, , \nonumber \\
    \Omega_{a \beta}{}^\gamma = &\, S^{-1}{\!}_a{}^b \, \mathring{e}_b{}^m 
    \big[ e_\beta{}^\nu \,\partial_m e_\nu{}^\gamma -\tfrac{1}{3}
    \delta_\beta{}^\gamma \,\partial_m \ln \Delta \big] \, ,
    \nonumber \\  
 \Omega_{ab}{}^\gamma =&\, 0 \, ,  \nonumber \\
 \Omega_{a\beta}{}^c =&\, \Delta^{1/3} S^{-1}{\!}_a{}^b \, e_\beta{}^\mu
 \big[ \mathring e_b{}^m \mathring e_n{}^d\,  S_d{}^c \mathring D_m 
 B_\mu{}^n  - {\cal D}_\mu S_b{}^c \big]  +\Delta^{1/3} e_\beta{}^\mu
 B_\mu{}^m\,  \mathring{\omega}{}_m{}^c{}_a \, ,  \nonumber \\ 
 \Omega_{ab}{}^c =&\, -2\,S^{-1}{\!}_{[a}{}^d \,S^{-1}{\!}_{b]}{}^e \,
 \mathring{e}_e{}^m \,\mathring{D}{}_m S_d{}^c - 2\,
 \mathring{\omega}{}_m{}^c{}_{[a}  \,S^{-1}{\!}_{b]}{}^d \, \mathring
 e_d{}^m  \,. 
\end{align}
Here we have defined ${\cal D}_\mu=\partial_\mu-B_\mu{}^m
\mathring{D}_m$, where $\mathring{D}_m$ is the derivative that is
covariant with respect to tangent-space transformations of the
background. Hence it contains the spin connection
$\mathring{\omega}_{m}{}^{ab}(y)$, 
\begin{equation}
  \label{eq:backgr-omega} 
  \mathring{\omega}_{m ab}=\, \tfrac12
  \mathring{e}_m{\!}^c(\mathring{\Omega}_{abc}
  -\mathring{\Omega}_{bca}-\mathring{\Omega}_{cab})\,,\qquad 
  \mathring{\Omega}_{ab}{}^c=2\,
  \mathring{e}_{[a}{}^m\mathring{e}_{b]}{}^n\, \partial_m
  \mathring{e}_{n}{}^c\,,
\end{equation}
and possibly the corresponding Christoffel connection, depending on
the tensor it acts on.  These results exhibit, up to dimension
dependent coefficients, the same structure as in the $11D$ case and we
refer to \cite{deWit:1986mz} for further details.

After substitution of the expressions \eqref{eq:Omega-decomp} into
\eqref{eq:gravitino-var} and some rearrangements, one obtains the
following result,  
\begin{align}
  \label{eq:gravitino-cov}
  \delta \psi_\mu{}^A=&\,D_\mu\epsilon^A
  -\tfrac16(\mathring{D}_m C_{\mu}{}^m)\epsilon^A
  +\tfrac{1}{12}\mathrm{i} \,(\gamma_\mu{}^{\beta\gamma} -4\,
  e_\mu{}^\beta\gamma^\gamma)\mathcal{H}_{\beta\gamma}{\!}^{AB}\,
  \Omega_{BC}\,\epsilon^C\nonumber\\  
  &\,-\tfrac43\mathrm{i} \,
  \bar\Omega^{AC}  \,\mathcal{V}_{CB}{}^m\, D_m\big(\gamma_\mu\,\epsilon^B\big)
  -\tfrac23\mathrm{i}\,\bar\Omega^{AC} D_m\big(\gamma_\mu\,
  \mathcal{V}_{CB}{}^m \big)  \epsilon^B \,, 
\end{align}
where we have now written the field $B_\mu{}^m$ as $C_\mu{}^m$, as
the above expression is the final result.  Here
$D_\mu$ and $D_m$ denote the full
$\mathrm{Spin}(4,1)\times\mathrm{USp}(8)$ covariant derivatives with
$\mathrm{USp}(8)$ connections $\mathcal{Q}_{\mu}$ and $\mathcal{Q}_m$,
such that
\begin{align}
  D_{\mu}\epsilon^A=&\,\mathcal{D}_\mu\epsilon^A
  -\tfrac14\hat\omega_\mu{}^{\alpha\beta}\gamma_{\alpha\beta}\epsilon^A
  -\mathcal{Q}_\mu{\!}^A{\!}_B\,\epsilon^B\,,\nonumber \\  
  D_m\epsilon^A=&\,\mathring{D}_m\epsilon^A-
  \mathcal{Q}_m{\!}^A{\!}_B\,\epsilon^B\,. 
\end{align}
Here the modified spin connection $\hat\omega_\mu{}^{\alpha\beta}$ is
defined by 
\begin{equation}
  \label{eq:mod-spin-conn}
  \hat\omega_{\mu}{}^{\alpha\beta} = \omega_\mu{}^{\alpha\beta}+
  \tfrac23\mathring{D}_mB_{\nu}{}^m\, e_\mu{}^{[\alpha}e^{\beta]\nu}\,,
\end{equation}
where $\omega_\mu{}^{\alpha\beta}$ is the regular torsion-free spin
connection expressed in terms of the f\"unfbein $e_\mu{}^\alpha$. The
two $\mathrm{USp}(8)$ connections, $\mathcal{Q}_\mu{\!}^A{}_B$ and
$\mathcal{Q}_m{\!}^A{}_B$, are equal to 
\begin{align}
  \label{eq:5+5-usp-conn}
  \mathcal{Q}_{\mu}{\!}^A{\!}_B =&\,\tfrac14\big[S^{-1\, ac}\, S_d{}^b\,
  \mathring{e}_c{}^m\,\mathring{e}_n{}^d\,\mathring{D}_mB_\mu{}^n
  -(S^{-1}\mathcal{D}_\mu
  S)^{ab}\big]\, \big(\Phi^\dagger\boldsymbol{\Gamma}_{ab}
  \Phi\big){}^A{\!}_B \nonumber\\ 
  &\,+\tfrac12\mathrm{i}\Delta^{-1/3}e_\mu{}^\alpha
  Q_{\alpha}\, \big(\Phi^\dagger\boldsymbol{\Gamma}_7
  \Phi\big){}^A{\!}_B \nonumber\\
  &\,-\tfrac{1}{24}\Delta^{-1/3}e_\mu{}^\alpha\varepsilon^{abcde}
  F_{\alpha abcd}\, \big(\Phi^\dagger\boldsymbol{\Gamma}_{e6}
  \Phi\big){}^A{\!}_B  \nonumber\\
  &\,+\tfrac{1}{8}\mathrm{i}\Delta^{-1/3}e_\mu{}^\alpha 
  \big(G_{\alpha ab}\,\big(\Phi^\dagger
  \mathbb{P}_+\boldsymbol{\Gamma}^{ab6} \Phi\big){}^A{\!}_B 
  -\bar G_{\alpha ab}\,\big(\Phi^\dagger 
  \mathbb{P}_-\boldsymbol{\Gamma}^{ab6} \Phi\big){}^A{\!}_B   \big)
  \nonumber\\
  &\,
  -  \big(\Phi^\dagger \, \partial_\mu \Phi\big){}^A{\!}_B \,, 
  \nonumber\\[2mm]
  \mathcal{Q}_m{\!}^A{\!}_B  =&\,
  -\tfrac14(S^{-1}\mathring{D}_{m}S)^{ab}\, 
  \big(\Phi^\dagger\boldsymbol{\Gamma}_{ab} \Phi\big){}^A{\!}_B
  +\tfrac12\mathrm{i}Q_m \,\big(\Phi^\dagger\boldsymbol{\Gamma}_7
  \Phi\big){}^A{\!}_B \nonumber\\
  &\,  +\tfrac{1}{24}\mathrm{i}
  \big( G_{mbc}  \,\big(\Phi^\dagger  \mathbb{P}_{+}
  \boldsymbol{\Gamma}^{bc6} \Phi\big){}^A{\!}_B 
  -\bar G_{mbc} \,\big(\Phi^\dagger
  \mathbb{P}_{-}\boldsymbol{\Gamma}^{bc6} \Phi\big){}^A{\!}_B \big) 
  \nonumber\\    
  &-\tfrac{1}{600}\varepsilon^{abcde}F_{abcde}
  \,\big(\Phi^\dagger\boldsymbol{\Gamma}_{m6} 
  \Phi\big){}^A{\!}_B -
  \big(\Phi^\dagger \, \partial_m \Phi\big){}^A{\!}_B\,.
\end{align}
The field strength $\mathcal{H}_{\alpha\beta}{}^{AB}$ spans the
$\overline{\mathbf{27}}$ of $\mathrm{USp}(8)$ and reads
\begin{align}
  \label{eq:def-H} 
  \mathcal{H}_{\alpha\beta}{}^{AB} =&\, 
  \mathrm{i}\Delta^{-1/3}\big[(S^{-1})_a{}^b \,\mathring{e}_b{}^m
  e_{[\alpha}{}^\mu\,\partial_m
  e_{\mu\beta]}
  -\Delta^{2/3}\,e_{[\alpha}{}^\mu
  e_{\beta]}{}^\nu\,\mathring{e}_m{}^b \,S_b{}^a\, 
  \mathcal{D}_{\mu}B_{\nu}{}^m\big] \nonumber\\
  &\,\qquad \times\big(\Phi^\dagger \boldsymbol{\Gamma}_{a6}
  \boldsymbol{\Gamma}_7\bar\Omega\bar\Phi\big){}^{AB}\nonumber\\ 
  &\,-\tfrac12\Delta^{-1/3}\big(G_{a\alpha\beta}\,\big(\Phi^\dagger
  \mathbb{P}_{+} \boldsymbol{\Gamma}^a
  \bar\Omega\bar\Phi\big){}^{AB}
  +\bar G_{a\alpha\beta}\,\big(\Phi^\dagger
  \mathbb{P}_{-} \boldsymbol{\Gamma}^a
  \bar\Omega\bar\Phi\big){}^{AB} \big)
 \nonumber\\ 
  &\,-\tfrac{1}{12}\Delta^{-1/3}\varepsilon_{\alpha\beta\gamma\delta\lambda}
  \big(G^{\gamma\delta\lambda}\,\big(\Phi^\dagger
  \mathbb{P}_{+} \boldsymbol{\Gamma}_6\bar\Omega\bar\Phi\big){}^{AB}
  -\bar G^{\gamma\delta\lambda}\,\big(\Phi^\dagger
  \mathbb{P}_{-} \boldsymbol{\Gamma}_6\bar\Omega\bar\Phi\big){}^{AB}
  \big)   \nonumber\\ 
  &\,-\tfrac16\mathrm{i}\Delta^{-1/3}\varepsilon^{abcde}
  F_{\alpha\beta abc} 
  \big(\Phi^\dagger\boldsymbol{\Gamma}_{de}
  \boldsymbol{\Gamma}_7\bar\Omega\bar\Phi\big){}^{AB} \,, 
\end{align}
 Finally we have used the identity
\begin{align}
  \label{eq:DV}
  D_m \mathcal{V}_{AB}{}^m = - \big[(S^{-1}\mathring{D}_m
  S)^{(ab)}\, e_a{}^m \,e_{n b}  +\tfrac13
   \partial_n\ln \Delta \big] \,\mathcal{V}_{AB}{}^n
  \,. 
\end{align}
With these definitions the local $\mathrm{USp}(8)$ covariance of the
gravitino supersymmetry variations has been established. 

Now we return to the supersymmetry transformations of the spin-$1/2$
fields. Upon combining the results \eqref{eq:grav-a-var} into the
covariant form $\delta\chi^{ABC}$, one obtains, after some rearrangements similar to
those used in $\delta\psi_\mu{\!}^A$, 
\begin{align} 
  \label{eq:delta-chi}
  \delta \chi^{ABC}  = &\, \tfrac12 \mathrm{i} \,\gamma^\mu  
  \mathcal{P}_\mu{}^{ABCD} \, \Omega_{DE} \, \epsilon^E  \nonumber\\
  &\,-\tfrac{3}{16} \,
  \gamma^{\alpha\beta} \big[ \mathcal{H}_{\alpha\beta}{}^{[ AB} \,\epsilon^{ C ]} -
  \tfrac13 \,\bar \Omega^{[AB } {\cal H}_{\alpha\beta}{}^{C]D}  \Omega_{DE}\,
  \epsilon^E \big]    \nonumber \\ 
  &\, -3\,\bar\Omega^{D[A}\big[ \bar\Omega^{B|E}\,\mathcal{V}_{DE}{}^m\, D_m
  \epsilon^{C]} 
  -\tfrac13\bar\Omega^{BC]}\,\mathcal{V}_{DE}{}^m
   \, D_m \epsilon^E\big] \nonumber\\
   &\,-\tfrac32 \bar\Omega^{D[A}
   \big[\bar\Omega^{B|E}\,D_m\mathcal{V}_{DE}{}^m\,
   \epsilon^{C]}-\tfrac13\bar\Omega^{BC]}\, D_m\mathcal{V}_{DE}{}^m
   \,\epsilon^E\big]\nonumber\\
  &\, -{2} \,  \mathcal{P}_m{}^{ABCD} \,{\cal V}_{DE}{}^m \epsilon^E \, .
\end{align} 
In this expression two new tensors appear, $\mathcal{P}_\mu{}^{ABCD}$
and  $\mathcal{P}_m{}^{ABCD}$, which transform in the
$\boldsymbol{42}$ representation of $\mathrm{USp}(8)$. These tensors
also appear in the so-called vielbein postulates, 
\begin{align} 
  \label{eq:vielbein-postulates}
  \mathring D_m  {\cal V}_{AB}{}^n - 2\,\mathcal{Q}_m{\!}^C{}_{ [ A}\, {\cal
    V}_{B ] C}{}^n + \Omega_{AC} \, \Omega_{BD} \, {\cal P}_{m}{}^{ CDEF}\,
  {\cal V}_{EF}{}^n =&\, 0 \, ,\nonumber\\[2mm]
  \mathcal{D}_\mu {\cal V}_{AB}{}^m + \tfrac13 \mathring D_n
  C_\mu{}^n\, \mathcal{V}_{AB}{}^m + \mathring D_n
  C_\mu{}^m \,\mathcal{V}_{AB}{}^n \qquad\qquad\qquad &\nonumber\\
  -2\,\mathcal{Q}_\mu{\!}^C{\!}_{ [A} \,\mathcal{V}_{B ] C}{}^m +
  \Omega_{AC}\, \Omega_{BD} \, \mathcal{P}_{\mu}{}^{ CDEF}\,
  \mathcal{V}_{EF}{}^m =&\,0 \,.
\end{align}
Note that these expressions are similar to the corresponding
postulates in $11D$ \cite{deWit:1986mz}. Such equations will apply to
all the generalized vielbeine, but we refrain from presenting further
results. Note that we have again written $B_\mu{}^m$ as $C_\mu{}^m$. 

Both the supersymmetry transformations \eqref{eq:gravitino-cov} and
\eqref{eq:delta-chi} thus take a manifestly $\mathrm{USp}(8)$ covariant
form. The two new tensors, $\mathcal{P}_\mu{}^{ABCD}$ and
$\mathcal{P}_m{}^{ABCD}$, are defined by (in the gauge where the phase
factor $\Phi$ is set to unity)
{\setlength\arraycolsep{1pt}
\begin{align} 
  \label{eq:AtensorExtAlt}
  \mathcal{P}_\mu{}^{ABCD}=&\,   \tfrac18 \Delta^{-1/3} (P_\mu 
  +\bar P_\mu ) \left[ (\mathbf{\Gamma}_{a } 
 \bar\Omega)^{[ AB}  (\mathbf{\Gamma}^{a} \bar\Omega)^{CD ]}
 +(\mathbf{\Gamma}_{6} \bar\Omega)^{[ AB}  (\mathbf{\Gamma}_6
 \bar\Omega)^{CD ]} +2 \, \bar{\Omega}^{[AB} \bar{\Omega}^{CD]}
   \right] \nonumber \\ 
  &\,  - \tfrac18  \Delta^{-1/3} (P_\mu - \bar P_\mu ) \left[
  (\mathbf{\Gamma}_{a} \bar\Omega)^{[ AB}  (\mathbf{\Gamma}^{a}
  \mathbf{\Gamma}_7 \bar\Omega)^{CD ]}  + (\mathbf{\Gamma}_{6}
  \bar\Omega)^{[ AB}  (\mathbf{\Gamma}_6 \mathbf{\Gamma}_7
  \bar\Omega)^{CD ]} \right]  \nonumber \\ 
& - \tfrac{3}{16} \left( (S^{-1} \mathcal{D}_\mu S)_{(ab)} - \delta_{c(a}
  e_{b)}{}^m e_n{}^c \mathring D_m B_\mu{}^n \right)
\nonumber \\
&\,\qquad\times  \left[ (\mathbf{\Gamma}^{ac} \mathbf{\Gamma}_7  
\bar\Omega )^{[ AB}  (\mathbf{\Gamma}^b{}_c \mathbf{\Gamma}_7
\bar\Omega )^{CD ]}  +(\mathbf{\Gamma}^{ a } \mathbf{\Gamma}_{ 6 }
\mathbf{\Gamma}_7 \bar\Omega )^{[ AB}  (\mathbf{\Gamma}^{b}
\mathbf{\Gamma}_{6} \mathbf{\Gamma}_7 \bar\Omega )^{CD ]}  \right]
\nonumber \\ 
&\, - \tfrac{3}{16} \left( \partial_\mu \ln \Delta - \mathring
  D_m B_\mu{}^m \right) \left[ (\mathbf{\Gamma}^{c}\mathbf{\Gamma}_{6}
  \mathbf{\Gamma}_7 \bar\Omega )^{[ AB} (\mathbf{\Gamma}_{c}
  \mathbf{\Gamma}_6 \mathbf{\Gamma}_7
  \bar\Omega )^{CD ]}  -\tfrac{10}{3} \, \bar{\Omega}^{[AB}
  \bar{\Omega}^{CD]}  \right]  \nonumber \\ 
&\,   -\tfrac1{32}  \, \Delta^{-1/3} e_\mu{}^\alpha 
F_{\alpha cdef} \varepsilon_a{}^{cdef} (\mathbf{\Gamma}_{ab} 
\mathbf{\Gamma}_7 \bar\Omega )^{[ AB}  (\mathbf{\Gamma}^b{} 
\mathbf{\Gamma}_6 \mathbf{\Gamma}_7 \bar\Omega )^{CD ]}    \nonumber \\
&\, - \tfrac{1}{32} \Delta^{-1/3} e_\mu{}^\alpha 
\varepsilon_{abcde} ( G_\alpha{}^{de} + \bar G_\alpha{}^{de} )  
  (\mathbf{\Gamma}^{ a} \mathbf{\Gamma}_7 \bar\Omega )^{ [ AB}  
 (\mathbf{\Gamma}^{b c } \mathbf{\Gamma}_7  \bar\Omega)^{CD ]}  \nonumber \\
&\,   + \tfrac{1}{16}  \mathrm{i}  \Delta^{-1/3} e_\mu{}^\alpha 
( G_{\alpha ab} - \bar G_{\alpha ab}  )  \nonumber\\
&\,\qquad\times\left[  (\mathbf{\Gamma}^{ab} 
\mathbf{\Gamma}_7 \bar\Omega )^{ [ AB}  
(\mathbf{\Gamma}_{6} \mathbf{\Gamma}_7  \bar\Omega)^{CD ]} 
+ 2(\mathbf{\Gamma}^{a} \mathbf{\Gamma}_7 \bar\Omega )^{ [ AB}  
(\mathbf{\Gamma}^{b } \mathbf{\Gamma}_{6 } \mathbf{\Gamma}_7
\bar\Omega)^{CD ]}  \right] \,,
\end{align} 
}
and
{\setlength\arraycolsep{1pt}
\begin{align} 
\label{eq:AtensorExtAlt2}
{\cal P}_m{}^{ABCD} =&\,  \tfrac18  (P_m +\bar P_m )   \left[
  (\mathbf{\Gamma}_{a } \bar\Omega)^{[ AB}  (\mathbf{\Gamma}^{a}
  \bar\Omega)^{CD ]} +(\mathbf{\Gamma}_{6} \bar\Omega)^{[ AB}
  (\mathbf{\Gamma}_6 \bar\Omega)^{CD ]} +2 \, \bar{\Omega}^{[AB}
  \bar{\Omega}^{CD]}  \right] \nonumber \\ 
&\,   - \tfrac18 (P_m - \bar P_m )   \left[ (\mathbf{\Gamma}_{a}
  \bar\Omega)^{[ AB}  (\mathbf{\Gamma}^{a} \mathbf{\Gamma}_7
  \bar\Omega)^{CD ]}  + (\mathbf{\Gamma}_{6} \bar\Omega)^{[ AB}
  (\mathbf{\Gamma}_6 \mathbf{\Gamma}_7 \bar\Omega)^{CD ]} \right]
\nonumber \\ 
&\, - \tfrac{3}{16}  (S^{-1} \mathring D_m S)_{(ab)}  \nonumber\\
&\, \qquad \times \left[
  (\mathbf{\Gamma}^{ac} \mathbf{\Gamma}_7 \bar\Omega )^{[ AB}
  (\mathbf{\Gamma}^b{}_c \mathbf{\Gamma}_7 \bar\Omega )^{CD ]}
  +(\mathbf{\Gamma}^{ a } \mathbf{\Gamma}_{ 6 } \mathbf{\Gamma}_7
  \bar\Omega )^{[ AB}  (\mathbf{\Gamma}^{b} \mathbf{\Gamma}_{6}
  \mathbf{\Gamma}_7 \bar\Omega )^{CD ]} \right] \nonumber \\ 
  &\, -\tfrac{3}{16}  \partial_m \ln \Delta  \left[
  (\mathbf{\Gamma}^{c}\mathbf{\Gamma}_{6} \mathbf{\Gamma}_7 \bar\Omega
  )^{[ AB}  (\mathbf{\Gamma}_{c} \mathbf{\Gamma}_6 \mathbf{\Gamma}_7
  \bar\Omega )^{CD ]}  -\tfrac{10}{3} \, \bar{\Omega}^{[AB}
  \bar{\Omega}^{CD]}  \right]  \nonumber \\ 
  &\,  -\tfrac{1}{800}  \, e_m{}^a F_{cdefg} \varepsilon^{cdefg}
  (\mathbf{\Gamma}_{ab} \mathbf{\Gamma}_7 \bar\Omega )^{[ AB}
  (\mathbf{\Gamma}^b{} \mathbf{\Gamma}_6 \mathbf{\Gamma}_7 \bar\Omega
  )^{CD ]}    \nonumber \\ 
  &\, -  \tfrac{1}{96} e_m{}^f \varepsilon_{abcde} ( G^{de}{}_f + \bar
  G^{de}{}_f )    (\mathbf{\Gamma}^{ a} \mathbf{\Gamma}_7 \bar\Omega )^{
    [ AB}  (\mathbf{\Gamma}^{b c } \mathbf{\Gamma}_7  \bar\Omega)^{CD ]}
  \nonumber \\ 
  &\,  + \tfrac{1}{48} \mathrm{i} \, e_m{}^c  \,  ( G_{abc} - \bar
  G_{abc} ) \nonumber\\
  &\, \qquad \times\left[  (\mathbf{\Gamma}^{ab} \mathbf{\Gamma}_7 \bar\Omega
    )^{ [ AB}  (\mathbf{\Gamma}_{6} \mathbf{\Gamma}_7  \bar\Omega)^{CD
      ]} + 2(\mathbf{\Gamma}^{a} \mathbf{\Gamma}_7 \bar\Omega )^{ [ AB}
    (\mathbf{\Gamma}^{b } \mathbf{\Gamma}_{6 } \mathbf{\Gamma}_7
    \bar\Omega)^{CD ]}  \right] \,. 
\end{align} 
} Note that the above formulae \eqref{eq:AtensorExtAlt} and
\eqref{eq:AtensorExtAlt2} are unique up to Fierz reordering.

\section{On the consistent truncation to $\boldsymbol{5D}$  
 $\boldsymbol{\mathrm{SO}(6)}$ gauged supergravity}
\label{sec:cons-trunc-maxim}
\setcounter{equation}{0} 
The results of this paper can be used to establish the full
consistency of the truncation of IIB supergravity compactified on the
sphere $S^5$ to $5D$ $\mathrm{SO}(6)$ gauged supergravity
\cite{Gunaydin:1985cu}, along the same lines that were followed
originally for the truncation of $11D$ supergravity compactified on
the sphere $S^7$ to $4D$ $\mathrm{SO}(8)$ gauged supergravity
\cite{deWit:1982ig,deWit:1983vq,deWit:1984nz,Nicolai:2011cy,
  deWit:2013ija,deWit:1986iy}. For IIB supergravity some partial
results have already appeared in the literature
\cite{Pilch:2000ue,Lee:2014mla} and they will be confirmed below from
the results of this paper. It is clear that additional results can be
obtained by a more complete analysis, but a full treatment is outside
the scope of this paper. The same holds for a study of more general
truncations along the lines pursued in \cite{Godazgar:2013oba} for
$11D$ supergravity.

It is worth stressing that this concept of a consistent truncation
goes beyond proving that solutions of $5D$ maximal $\mathrm{SO}(6)$
gauged supergravity can be uplifted to the IIB theory. Rather,
starting from the fully supersymmetric solution with
$\mathrm{AdS}_5\times S^5$, one sweeps out the full field
configuration space of $5D$ maximal supergravity in the ten-dimensional
field configuration space. This is done by writing the $10D$ fields as
functions of the $5D$ fields, involving $y$-dependent functions,
mostly constructed from the $S^5$ Killing spinors, in such a way that
the $10D$ supersymmetry transformations remain consistent upon
extracting these $y$-dependent factors. In the case at hand these
eight independent, pseudo-real, Killing spinors, $\eta^A(y)$, satisfy
\begin{equation}
  \label{eq:Killing-spinor}
  \big(\mathring{D}_m +\tfrac12  m_5 \,\mathring{e}{}_m{\!}^a \,
  \boldsymbol{\Gamma}_{a6}\big)\eta(y)  =0 \,.
\end{equation}
Here $m_5$ denotes the inverse $S^5$ radius which is related to the
background value of the field strength $F_{mnpqr}$ by $m_5 =
\tfrac1{120}\,\mathring{e}\, \varepsilon^{mnpqr}
\mathring{F}_{mnpqr}$. Furthermore $\mathring{D}_m$ equals the $S^5$
background covariant derivative and $\mathring{e}{}_m{\!}^a$ is the
globally defined f\"unfbein on $S^5$. The Killing spinor equation
\eqref{eq:Killing-spinor} is motivated by the fact that it
characterizes the supersymmetry of the $\mathrm{AdS}_5\times S^5$
solution of IIB supergravity. Note that all Killing spinors in this
section will be commuting.

In view of what follows it is useful to first discuss these Killing
spinors in more detail. Since \eqref{eq:Killing-spinor} is a
first-order differential equation it allows for eight independent
solutions. However, in five Euclidean dimensions, the Clifford algebra
associated with the $\mathrm{SO}(5)$ gamma matrices has an
automorphism group equal to $\mathrm{Sp}(1)\cong
\mathrm{SU}(2)$. Consequently one can choose six independent spinors
that are not related by the action of the automorphism group, so that
the orbit that is then swept out under the action of the
$\mathrm{SU}(2)$ automorphism group will yield the two remaining
independent spinors. Bilinears constructed from the Killing spinors
that involve only the original $\mathrm{SO}(5)$ gamma matrices will
necessarily be invariant under the automorphism group and therefore
the number of independent spinor bilinears of this type will
constitute $6\otimes 6$ independent bilinears which decompose into 
$15$ anti-symmetric and $21$ symmetric components. This argument,
which incidentally also plays a role when analyzing the number of
degrees of freedom of the generalized vielbeine in section
\ref{sec:gen-vielbeine}, explains why the bilinears produce precisely
$15$ independent Killing vectors. More specifically it follows that
\begin{equation}
  \label{eq:S6-killing-vectors}
  \mathring{e}{}_a{}^m\, \bar\eta_1 (y)\, \boldsymbol{\Gamma}^{a6}
  \boldsymbol{\Gamma} _7
  \,\eta^2(y)  = \sum_{[{\hat a \hat b}]}  C^{ \hat a \hat b} \,  K^m{}_{\hat
    a\hat b} (y)  \,,
\end{equation}
where $\eta^{1,2}$ are two possible Killing spinors (with
$\bar\eta\equiv \eta^\dagger$) and the indices $\hat a,\hat{b},
\ldots$ denote the components of the defining representation of the
$\mathrm{SO}(6)$; in this background this $\mathrm{SO}(6)$ corresponds
to the isometry group of the sphere $S^5$. The fifteen Killing vectors
are labeled with anti-symmetric pairs $[{\hat a\hat b}]$, and the
$C^{{\hat a\hat b}}$ are constants. To prove this relation one can
write the gamma matrices in terms of the original $\mathrm{SO}(5)$
gamma matrices and/or one can prove directly that the left-hand side
of \eqref{eq:S6-killing-vectors} satisfies the Killing equation by
virtue of \eqref{eq:Killing-spinor}.

Taking the derivative of the Killing vectors one finds another tensor
that is also anti-symmetric in $[{\hat a\hat b}]$ (note that indices are
lowered/raised with the $S^5$ metric $\mathring{g}_{mn}$ and its
inverse),
\begin{align}
  \label{eq:D-Killing}
  \mathring{D}_m K_{n\,{\hat a\hat b}\,} = m_5\,  K_{mn\,{\hat a\hat b}} \,. 
\end{align}
In five dimensions this tensor is known as a Killing tensor. It
satisfies the equation
\begin{equation}
  \label{eq:Killing potential}
  \mathring{D}_m K_{np\,{\hat a\hat b}}  = -2\, m_5 \, \mathring{g}_{m[n} \,
  K_{p]\, {\hat a\hat b}}\, .
\end{equation}
From the previous results one then derives 
\begin{equation}
  \label{eq:S6-killing-tensors}
  \mathring{e}{}_m{}^a\,\mathring{e}{}_n{}^b\,  \bar\eta_1 (y)\,
  \boldsymbol{\Gamma}_{ab} 
  \boldsymbol{\Gamma} _7
  \,\eta^2(y)  = \sum_{[{\hat a \hat b}]}  C^{{\hat a \hat b}} \,
  K_{mn \,{\hat a\hat b}} (y)  \,.
\end{equation}

After these observations we turn to the consistent truncation
ans\"atze for the $10D$ fields. We start from eight independent
Killing spinors, now labeled by indices $i,j,\ldots=1,2,\ldots,8$,
such that these spinors form an orthonormal basis in the
$\mathrm{USp}(8)$ spinor space and are subject to a pseudo-reality
condition,
\begin{equation}
  \label{eq:complete-spinor}
  \bar \eta^i(y) \,\eta_j(y)= \delta^i{\!}_j \,, \qquad
  \bar\eta^i{\!}_A = \bar\Omega^{ij} \,\Omega_{AB} \, \eta_j{\!}^B\,,  
\end{equation}
where $\bar\Omega^{ij}$ and $\Omega_{AB}$ are the symplectic matrices
used before. The truncation for the fermions, the supersymmetry
parameters and the space-time vielbein $e_\mu{}^\alpha$ are then assumed to
take the form,\footnote{ 
  The phase factor $\Phi$ is only implicit in the formulae below, but
  it actually plays a crucial role to ensure that consistency is
  achieved (see e.g. \cite{deWit:1983vq}). }  
\begin{align}
  \label{eq:truncation-formulae}
  \psi_\mu{}^A(x,y) =&\, \psi_\mu{}^i(x) \,\eta_i{\!}^A(y)\,, \nonumber\\
  \epsilon^A(x,y) =&\, \epsilon^i(x)\,\eta_i{\!}^A(y)\,, \nonumber\\
  \chi^{ABC}(x,y) =&\, \chi^{ijk}(x) \,\eta_i{\!}^A(y)\,
  \eta_j{\!}^B(y)\,  \eta_k{\!}^C(y) \,,  \nonumber\\
  e_\mu{}^\alpha(x,y) =&\, e_\mu{}^\alpha(x)\,. 
\end{align}
Making this assumption will obviously restrict the $\mathrm{USp}(8)$ R-symmetry
transformations to 
\begin{equation}
  \label{eq:2trunc-R-symm}
    U^A{\!}_B(x,y) = U^i{}_j (x) \,\eta_i{\!}^A(y)\,\bar\eta^j{\!}_B(y)\,, 
\end{equation}
and leaves the group structure intact by virtue of the conditions
\eqref{eq:complete-spinor}. Observe that the supersymmetry
transformations for $e_\mu{}^\alpha$ are consistent under this
truncation. However, for the other bosons the truncation ansatz is
more subtle. 

To derive the truncation ans\"atze for the remaining bosons one first
considers their supersymmetry variations into the fermions, defined
according to \eqref{eq:truncation-formulae}. For instance, consider
\eqref{eq:VecVar}, which will now take the form,
\begin{equation}  
  \label{eq:VecVar-trunc}
  \delta C_{\mu}{\!}^{M}(x,y) =  2 \,\big[\mathrm{i}\bar\Omega^{ik}\,
  \bar\epsilon_{k}(x)\,\psi_\mu{\!}^j(x) 
  +\bar\epsilon_k(x)\,\gamma_\mu\chi^{ijk}(x)\big]
  \mathcal{V}_{ij}{\!}^M(x,y) \,, 
\end{equation}
where
\begin{equation}
  \label{eq:vec-var-5}
  \mathcal{V}_{ij}{\!}^{M}(x,y) = \eta_i{\!}^A(y)\,  \eta_j{\!}^B(y)\,
  \mathcal{V}_{AB}{}^{M}(x,y) \,.  
\end{equation}
The consistency of the truncation now requires that the $y$-dependence
of $C_\mu{}^M$ and $\mathcal{V}_{ij}{}^M$ will match. 

Before deriving some of the additional truncation results, let us
first compare the situation regarding the compactification on the
torus $T^5$ and the sphere $S^5$. In the torus truncation all the
fields $C_\mu{}^M$ will appear and will be independent of the torus
coordinates $y^m$. Consequently the generalized vielbeine
$\mathcal{V}_{ij}{}^M$ will also be $y$-independent and they will be
precisely equal to the corresponding quantities $U_{ij}{}^M(x)$ that
are a representative of the $\mathrm{E}_{6(6)}/\mathrm{USp}(8)$ coset
space.\footnote{ 
  Here we deviate from the notation used in \cite{deWit:2004nw} where
  the $U_{ij}{}^M(x)$ are denoted also by $\mathcal{V}_{ij}{}^M$.
 } 
The tensor fields $C_{\mu\nu\,M}$ can be gauged away in the torus
truncation where they carry no additional information and they are
simply dual to the vector fields.

The situation for the $S^5$ compactification is different, as in this
case the various `physical' fields reside in both the $C_\mu{}^M$ and
$C_{\mu\nu\,M}$ \cite{Gunaydin:1985cu}. More precisely, in this case
there are fifteen vector fields transforming in the adjoint
representation of the $\mathrm{SO}(6)$ subgroup of $\mathrm{E}_{6(6)}$
and twelve tensor fields transforming as a direct product of the
vector representation of the same $\mathrm{SO}(6)$ subgroup and the
doublet represention of the $\mathrm{SU}(1,1)$ subgroup of
$\mathrm{E}_{6(6)}$. The remaining vector and tensor fields in the
sphere truncation are the duals of these $15\oplus12$ fields, which
can be gauged away in the embedding tensor approach. This
decomposition in terms of the expected vector and tensor fields must
be reflected in the truncation ans\"atze for the vectors and tensors.

It is important to realize that the fields $C_\mu{}^M$ and
$C_{\mu\nu\,M}$ are gauge fields, which excludes field-dependent
multiplicative redefinitions. Given that the $y$-dependence should be
extracted in the form of the geometric quantities associated with the
sphere, it is rather obvious what the truncation ans\"atze should
be. Let us first demonstrate this for the $\mathrm{SU}(1,1)$ invariant
vector and tensor fields, $C_\mu{}^m$, $C_\mu{}^{mn}$, $C_{\mu\nu\,m}$
and $C_{\mu\nu\,mn}$, each of which can be decomposed into the fifteen
Killing vectors or tensors according to
\begin{align}
  \label{eq:KK-vector-tensor}
  C_\mu{}^m (x,y) =&\, K^{m}{}_{\hat a\hat b}(y)\, A_{\mu}{}^{{\hat a\hat b}} (x) \,,
  \nonumber\\
  C_\mu{}^{mn} (x,y) =&\, K^{mn}{}_{\hat a\hat b}(y)\, \tilde
  A_\mu{}^{\hat a\hat b} (x) \,,\nonumber\\
  C_{\mu\nu\,m} (x,y) =&\, K_{m}{}^{\hat a\hat b}(y)\, 
      B_{\mu\nu\,{\hat a\hat b}} (x) \,,\nonumber\\
  C_{\mu\nu\,mn} (x,y) =&\, K_{mn}{}^{\hat  a\hat b}(y)\, \tilde
  B_{\mu\nu\,{\hat a\hat b}} (x) \,.
\end{align}
However, as explained above, in $5D$ one has only fifteen vector and
fifteen tensor fields in the $\mathrm{SU}(1,1)$ invariant sector, so
that one must assume that $A_{\mu}{}^{{\hat a\hat b}} (x) $ and
$\tilde A_{\mu}{}^{{\hat a\hat b}} (x)$ are identical up to a possible
multiplicative constant; the same holds for the tensor fields
$B_{\mu\nu\,{\hat a\hat b}} (x) $ and $\tilde B_{\mu\nu\,{\hat a\hat b}}
(x)$. However, here and in the following we will not be concerned
about numerical factors, also in view of the fact that we have
not adopted specific normalizations for the Killing vectors and
tensors.

A similar decomposition applies to the generalized vielbeine
$\mathcal{V}_{ij}{}^m$, $\mathcal{V}_{ij}{}^{mn}$,
$\mathcal{V}_m{}^{ij}$ and $\mathcal{V}_{mn}{}^{ij}$ which appear in
the variation of the above fields,
\begin{align}
  \label{eq:vielbein-trunc}
   \mathcal{V}_{ij}{\!}^{m}(x,y) =&\, U_{ij}{\!}^{\hat a\hat b}(x)\,
   K^{m}{}_{\hat a\hat b}(y) \,,   \nonumber\\
   \mathcal{V}_{ij}{\!}^{mn}(x,y)  =&\,  U_{ij}{\!}^{\hat a\hat b}(x)\,
   K^{mn}{}_{\hat a\hat b}(y) \,,  \nonumber\\
   \mathcal{V}_{m}{}^{ij}(x,y)  =&\, U_{\hat a\hat b}{}^{ij}(x)\,
   K_{m}{}^{\hat a\hat b}(y) \,,  \nonumber\\
   \mathcal{V}_{mn}{}^{ij}(x,y)  =&\,  U_{\hat a\hat b}{}^{ij}(x)\,
   K_{mn}{}^{\hat a\hat b}(y) \,,  
\end{align}
where, as we have explained above, $U_{ij}{\!}^{\hat a\hat b}(x)$ and
$U_{\hat a\hat b}{}^{ij}(x)$ are the (unique) components of the
$\mathrm{E}_{6(6)}/\mathrm{USp}(8)$ coset space satisfying 
\begin{equation}
  \label{eq:speudoreal-U}
  U_{\hat a\hat b}{}^{ij}(x)\, U_{ij}{\!}^{\hat c\hat d}(x) = 
  2\,\delta_{\hat a\hat b}{\!}^{\hat c\hat d}\,. 
\end{equation}
Note that the combined equations \eqref{eq:KK-vector-tensor} and
\eqref{eq:vielbein-trunc} ensure that the corresponding supersymmetry
transformations are consistent under the truncation. 

Subsequently we consider the following identities that follow from direct
calculation using the generalized vielbeine presented in section
\ref{sec:gen-vielbeine}, after converting the $\mathrm{USp}(8)$
indices according to \eqref{eq:vec-var-5},
\begin{align}
  \label{eq:so(6)-sector-identites}
     \bar{\mathcal{V}}^{ik\,m} \,\mathcal{V}_{kj}{}^n
        +\bar{\mathcal{V}}^{ik\,n} \,\mathcal{V}_{kj}{}^m =&\, -\tfrac14
              \delta^i{}_j \,\bar{\mathcal{V}}^{kl\,m}
              \,\mathcal{V}_{kl}{}^n\,, \nonumber\\[2mm]
  \bar\Omega^{ik}\,\bar\Omega^{jl} \,\mathcal{V}_{ij}{}^m \,\mathcal{V}_{kl}{}^n=&\, \tfrac12 
  \Delta^{-2/3} g^{mn} \,,\nonumber\\[2mm]
  \bar\Omega^{ik}\,\bar\Omega^{jl} \,\mathcal{V}_{ij}{}^m \,\mathcal{V}_{kl}{}^{np}  =&\,  
  \tfrac{32}{15} \sqrt{5} \,\mathring{e}^{-1}
  \varepsilon^{npqrs} \big[A_{qrst} +\tfrac3{16}
  \mathrm{i}\varepsilon_{\alpha\beta}  \,A^\alpha{\!}_{qr}
  A^\beta{\!}_{st}\big]  \,
  \bar{\mathcal{V}}^{ij\,m} \,\mathcal{V}_{ij}{}^t\,,
\end{align}
where $g^{mn}$ is the full (inverse) internal metric, which depends on
the scalar fields. It is therefore different from the $S^5$ inverse
metric $\mathring{g}^{mn}(y)$, unless the scalar fields take their
background values. We remind the reader that the generalized vielbeine
are pseudo-real so that the complex conjugate equals
$\bar{\mathcal{V}}^{ij\,m}\equiv \big(\mathcal{V}_{ij}{}^m\big)^\ast =
\bar\Omega^{ik}\,\bar\Omega^{jk} \,\mathcal{V}_{kl}{}^m$. Hence it
follows that
\begin{align}
  \label{eq:full-metric}
  \Delta^{-2/3}\, g^{mn}(x,y)  = 2\,\bar\Omega^{ik}\,\bar\Omega^{jl} \,
  U_{ij}{\!}^{\hat a\hat b}(x)\,  
  U_{kl}{\!}^{\hat c\hat d}(x) \; K^{m}{}_{\hat a\hat b}(y)\,
  K^{n}{}_{\hat c\hat d}(y)\,. 
\end{align}
with $\Delta^2= \det[g(x,y)]/\det[\mathring{g}(y)]$. This result is
rather generic and was first found for $11D$ supergravity compactified
on $S^7$ \cite{deWit:1984nz} with the prefactor $\Delta^{-1}$. For IIB
supergravity compactified on $S^5$ the above result was established in
\cite{Pilch:2000ue,Lee:2014mla}.

The next step is to study the consequences of the third identity
\eqref{eq:so(6)-sector-identites}. Substitution of the generalized
vielbeine leads to the equation 
\begin{align}
  \label{eq:A-mnpq-identity}
    \Delta^{-2/3} \big[A_{mnpq} +\tfrac3{16}
  \mathrm{i}\varepsilon_{\alpha\beta}  \,A^\alpha{\!}_{[mn}
  A^\beta{\!}_{p]q}\big] = &\, \tfrac1{64} \sqrt{5} \, 
  \bar\Omega^{ik}\,\bar\Omega^{jl} \,
  U_{ij}{\!}^{\hat a\hat b}(x)\,  
  U_{kl}{\!}^{\hat c\hat d}(x) \, g_{qr}(x,y) \nonumber\\
  &\,\times \mathring{e} \,\varepsilon_{mnptu}  K^{r}{}_{\hat a\hat b}(y)\, 
  K^{tu}{}_{\hat c\hat d}(y)\,. 
\end{align}
This identity has been derived in the context of generalized geometry
\cite{Lee:2014mla} where the corresponding reduction manifold admits a
generalized parallelization. The derivation above follows the same
approach as the one followed in the context of $11D$ supergravity
\cite{deWit:2013ija}, where it gave rise to the non-linear ansatz of
the internal tensor $A_{mnp}$.  One term on the left-hand side should
be modified in view of the fact that there is a non-zero background
four-form potential $\mathring{A}_{mnpq}$ because the
five-form field strength is non-vanishing in this background. The term
$A_{mnpq}$ on the left-hand side should therefore be replaced by
$A_{mnpq} -\mathring{A}_{mnpq}$.

Subsequently we continue to the twelve vector and twelve tensor fields
that transform under $\mathrm{SU}(1,1)$, namely $C_{\mu\,\alpha}$,
$C_{\mu\,\alpha m}$, $C_{\mu\nu}{}^\alpha$ and $C_{\mu\nu}{}^{\alpha
  m}$, which should be decomposed into the twelve vector and and
twelve tensor fields that one expects in $5D$. However, in view of
their number, it is not possible to expand these fields in terms of
Killing vectors or tensors. Therefore we introduce the
$\mathrm{SO}(6)$ vector fields $Y^{\hat a}(y)$ that satisfy $Y^{\hat
  a}(y)\, Y_{\hat a}(y)= 1$, whose parametrization in terms of the
$y^m$ is based on the same $\mathrm{SO}(6)/\mathrm{SO}(5)$ coset
representative as all other geometric quantities of $S^5$, such as the
metric and the Killing vectors and tensors (see, e.g. \cite{deWit:1984nz}). In that case one can
parametrize the remaining vector and tensor fields in terms of the
twelve expected $5D$ fields,
\begin{align}
  \label{eq:KK-vector-tensor-alpha}
  C_{\mu\,\alpha} (x,y) =&\, Y^{\hat a}(y)\, A_{\mu\,\alpha\hat a} (x) \,,
  \nonumber\\
 C_{\mu\,\alpha m} (x,y) =&\,\partial_m  Y^{\hat a}(y) \, 
  A_{\mu\,\alpha\hat a} (x) \,,\nonumber\\
 C_{\mu\nu}{}^{\alpha} (x,y) =&\, Y_{\hat a}(y)\, 
     B_{\mu\nu}{}^{\alpha \hat a} (x) \,,\nonumber\\
 C_{\mu\nu}{}^{\alpha m} (x,y) =&\, \mathring{g}^{mn}\,\partial_n  Y_{\hat a}(y) \, 
  B_{\mu\nu}{}^{\alpha \hat a}(x) \,.
\end{align}
A similar decomposition now applies to the generalized vielbeine
$\mathcal{V}_{ij\,\alpha}$, $\mathcal{V}_{ij \,\alpha m}$,
$\mathcal{V}^{\alpha\,ij}$ and $\mathcal{V}^{\alpha m\,ij}$ which appear in
the variation of the above fields,
\begin{align}
  \label{eq:vielbein-trunc-alpha}
   \mathcal{V}_{ij\,\alpha}(x,y) =&\, U_{ij\,\alpha\hat a}(x)\, Y^{\hat a}(y)
   \,,   \nonumber\\
   \mathcal{V}_{ij\, \alpha m}(x,y)  =&\,  U_{ij\,\alpha\hat a}(x) \,\partial_m 
   Y^{\hat a}(y)\,,  \nonumber\\
   \mathcal{V}^{\alpha \,ij}(x,y)  =&\, U^{\alpha\hat a\,ij}(x)\, 
   Y_{\hat a}(y) \,,  \nonumber\\ 
   \mathcal{V}^{\alpha m\,ij}(x,y)  =&\,  U^{\alpha\hat
     a\,ij}(x)\,\mathring{g}^{mn}\,\partial_n  Y_{\hat a}(y)  \,,  
\end{align}
where $U_{ij\,\alpha\hat a}(x)$ and $U^{\alpha\hat a\,ij}(x)$ are
again related to specific components of the
$\mathrm{E}_{6(6)}/\mathrm{USp}(8)$ coset space that appear in the
$5D$ theory. They satisfy 
\begin{equation}
  \label{eq:speudoreal-U}
  U^{\alpha\hat a\,ij}(x)\, U_{ij \,\beta\hat b}(x) =
  \delta^\alpha{\!}_\beta\, 
  \delta^{\hat a}{}_{\hat b} \,. 
\end{equation}

Now we consider the following identities that can be derived for the
generalized vielbeine, 
\begin{align}
  \label{eq:V-Valpha-identities}
  \bar\Omega^{ik}\,\bar\Omega^{jl} \, {\mathcal{V}}_{ij}{}^{m}
  \,\mathcal{V}_{kl\,\alpha n} =&\, \mathrm{i}  
  \varepsilon_{\alpha\beta} A^\beta{\!}_{np} \,
  \bar{\mathcal{V}}^{ij\,m} \,\mathcal{V}_{ij}{}^p\,, \nonumber \\
 \varepsilon_{\alpha\gamma}\, \Omega_{ik}\,\Omega_{jl}
 \,{\mathcal{V}}^{\gamma}{}^{ij}\, \mathcal{V}^{\beta\,kl} =&\,
 \tfrac5{4} 
  \Delta^{-4/3}  \big(\delta_\alpha{}^\beta-
  2\,\phi_\alpha\phi^\beta\big) \,.
\end{align}
From these identities we can derive the following results upon 
substituting the above truncation ans\"atze, 
\begin{align}
  \label{eq:2-tensor}
  \Delta^{-2/3}\, A^\alpha{\!}_{mn} = 2\mathrm{i}
  \,\varepsilon^{\alpha\beta} \, \bar\Omega^{ik}\,\bar\Omega^{jl}\, 
  U_{ij}{}^{\hat a \hat b}(x) \,U_{kl\,\beta\hat c}(x) \, K^p{}_{\hat
    a\hat b} (y) \,g_{p[m}(x,y) \, \partial_{n]} Y^{\hat c}(y) \,,  \nonumber\\[2mm]
  \Delta^{-4/3}  \big(\delta_\alpha{}^\beta-
  2\,\phi_\alpha\phi^\beta\big) = \tfrac45 \varepsilon_{\alpha\gamma}\, 
  \Omega_{ik}\,\Omega_{jl}\,
  U^{\gamma \hat a\, ij}(x) \,U^{\beta\hat b\,kl}(x) \, 
  Y_{\hat a }(y) \,Y_{\hat b}(y) \,. 
\end{align}
The first result has recently been derived based on generalized
geometry \cite{Lee:2014mla}, while the second result 
has been obtained long ago (under some mild assumptions) in
\cite{Pilch:2000ue} by using the same strategy as in this section. 

It is clear that so far we have probed only part of the possible
identities that can be derived based on the results of this paper. At
the same time, the mutual consistency of the various implications
should also be carefully investigated, in the same way as this was
done for $11D$ supergravity.  It should be interesting to pursue these
questions further.

\subsection*{Acknowledgements}
We acknowledge helpful discussions with Nicolas Boulanger, Hadi
Godazgar, Mahdi Godazgar, Olaf Hohm, Hermann Nicolai, Krzysztof Pilch,
Henning Samtleben, Daniel Waldram and Peter West. The work of F.C. and
B.d.W. is supported by the ERC Advanced Grant no. 246974, {\it
  ``Supersymmetry: a window to non-perturbative physics''}.  O.V. is
supported by a Marie Curie fellowship and is grateful to the CPHT of
\'Ecole Polytechnique for managing the administration of his
grant. O.V. is also supported in part by DOE grant de-sc0007870.
B.d.W. thanks the Center for the Fundamental Laws of Nature of Harvard
University for hospitality extended to him during the course of this
work.

\begin{appendix}
\section{Decomposition of gamma matrices and spinors}
\label{App:red-spinors-gamma-matrices}
\setcounter{equation}{0}
We start from $32\times32$ hermitian gamma matrices $\breve\Gamma_A$,
where $A=1,2,\ldots 10$, satisfying the Clifford algebra
anti-commutation relation, $\{\breve\Gamma_A,\,\breve\Gamma_B\} =
2\,\delta_{AB}\,\oneone_{32}$, and proceed in a way that is
independent of a specific representation for these gamma matrices.
The hermitian chirality operator, $\breve\Gamma_{11}$, is defined by
\begin{equation}
  \label{eq:Gamma-11}
  \breve\Gamma_{11} = \mathrm{i} \breve\Gamma_1\,\breve\Gamma_2\cdots
  \breve\Gamma_{10}\,, 
\end{equation}
and satisfies  
\begin{equation}
  \label{eq:prop-Gamm-11}
  \breve\Gamma_{11}{}^2 =\oneone_{32} \,, \qquad \{\breve\Gamma_A,
  \breve\Gamma_{11}\} = 0   \,. 
\end{equation}
Moreover we note the identity, 
\begin{equation}
  \label{eq:chiral-dual}
  \breve\Gamma^{ABCDE} = \tfrac1{120} \mathrm{i}
  \varepsilon^{ABCDEFGHIJ} \,\breve \Gamma_{FGHIJ} \,\breve\Gamma_{11}
  \,. 
\end{equation}
When considering compactifications from ten- to five-dimensional
space-times, the $10D$ tangent space is decomposed accordingly into a
direct product of two five-dimensional spaces, one corresponding to a
five-dimensional space-time and one corresponding to a
five-dimensional internal space. Since we are dealing with spinor
fields, it is then important to identify the gamma matrices
appropriate to this product space in terms of the original $10D$ gamma
matrices. 

To do so one first decomposes the gamma matrices into two sets,
$\breve\Gamma_\alpha$ with $\alpha=1,2,\dots,5$ and $\breve\Gamma_{a+5}$ with
$a=1,2,\ldots,5$.~\footnote{
  At this stage there is no difference between upper and lower
  indices, so that we are dealing with a positive Euclidean
  metric.}  
Subsequently one introduces hermitian matrices associated with the
two five-dimensional sectors,
\begin{equation}
  \label{eq:sub-chiral}
  \tilde\gamma= \breve\Gamma_1\,\breve\Gamma_2\,\breve\Gamma_3
  \,\breve\Gamma_4\,\breve\Gamma_5\,,\qquad 
  \tilde\Gamma= \breve\Gamma_6\,\breve\Gamma_7\,\breve\Gamma_8\,
  \breve\Gamma_9\,\breve \Gamma_{10}\,, 
\end{equation}
which satisfy the following properties,
\begin{align}
  \label{eq:tilde-prop}
  \tilde\gamma^2=&\, \oneone_{32} \,,\quad
  \tilde\Gamma^2= \oneone_{32}\,,\quad
  \{\tilde\gamma,\tilde\Gamma\}=0\,, 
  \quad
  \breve\Gamma_{11} = \mathrm{i} \tilde\gamma \,\tilde\Gamma \,. 
\end{align}
Subsequently one defines two sets of mutually commuting {\it hermitian}
gamma matrices,
\begin{equation}
  \label{eq:gamma-sets}
   \hat\gamma_\alpha =\mathrm{i}\breve\Gamma_\alpha\, \tilde\Gamma\,, \qquad
  \hat\Gamma_a = \mathrm{i}\breve\Gamma_{a+5}\,\tilde\gamma\,,
\end{equation}
so that
$\{\hat\gamma_\alpha,\hat\gamma_\beta\}=2\,\delta_{\alpha\beta}
\oneone_{32}$, $\{\hat\Gamma_a,\hat\Gamma_b\}=2\,\delta_{ab}
\oneone_{32}$, and $[\hat\gamma_\alpha,\hat \Gamma_a]=0$.  The 
matrices $\hat\gamma_\alpha$ will refer to the five-dimensional
space-time (to account for the signature one may write one of the five
gamma matrices, say $\hat\gamma^1$ as $\mathrm{i}\hat\gamma^0$) and
the matrices $\hat\Gamma_a$ to the five-dimensional internal
space. The matrices $\hat\gamma_\alpha$ and $\hat\Gamma_a$
commute with $\breve\Gamma_{11}$, as one can easily verify from the
above equations. It is important to note that
\begin{align}
  \label{eq:5-gamma}
  \hat\gamma_{[\alpha} \hat\gamma_\beta \hat\gamma_\gamma
  \hat\gamma_\delta \hat\gamma_{\tau]} =&\,
  \varepsilon_{\alpha\beta\gamma\delta\tau} \, \Gamma_{11}\,,
  \nonumber\\
  \hat\Gamma_{[a} \hat\Gamma_b \hat\Gamma_c
  \hat\Gamma_d \hat\Gamma_{e]} =&\, - 
  \varepsilon_{abcde} \, \Gamma_{11}\,,
\end{align}
where $\varepsilon_{12345}= +1$. Obviously, by choosing an explicit
representation for the $10D$ gamma matrices, one obtains explicit
expressions for the various matrices that we have defined above which
will reflect their properties.

Let us now consider the charge conjugation matrix. In ten dimensions
there exist two possible options for the charge conjugation matrix,
denoted by $\breve C_\pm$, satisfying 
\begin{equation}
  \label{eq:transprose-Gamma}
  \breve C_\pm\,\breve\Gamma_A \breve C_\pm^{-1} = \pm
  \breve\Gamma_A{}^\mathrm{T} \,,\quad 
  \breve C_\pm{}^\mathrm{T}= \pm \breve C_\pm\,, \quad \breve
  C_\pm{}^\dagger = \breve C_\pm^{-1}\,, 
\end{equation}
which lead to the following results,
\begin{equation}
  \label{eq:pseudo-chir-symmetry}
  \breve C_\pm\breve\Gamma_{11} \breve C_\pm^{-1} =
  -\breve\Gamma_{11}{}^\mathrm{T}\,,\quad  
 \breve  C_\pm\tilde \gamma \breve C_\pm^{-1} =\pm\tilde\gamma{}^\mathrm{T} \,,\quad
  \breve C_\pm\tilde\Gamma \breve C_\pm^{-1} =\pm \tilde\Gamma{}^\mathrm{T}\,. 
\end{equation}
From the first equation \eqref{eq:pseudo-chir-symmetry}, it follows
that $\breve C_\pm$ satisfy
\begin{equation}
  \label{eq:two-C}
  (\breve C_\pm\breve\Gamma_{11})^\mathrm{T} = \breve\Gamma_{11}{}^\mathrm{T}\,
  \breve C_\pm{}^\mathrm{T}= \mp (\breve C_\pm\,\breve\Gamma_{11})\,,
\end{equation}
so that the two options for the charge conjugation matrix can simply
be related by multiplication with $\Gamma_{11}$.  Furthermore we note
that both $\breve C_\pm\tilde\Gamma$ and $\breve C_\pm\tilde\gamma$ are symmetric
and unitary matrices. Up to a phase factor, these can act as the
charge conjugation matrices in the $5D$ context, as is demonstrated by 
\begin{equation}
  \label{eq:gamma-equiv}
  (\breve C_\pm \tilde\Gamma)
  \hat\gamma_\alpha  (\breve C_\pm \tilde\Gamma)^{-1} =
  \hat\gamma_\alpha{}^\mathrm{T} \,,\quad 
  (\breve C_\pm \tilde\Gamma) \hat\Gamma_a  (\breve C_\pm \tilde\Gamma)^{-1} = 
  \hat\Gamma_a{}^\mathrm{T} \,. 
\end{equation}
Similar relations hold for $(\breve C_\pm\tilde\gamma)$. 

To appreciate the significance of this result, let us consider the definition of the
Dirac conjugate in the $5D$ context, defined by $\psi^\dagger
\mathrm{i}\hat\gamma^0$, where $\hat\gamma^0$ was related to
$\hat\gamma^1$ as explained below \eqref{eq:gamma-sets}. From these
relations it follows straightforwardly that the $5D$ Dirac conjugate
$\bar\psi\big\vert_{5D}$ is related to the $10D$ conjugate according to
\begin{equation}
  \label{eq:new-D-conj}
  \bar \psi\big\vert_{5D}  = \mathrm{i} \bar\psi\big\vert_{10D} \,\tilde\Gamma\,.   
\end{equation}
Consequently, identifying the Majorana conjugate defined in
\eqref{eq:doubling} in the $10D$ context with the one in the $5D$
context, one concludes that the charge conjugation matrix in the
$5D$ context equals
\begin{equation}
  \label{eq:charge-conj-10to5}
  \hat C= \mathrm{i}\tilde\Gamma^\mathrm{T}  \breve C_\pm= \pm\mathrm{i} \breve
  C_\pm \,\tilde\Gamma\,,
\end{equation}
so that $\hat C^{-1}\big[\bar\psi\vert_\mathrm{5D}\big]^\mathrm{T} =
\psi^\mathrm{c}$, and likewise $\psi^\mathrm{T} =
\bar\psi^\mathrm{c}\vert_\mathrm{5D} \,\hat C^{-1}$.  
As a consequence the two commuting sets of $32\times 32$ gamma matrices,
$\hat\gamma_\alpha$ and $\hat\Gamma_a$, satisfy the relations known
from five dimensions,
\begin{equation}
  \label{eq:hat-gamma-c}
  \hat C \hat\gamma_\alpha \hat C^{-1} =\hat
  \gamma_\alpha{}^\mathrm{T}\,, \qquad   
  \hat C \hat\Gamma_a \hat C^{-1} = \hat\Gamma_a{}^\mathrm{T}\,,
  \qquad \hat C^\mathrm{T}= \hat C\,,\qquad \hat C^\dagger= \hat
  C^{-1}\,.  
\end{equation}
This leads to the rearrangement formula,
\begin{equation}
  \label{eq:rearranging}
  \bar\chi\,\Gamma \psi = - \bar\psi^\mathrm{c} \,\hat
  C^{-1}\Gamma^\mathrm{T} \hat C \,\chi^\mathrm{c}\,, 
\end{equation}
where $\Gamma$ denotes any matrix in the spinor space, which in all
cases of interest takes the form a product of gamma matrices
$\hat\Gamma^a$ and $\hat\gamma_\alpha$.  Observe that the new charge
conjugation matrix is not anti-symmetric, as one might expect on the
basis of a single irreducible $5D$ Clifford algebra representation. We
return to this issue shortly.

In this paper we discuss the type-IIB theory where the spinor fields
are chiral and complex. Therefore the above formulae have to be
projected on an eigenspace of $\Gamma_{11}$ and the effective $5D$
gamma matrices defined in \eqref{eq:gamma-sets} are consistent with
the $10D$ chirality constraint on the spinor fields, because they are
proportional to an even number of the original $10D$ gamma
matrices. However, it is important to realize that IIB supergravity
contains independent spinor fields of {\it opposite} chirality, namely
$\psi_M$ and $\lambda$. This leads to a subtlety in view of 
\eqref{eq:5-gamma}, which indicates that different chirality spinors
involve inequivalent gamma matrix representations in $5D$. However,
one has to keep in mind that the chirality assignment can easily be
changed in the $5D$ context by redefining the spinors by
multiplication with one of the matrices \eqref{eq:sub-chiral}.

Let us now assume that we are starting from $10D$ with fermion fields
of {\it positive} chirality. Hence we can choose a Weyl basis where
$\breve\Gamma_{11}$ is diagonal and make use of the fact that it
commutes with the mutually commuting gamma matrices $\hat\gamma_\alpha$ and
$\hat\Gamma_a$. Hence we write
\begin{equation}
  \label{eq:direct-prod-gamma}
  \hat\gamma_\alpha = \sigma_3\otimes\gamma_\alpha \otimes \oneone_{4}\,, \qquad 
  \hat\Gamma_a = \sigma_3\otimes \oneone_{4} \otimes \Gamma_a\,,
\end{equation}
where $\breve\Gamma_{11} =\sigma_3\otimes\oneone_{16}$ and
$\gamma_\alpha$ and $\Gamma_a$ are $4\times 4$ matrices. It then
follows from \eqref{eq:5-gamma} that they define irreducible
representations of the respective Clifford algebras, as
\begin{equation}
  \label{eq:gamma-representations}
    \gamma_{[\alpha} \gamma_\beta \gamma_\gamma
  \gamma_\delta \gamma_{\tau]} =
  \varepsilon_{\alpha\beta\gamma\delta\tau} \,\oneone_{4}\,, \qquad
  \Gamma_{[a} \Gamma_b \Gamma_c
  \Gamma_d \Gamma_{e]} = - 
  \varepsilon_{abcde} \,\oneone_{4}\,.
\end{equation}
The $10D$ chiral spinors thus transform under the direct product group
$\mathrm{Spin}(1,4)\times\mathrm{USp}(4)$, whose generators are
provided by the anti-symmetrized products of gamma matrices,
$\gamma_{\alpha\beta}$ and $\Gamma_{ab}$,
respectively. Correspondingly the charge conjugation matrix $\hat C$
can be written (adjusting possible phase factors) as the direct
product of the two $5D$ {\it anti-symmetric} charge conjugation matrices, 
\begin{equation}
  \label{eq:charge-conj-2}
  \hat C_{(16)} = C \otimes  \Omega_{(4)} \,,
\end{equation}
where $C$ denotes the anti-symmetric charge
conjugation matrix for a $5D$ space-time spinor and $\Omega_{(4)}$ is
the symplectic matrix that is invariant under the $\mathrm{USp}(4)$
R-symmetry. In this case we may write \eqref{eq:hat-gamma-c} as
\begin{equation}
  \label{eq:charge-conj-55}
  C\,\gamma_\alpha C^{-1} = \gamma_\alpha{}^\mathrm{T}\,, \qquad 
  \Omega_{(4)}\,\Gamma_a \Omega_{(4)}^{-1} = \Gamma_a{}^\mathrm{T}\,. 
\end{equation}

However, the chiral spinors are complex which implies that the fields
$(\psi,\psi^\mathrm{c})$, which constitute the 32-component spinor
$\Psi$, can again be rearranged in a pseudo-real form as in
\eqref{eq:pseudo-real-32}. The doubling of field components enables
one to realize the extension of the R-symmetry group from
$\mathrm{USp}(4)\times\mathrm{U}(1)$ to $\mathrm{USp}(8)$.  It then
follows from \eqref{eq:pseudo-real-32} that the extended
$\mathrm{USp}(8)$ invariant tensor must take the form
\begin{equation}
  \label{eq:symplectic-8}
  \Omega= \Omega_{(4)} \otimes \sigma_1\,. 
\end{equation}
Consequently, \eqref{eq:pseudo-real-32} and
\eqref{eq:charge-conj-2} imply the symplectic Majorana condition,
\begin{equation}
  \label{eq:sympl-majorana-5D}
  C^{-1}\bar\Psi^\mathrm{T}  = \Omega\,\Psi\,,  
\end{equation}
where $\Omega$ is an $8\times 8$ anti-symmetric matrix. Both matrices
$C$ and $\Omega$ are anti-symmetric and unitary.

We close this appendix with some additional definitions that will be
useful in the next appendix \ref{App:R-symm-assignm-fermions}. First of
all we write the anti-symmetric tensor $\Omega_{(4)}$ as
$\Omega_{(4)}{}_{IJ}$ and its complex conjugate as
$\bar\Omega_{(4)}{}^{IJ}$, so that
$\Omega_{(4)}{}_{IJ}\,\bar\Omega_{(4)}{}^{JK} =-\delta_I{}^K$, where $I,J,K=1,\ldots,4$. The
gamma matrices $\Gamma_a$ are then written as $\Gamma_a{}^I{}_J$, so
that
\begin{align}
  \label{eq:Omega-4}
  \Omega_{(4)}{}^\mathrm{T} = -\Omega_{(4)}\,,\quad
  (\Omega_{(4)}\,\Gamma_a)^\mathrm{T} =
  -(\Omega_{(4)}\,\Gamma_a)\,,\quad
  (\Omega_{(4)}\,\Gamma_{ab})^\mathrm{T} = 
   (\Omega_{(4)}\,\Gamma_{ab} )\,,
\end{align}
with similar relations for $(\Gamma_a\,\bar \Omega_{(4)})^{IJ}$ and
$(\Gamma_{ab}\,\bar \Omega_{(4)})^{IJ}$.  The six matrices
$\Omega_{(4)}{}_{IJ}$ and $(\Omega_{(4)}\,\Gamma_a)_{IJ}$ form a
complete set of $4\times4$ anti-symmetric matrices, and the ten
matrices $(\Omega_{(4)}\,\Gamma_{ab})_{IJ}$ a complete set of
$4\times4$ symmetric matrices  This leads to the
completeness relations
\begin{align}
  \label{eq:complete-anti-symm}
  \Omega_{(4)}{}_{IJ} \;\bar \Omega_{(4)}{}^{KL} +
  (\Omega_{(4)}\,\Gamma_a)_{IJ}\; (\Gamma^a\, \bar \Omega_{(4)})^{KL}
  = 4\, \delta_{[I}{\!}^{K}\, \delta_{J]}{\!}^L \,, \nonumber\\
  (\Omega_{(4)}\,\Gamma_{ab})_{IJ}\; (\Gamma^{ab}\, \bar \Omega_{(4)})^{KL}
  = 8\, \delta_{(I}{\!}^{K}\, \delta_{J)}{\!}^L \,. 
\end{align}

\section{The R-symmetry group and the fermion representations}
\label{App:R-symm-assignm-fermions}
\setcounter{equation}{0}
In the previous appendix we considered a $10D$ chiral spinor and
described its properties in the context of a product of a
five-dimensional space-time and a five-dimensional internal space. The
gamma matrices and the charge conjugation matrices were decomposed
accordingly. The $10D$ spinors then transform under a
subgroup of the original $\mathrm{Spin}(1,9)$ transformations
consisting of the $\mathrm{Spin}(1,4)$ group associated with the $5D$
space-time and the group $\mathrm{USp}(4)$ associated with the
internal space.

However, $\mathrm{USp}(4)$ is not the full automorphism group (or
R-symmetry group) of the eight symplectic Majorana spinors. This group
is actually equal to $\mathrm{USp}(8)$, which consists of the unitary
transformations that leave the symplectic and unitary tensor $\Omega$,
invariant. The generators of this group can be easily identified in
terms of direct products of the $4\times 4$ gamma matrices $\Gamma_a$,
defined in \eqref{eq:direct-prod-gamma}, their anti-symmetrized
products $\Gamma_{ab}$ and the unit matrix $\oneone_4$, and the
$2\times2$ matrices $(\oneone_2, \sigma_1,\sigma_2,\sigma_3)$. As a
result one derives all the 36 generators of the Lie algebra
$\mathfrak{usp}(8) =\mathfrak{su}(8) \cap \mathfrak{sp}(8,
\mathbb{R})$, by constructing the complete set of traceless and
anti-hermitian matrices that preserve the symplectic form $\Omega$,
\begin{align} 
  \label{eq:usp8gen}
& T \equiv \mathrm{i} \oneone_4 \otimes \sigma_3 , \qquad 
T_a \equiv \mathrm{i} \Gamma_a \otimes \sigma_3 , \nonumber \\
& T_{ab}{}^0 \equiv  \Gamma_{ab} \otimes \oneone_2 , \qquad 
T_{ab}{}^1 \equiv \Gamma_{ab} \otimes \sigma_1 , \qquad 
T_{ab}{}^2 \equiv \Gamma_{ab} \otimes \sigma_2 \,.
\end{align}
As expected these matrices close under commutation,
\begin{align}
  \label{eq:usp-commutators}
  &[T , T_{ab}{}^1] = -2\, T_{ab}{}^2 , \quad 
  [T , T_{ab}{}^2] = 2\, T_{ab}{}^1 , \quad  [T_a , T_b] = -2\,T_{ab}{}^0 , \nonumber \\[8pt]
  & [T_a , T_{bc}{}^0] = 4\, \delta_{a[b} \,T_{c]} , \quad 
  [T_a , T_ {bc}{}^1 ] = \varepsilon_{abcde}\, T^{de\,2} , \quad 
  [T_a , T_{bc}{}^{2} ] = -\varepsilon_{abcde}\, T^{de\,1} , \nonumber \\[8pt]
  & [T_{ab}{}^0, T^{cd\,0}] = -8\,\delta_{[a}{}^{[c}\, T_{b]}{}^{d]}{}^0  , \quad 
  [T_{ab}{}^0 , T^{cd\,1} ] = -8\, \delta_{[a}{}^{[c}\, T_{b]}{}^{d]}{}^1  , \quad 
  [T_{ab}{}^0 , T^{cd\,2} ] = -8\, \delta_{[a}{}^{[c}\, T_{b]}{}^{d]}{}^2 , \nonumber \\[8pt]
  & [T_{ab}{}^1 , T^{cd\,1}] = -8 \,\delta_{[a}{}^{[c}\, T_{b]}{}^{d]}{}^0 ,
  \quad 
  [T_{ab}{}^2 , T^{cd \,2} ] = -8\,\delta_{[a}{}^{[c}\, T_{b]}{}^{d]}{}^0 , 
   \nonumber \\[8pt] 
  & [T_{ab}{}^1 , T^{cd\,2} ] = -2\, \varepsilon_{abcde}\, T^e .
\end{align}
Observe that the generators are anti-hermitian and the structure
constants are real, in agreement with $\mathfrak{usp}(8)$ being a real
form. The $T_{ab}{}^0$ are the generators of the group
$\mathrm{USp}(4)\cong\mathrm{SO}(5)$. When extended with the
generators $T_a$ one obtains the group
$\mathrm{SU}(4)\cong\mathrm{SO}(6)$ which obviously commutes with the
generator $T$. As we will exhibit later, $T$ corresponds to the
$\mathrm{SO}(6)$ chirality operator. The latter commutes with the
$\mathrm{U}(1)$ transformations of the original $10D$ theory. Clearly
$\mathrm{SU}(4)\times\mathrm{U}(1)$ is a maximal subgroup of
$\mathrm{USp}(8)$.

A chiral $10D$ spinor $\Psi$ can be decomposed into eight $5D$
symplectic Majorana spinors $\psi{\!}^{A}$, where $A=1,\ldots,8$. Note
that from now on we employ indices $A,B,\ldots$ to label the
symplectic Majorana spinors. The same indices were previously used in
the $10D$ theory (in particular in section \ref{sec:results-2B-sugra}
and appendix~\ref{App:red-spinors-gamma-matrices}) to denote the $10D$
tangent-space components. This should not give rise to confusion in
view of the fact that the $10D$ tangent space will no longer play a
role in what follows.  In view of the direct-product structure
indicated in \eqref{eq:usp8gen} the indices $A$ can be written as
index pairs $A=(I\alpha)$, where $I=1,\ldots,4$ are $\mathrm{USp}(4)$
indices and $\alpha=+,-$. Here $\alpha=+$ ($\alpha=-$) indicates that
we are dealing with a chiral (anti-chiral) $\mathrm{SO}(6)$ spinor
with positive (negative) $\mathrm{U}(1)$
charge\footnote{
  We ignore the various redefinitions of the spinors that are
  considered in section~\ref{sec:first-field-redefin}. These
  redefinitions should be performed before making the decompositions
  described in this appendix, but their precise details are not
  relevant here.  }. 
Based on this direct-product structure the eight $5D$ gravitini
$\psi_\mu{\!}^A$ transform under the $\mathrm{USp}(8)$ R-symmetry
group with generators that can be read off directly from
\eqref{eq:usp8gen}.  It is thus clear that that each of the
$\psi_\mu{\!}^A$ decomposes into two components of opposite
$\mathrm{SO}(6)$ chirality which therefore carry opposite values of the
$\mathrm{U}(1)$ charge. This fact enables us to unambiguously identify
the various chiral fermionic components on the basis of this
charge. Furthermore we note that the symplectic Majorana constraint
\eqref{eq:sympl-majorana-5D} relates fermion fields of opposite
$\mathrm{U}(1)$ charges, which is consistent with the form of the
symplectic matrix $\Omega$ defined in \eqref{eq:symplectic-8}. For
instance, for the gravitini we have
\begin{equation}
  \label{eq:gravitini}
  C^{-1} \bar\psi_{\mu\,I+}{}^\mathrm{T} = (\Omega_{(4)})_{IJ}   \, \psi_\mu{}^{J-} \,, 
\end{equation}
where $C$ denotes the charge conjugation matrix associated with the
five-dimensional space-time. 

Let us now turn to the spin-1/2 fermions which originate from the
fields $(\psi_a,\psi_a{\!}^\mathrm{c})$ and $\lambda,
\lambda^\mathrm{c}$ and constitute 48 independent $5D$ symplectic
Majorana spinors. From $5D$ maximal supergravity we know that these
spinors can be written as a symplectic traceless, fully anti-symmetric
three-rank $\mathrm{USp}(8)$ tensor $\chi^{ABC}$. This is consistent
with the fact that the spin-$1/2$ fields carry $\mathrm{U}(1)$ charges
$\pm1/2$ and $\pm3/2$. We intend to determine the (linear) relation
between the components of $\chi^{ABC}$ and the fields $\psi_a{}^A$ and
$\lambda^A$ by making use of the fact that these fields do all
transform consistently under the action of the maximal subgroup
$\mathrm{SU}(4)\times\mathrm{U}(1)$ of $\mathrm{USp}(8)$. To see how
this works let us present the branching of $\psi_\mu{}^A$ and
$\chi^{ABC}$ under the $\mathrm{SU}(4)\times\mathrm{U}(1)$ subgroup,
\begin{align}
  \label{eq:8+48-branching}
  \boldsymbol{8}\;
  \stackrel{\mathrm{SU}(4)\times\mathrm{U}(1)}{\longrightarrow} \; &\,
  \big(\boldsymbol{4},\tfrac12 \big)
  \oplus \big (\overline{\boldsymbol{4}},-\tfrac12 \big) \,, \nonumber\\[1mm]
  \boldsymbol{48} \;
  \stackrel{\mathrm{SU}(4)\times\mathrm{U}(1)}{\longrightarrow} \; &\,
  \big(\overline{\boldsymbol{4}},\tfrac32 \big) \oplus \big
  (\boldsymbol{4},-\tfrac32\big) \oplus
  \big(\boldsymbol{20},\tfrac12\big) \oplus
  \big(\overline{\boldsymbol{20}},-\tfrac12 \big)\,.
\end{align}
The chiral representations on the right-hand side are now
unambiguously identified by the corresponding $\mathrm{U}(1)$ charge,
so that they must correspond to the fields $\psi_\mu$,
$\psi_\mu{}^\mathrm{c}$, and $\lambda$, $\lambda^\mathrm{c}$, $\psi_a$
and $\psi_a{}^\mathrm{c}$, respectively.\footnote{
  A vector-spinor in odd dimension $d$ can consistently transform
  under $\mathrm{SO}(d+1)$ by describing it as an irreducible chiral
  vector-spinor in $d+1$ dimensions. }  
To determine the precise relation, we again write the indices of
the symplectic Majorana field $\chi^{ABC}$ by employing the direct-product
representation introduced before, with $A=I\alpha$, $B=J\beta$ and
$C=K\gamma$.  Since $\alpha,\beta,\gamma$ take only two possible index
values, at least two of them must be equal. Hence we may distinguish the
fields $\chi^{I\pm\,J\pm\,K\pm}$, which must be fully anti-symmetric
in the indices $I,J,K$, and thus correspond to $4+4$ symplectic
Majorana fields, and the fields $\chi^{I\pm\,J\pm\,K\mp}$, which are
anti-symmetric in the indices $I,J$, and thus define $24+24$
fields. The remaining fields $\chi^{I\alpha\,J\beta\,K\gamma}$ follow
then from imposing the overall anti-symmetry. However, unlike the
fields $\chi^{I\pm\,J\pm\,K\pm}$, the fields $\chi^{I\pm\,J\pm\,K\mp}$
are not manifestly traceless with respect to contractions with the
symplectic matrix $\Omega$. This implies that one must impose the
additional condition
\begin{equation}
  \label{eq:constraint-chi-1/2}
  \chi^{I\pm\,J\pm\,K\mp}\,(\Omega_{(4)})_{JK} =0\,, 
\end{equation}
which reduces the number of independent spinors in this sector to
$20+20$, as it should. 

Let us first analyze the correspondence for the spinors $\chi^{ABC}$
with positive $\mathrm{U}(1)$ charge $+\tfrac32$, which must be
linearly related to the $10D$ spinor $\lambda$. The former must be
given by $\chi^{I+\,J+\,K+}$, which must necessarily be fully
anti-symmetric in $\mathrm{USp}(4)$ indices.  From \eqref{eq:Omega-4}
one then concludes that $\chi^{I+\,J+\,K+}$ can be decomposed into
two terms, namely $(\bar\Omega_{(4)})^{[IJ}\,
(\lambda^\mathrm{c}\,)^{K]}$ and $(\Gamma^a\bar\Omega_{(4)})^{[IJ}
(\Gamma_a \lambda^\mathrm{c}\,)^{K]}$.  However, the first
completeness relation \eqref{eq:complete-anti-symm} leads to
\begin{equation}
  \label{eq:dependent-rel}
  (\Gamma^a\bar\Omega_{(4)})^{IJ}   (\Gamma_a\psi)^{K} = -4\,
  (\bar\Omega_{(4)})^{K[I} \, \psi^{J]}  -(\bar\Omega_{(4)})^{IJ}\, \psi^K\,,
\end{equation}
for an arbitrary $\mathrm{USp}(4)$ spinor $\psi$, so that the two terms are in fact
related. Hence we may adopt the following ansatz,
\begin{equation}
  \label{eq:chi+++}
  \chi^{I+\,J+\,K+} = c_{3/2}\, (\bar\Omega_{(4)})^{[IJ}\,
  \lambda^{K]} \,,
\end{equation}
where $c_{3/2}$ is a complex proportionality factor which is
undetermined at this stage. The fields with charge $-\tfrac32$ are
then defined through the symplectic Majorana condition,
\begin{align}
  \label{eq:chi---}
  \chi^{I-\,J-\,K-} \equiv&\, -(\bar\Omega_{(4)})^{lL}\,
  (\bar\Omega_{(4)})^{JM}\, (\bar\Omega_{(4)})^{KN}\, C^{-1}\,
  \bar\chi_{L+\,M+\,N+} {}^\mathrm{T} \nonumber\\
  = &\,  \bar c_{3/2} \, (\bar\Omega_{(4)})^{[IJ}\,
  (\lambda^\mathrm{c})^{K]} \,.
\end{align}

The relation between the spinors $\chi^{I+\,J+\,K-}$ and $\psi_a$ with
$\mathrm{U}(1)$ charge $+\tfrac12$ is more subtle. First consider
the following ansatz,
\begin{equation}
  \label{eq:chi++-}
  \chi^{I+\,J+\,K-} = c_{1/2}\, \big[  (\Gamma^a\bar\Omega_{(4)})^{IJ}\,
  (\hat\psi_a)^{K}  - (\bar\Omega_{(4)})^{IJ}\,
  (\Gamma^a\,\hat\psi_a)^{K}\big] \,,
\end{equation}
where $\hat \psi_a= \psi_a + \alpha \, \Gamma_a\Gamma^b\psi_b$ with
$\alpha$ an undetermined parameter, so that we are now dealing with two
new parameters, $c_{1/2}$ and $\alpha$. The linear combination in \eqref{eq:chi++-} is chosen such that
the $\mathrm{USp}(8)$ constraint \eqref{eq:constraint-chi-1/2} is
satisfied. An alternative version of \eqref{eq:chi++-}, which is the
one that we will actually use, is
\begin{align}
  \label{eq:chi++-2}
  \chi^{I+\,J+\,K-} =&\, c_{1/2}\, \big[  (\Gamma^a\bar\Omega_{(4)})^{IJ}\,
  (\psi_a)^{K}  - (\bar\Omega_{(4)})^{IJ}\,
  (\Gamma^a\,\psi_a)^{K}\big] \nonumber\\
  &\, +c_{1/2}' \,\big[  (\bar\Omega_{(4)})^{IJ}\, (\Gamma^a\psi_a)^K
  +\tfrac23  (\bar\Omega_{(4)})^{K[I}\,(\Gamma^a\psi_a)^{J]}\big]  \,,
\end{align}
but also this expression can be rewritten by making use of the identity
\begin{equation}
  \label{eq:vector-dependentl}
  (\Gamma^a\,\bar\Omega_{(4)})^{[IJ}\, (\psi_a)^{K]}=
  -  (\bar\Omega_{(4)})^{[IJ}   (\Gamma^a\,\psi_a)^{K]}  \,. 
\end{equation}

As before we define the spinor components with $\mathrm{U}(1)$ charge
$-\tfrac12$ by
\begin{align}
  \label{eq:chi--+}
  \chi^{I-\,J-\,K+} \equiv&\, -(\bar\Omega_{(4)})^{lL}\,
  (\bar\Omega_{(4)})^{JM}\, (\bar\Omega_{(4)})^{KN}\, C^{-1}\,
  \bar\chi_{L+\,M+\,N-} {}^\mathrm{T} \nonumber\\
  = &\, \bar c_{1/2} \, \big[ (\Gamma^a\bar\Omega_{(4)})^{IJ}\,
  (\psi_a{}^\mathrm{c})^{K} - (\bar\Omega_{(4)})^{IJ}\,
  (\Gamma^a\,\psi_a{}^\mathrm{c})^{K}\big] \nonumber\\
  &\, +\bar c'{\!}_{1/2} \,\big[ (\bar\Omega_{(4)})^{IJ}\,
  (\Gamma^a\psi_a{}^\mathrm{c})^K +\tfrac23
  (\bar\Omega_{(4)})^{K[I}\,(\Gamma^a\psi_a{}^\mathrm{c})^{J]}\big]
  \,.
\end{align}
Hence we have obtained the linear relation between $\chi^{ABC}$ and
the original $10D$ spinors, depending on three unknown complex
constants, $c_{3/2}$, $c_{1/2}$ $c'{\!}_{1/2}$. Their values are
determined in section \ref{sec:gen-vielbeine}, as we will be
discussing at the end of this appendix.

We will now merge the chiral and anti-chiral spinors with opposite
$\mathrm{U}(1)$ charges into eight-component symplectic Majorana
spinors. In that case it is convenient to introduce $\mathrm{SO}(6)$
gamma matrices and chiral projection operators. The $8\times 8$ gamma
matrices $(\boldsymbol{\Gamma}_{\hat a})^A{\!}_B$, where $\hat
a=1,\ldots,6$, are defined in terms of direct products of $4\times4$
and $2\times2$ matrices, just as in \eqref{eq:usp8gen},
\begin{equation}
  \label{eq:def-gamma-6}
  \boldsymbol{\Gamma}_a\equiv \Gamma_a\otimes \sigma_1\,,\qquad
  \boldsymbol{\Gamma}_6\equiv \oneone_4\otimes\sigma_2\,. 
\end{equation}
These (hermitian) gamma matrices satisfy the Clifford property 
\begin{equation}
  \label{eq:clifford-6}
  \big\{ \boldsymbol{\Gamma}_{\hat a}~,\boldsymbol{\Gamma}_{\hat b}\big\}  =
  2\,\delta_{\hat a \hat b} \, \oneone_8\,.
\end{equation}
and satisfy the following charge-conjugation properties,
\begin{equation}
  \label{eq:charge-conj-6}
  \Omega \,\boldsymbol{\Gamma}_{\hat a}\, \Omega^{-1} =
  \boldsymbol{\Gamma}_{\hat a}{}^\mathrm{T} \,,\;\mbox{with}\,\;
  \Omega^\mathrm{T} = -\Omega\,,\quad \Omega^{-1} = -\bar\Omega\,, 
\end{equation}
where the anti-symmetric charge conjugation matrix $\Omega_{AB}$ was
defined in \eqref{eq:symplectic-8}. The chirality operator
$\boldsymbol{\Gamma}_7$ is obtained in the standard way,
\begin{equation}
  \label{eq:chiral-6}
  \boldsymbol{\Gamma}_{[\hat a} \, \boldsymbol{\Gamma}_{\hat b}
  \cdots\boldsymbol{\Gamma}_{\hat f]} =-\mathrm{i}  \varepsilon_{\hat a\hat
    b\hat c\hat d\hat e\hat f}  \,
  \boldsymbol{\Gamma}_7 \,, \quad \mbox{where}\,\;
  \boldsymbol{\Gamma}_7 = \oneone_4\otimes\sigma_3\,. 
\end{equation}
Observe that $\boldsymbol{\Gamma}_7$ is hermitian and behaves under
charge conjugation as $\Omega \,\boldsymbol{\Gamma}_7\, \Omega^{-1} =-
\boldsymbol{\Gamma}_7{}^\mathrm{T}$. Furthermore
$\boldsymbol{\Gamma}_7$ coincides with the $\mathrm{U}(1)$ charge that
was already present in the original $10D$ theory. 

The gamma matrices $\boldsymbol{\Gamma}_{\hat a}$ and their multiple
anti-symmetrized products define a complete basis for matrices in the 8-dimensional
spinor space. They can conveniently be decomposed into 28 anti-symmetric
matrices $\Omega$, $\Omega \boldsymbol{\Gamma}_{\hat a}$, $\Omega
\boldsymbol{\Gamma}_{\hat a}\boldsymbol{\Gamma}_7$ and $\Omega
\boldsymbol{\Gamma}_{\hat a\hat b}\boldsymbol{\Gamma}_7$, and 36 symmetric
matrices $\Omega \boldsymbol{\Gamma}_7$, $\Omega
\boldsymbol{\Gamma}_{\hat a\hat b}$ and
$\Omega\boldsymbol{\Gamma}_{\hat a\hat b\hat c}$. The
latter are related to the anti-hermitian generators of
$\mathrm{USp}(8)$ that were already defined in \eqref{eq:usp8gen},
\begin{equation}
  \label{eq:usp-gamma-6}
  \begin{array}{rcl}
    T \!\!\!&=& \!\!\!\mathrm{i} \boldsymbol{\Gamma}_7\,,\\
     ~&~&~
    \end{array}   
    \qquad
   \begin{array}{rcl}
     T_a \!\!\! &=&\!\!\! \boldsymbol{\Gamma}_{a6}\,,\\
    T_{ab}{}^0 \!\!\!&=&\!\!\! \boldsymbol{\Gamma}_{ab}\,,
    \end{array}
    \qquad
  \begin{array}{rcl}
    T{\!\!}_{ab}{}^1  \!\!\! &=&\!\!\! \tfrac16\varepsilon_{abcde6} \,
  \boldsymbol{\Gamma}^{cde}\,,\\ 
    T{\!\!}_{ab}{}^2 \!\!\!&=&\!\!\! \boldsymbol{\Gamma}_{ab6}\,.     \end{array}
\end{equation}

We have now obtained a parametrization of the relation between the
fields $\chi^{ABC}$ and the fields $\lambda$, $\lambda^\mathrm{c}$,
$\psi_a$ and $\psi_a{}^\mathrm{c}$ originating from the $10D$ theory
in terms of (anti-)chiral components. This relation is in accordance
with the $\mathrm{SU}(4)\times\mathrm{U}(1)$ branching of the spinor
fields presented in \eqref{eq:8+48-branching}. The resulting
expressions for given charges were given in \eqref{eq:chi+++},
\eqref{eq:chi---}, \eqref{eq:chi++-2}, \eqref{eq:chi--+}, which can be
converted in terms of the $\mathrm{SO}(6)$ gamma matrices
$\boldsymbol{\Gamma}_{\hat a}$. Since we have established this
relation for chiral and anti-chiral components separately, it is
convenient to introduce chiral projection operators
\begin{equation}
  \label{eq:so(6)-proj}
  \mathbb{P}_\pm=
  \tfrac12\big(\oneone\pm\boldsymbol{\Gamma}_7 \big) \,. 
\end{equation}
The spinor $\chi^{ABC}$ is subsequently decomposed in tri-spinors with
all possible chiralities,  
\begin{align}
  \label{eq:fulll-chi}
  \chi^{ABC} = \chi^{ABC}_{+++} + \chi^{ABC}_{---} +
  \chi^{ABC}_{++-}  +\chi^{ABC}_{+-+}+\chi^{ABC}_{-++}
  +\chi^{ABC}_{--+} +\chi^{ABC}_{-+-} +\chi^{ABC}_{+--} \, .
\end{align}
For the spinors with $\mathrm{U}(1)$ charge equal to $+3/2$ and $+1/2$
we derive, respectively, 
\begin{align}
  \label{eq:so(6)-ansatz}
  \chi^{ABC}{}_{(+++)} = &\, \mathrm{i}\,
  c_{3/2}\,\mathbb{P}_+{\!}^A{\!}_D \,\mathbb{P}_+{\!}^B{\!}_E
  \,\mathbb{P}_+{\!}^C{\!}_F  
  \,\big[\boldsymbol{\Gamma}_7 \boldsymbol{\Gamma}_6
  \,\bar\Omega\big]^{[DE}
   \,\lambda^{F]}\,, \nonumber\\
   \chi^{ABC}{}_{(++-)} = &\, \mathrm{i}\, c_{1/2}
   \,\mathbb{P}_+{\!}^A\!{}_D \,\mathbb{P}_+{\!}^B{}_E
   \,\mathbb{P}_-{\!}^C{\!}_F \Big[ \big[ \boldsymbol{\Gamma}^a \,
   \bar\Omega\big]^{DE} \, \big( \boldsymbol{\Gamma}_7
   \boldsymbol{\Gamma}_6 \,\psi_a\big){}^{F} -
   \big[\boldsymbol{\Gamma}_7 \boldsymbol{\Gamma}_6 \,
   \bar\Omega\big]^{DE} \,\big(
   \,\boldsymbol{\Gamma}^{a}\,\psi_a\big)^F \Big]
   \nonumber\\
   &\,+ \mathrm{i}\, c'{\!}_{1/2} \,\mathbb{P}_+{\!}^A{\!}_D
   \,\mathbb{P}_+{\!}^B{\!}_E \,\mathbb{P}_-{\!}^C{\!}_F
   \Big[\big[\boldsymbol{\Gamma}_7 \boldsymbol{\Gamma}_6 \,
   \bar\Omega\big]^{DE} \,\big(
   \,\boldsymbol{\Gamma}^{a}\,\psi_a\big)^F -\tfrac23\,
   \bar\Omega^{F[D}
   \,\big[\boldsymbol{\Gamma}_7\boldsymbol{\Gamma}^{a6}
   \psi_a\big]^{E]} \Big]\,,
\end{align}
where the spinors $\lambda$ and $\psi_a$ are now 8-component spinors
consisting of $(\lambda,\lambda^\mathrm{c})$ and $(\psi_a,
\psi_a{\!}^\mathrm{c})$. The labels $(+++)$ and $(++-)$ on the
left-hand side indicate how the indices are contracted with the chiral
projectors. Note that the combinations $(+-+)$ and $(-++)$ are related
upon interchanging the indices $A,B,C$ correspondingly. The
corresponding spinors with charges $-3/2$ and $-1/2$ read the same
with $c_{3/2}$, $c_{1/2}$ and $c'{\!}_{1/2}$ replaced by their complex
conjugates and with opposite projectors.  

Confronting the above decompositons to the equations
\eqref{eq:psi-lambda=chi} uniquely determines the 
three constants to $c_{3/2} = -\tfrac3{4}$, $c_{1/2}= -\tfrac1{4}$ and
$c^\prime{\!}_{1/2} = -\tfrac12$. The corresponding expression for
$\chi^{ABC}$ equals 
\begin{align}
  \label{eq:solution-for-chi}
  \chi^{ABC}=&\, -\tfrac3{8} \mathrm{i} \Big[
  \big(\boldsymbol{\Gamma}_6\,\bar\Omega\big)^{[AB}
  \,\big(\boldsymbol{\Gamma}_7 \lambda\big)^{C]} +
  \big(\boldsymbol{\Gamma}_7
  \boldsymbol{\Gamma}_6\,\bar\Omega\big)^{[AB} \,\lambda^{C]} \Big]
  \nonumber\\
  &\, -\tfrac38\mathrm{i}
  \Big[\big(\boldsymbol{\Gamma}^a\,\bar\Omega\big)^{[AB}
  \,\big(\boldsymbol{\Gamma}_7\boldsymbol{\Gamma}_6 \psi_a \big)^{C]} 
  - \big(\boldsymbol{\Gamma}_7 \boldsymbol{\Gamma}^a\,
  \bar\Omega\big)^{[AB}
  \,\big(\boldsymbol{\Gamma}_6 \psi_a \big)^{C]} 
  \Big]\nonumber\\
  &\, -\tfrac38\mathrm{i} \Big[
  \big(\boldsymbol{\Gamma}_7\boldsymbol{\Gamma}_6\,\bar\Omega\big)^{[AB}   
  \,\big(\boldsymbol{\Gamma}^a \psi_a \big)^{C]} 
  - \big(\boldsymbol{\Gamma}_6 \,\bar\Omega\big)^{[AB}
  \,\big(\boldsymbol{\Gamma}_7 \boldsymbol{\Gamma}^a  \psi_a
  \big)^{C]}  \Big]\nonumber\\
  &\, -\tfrac12\mathrm{i} \,
     \bar\Omega^{[AB} \,\big(\boldsymbol{\Gamma}_7
     \boldsymbol{\Gamma}_6 \boldsymbol{\Gamma}^a\psi_a \big){}^{C]}
     \,. 
\end{align} 
Here we shoud stress that this form of the solution is not unique as
it can be rewritten by Fierz reordering. In section
\ref{sec:gen-vielbeine} we have presented an equivalent but shorter
expression.

\end{appendix}

\providecommand{\href}[2]{#2}

\end{document}